\pdfoutput=1
\RequirePackage{ifpdf}
\ifpdf % We are running pdfTeX in pdf mode
\documentclass[pdftex]{sigma}
\else
\documentclass{sigma}
\fi

\usepackage{xspace}
\usepackage{easybmat}

\usepackage{ytableau}
\usepackage{tikz}
\usetikzlibrary{decorations.pathreplacing}

\newcommand{\lamdot}{\textbullet}

\newcommand{\J}{J}
\newcommand{\F}{{F}}

\newcommand{\rr}{\mathfrak{r}}
\newcommand{\BR}{\mathbb{R}}

\newcommand{\K}{K}
\newcommand{\Kb}{\bar K}

\newcommand{\ID}{I}
\newcommand{\BC}{\mathbb{C}}

\newcommand{\fg}{\mathfrak{g}}
\newcommand{\fn}{\mathfrak{n}}
\newcommand{\fh}{\mathfrak{h}}
\newcommand{\fb}{\mathfrak{b}}
\newcommand{\fgl}{\mathfrak{gl}}

\newcommand{\ep}{\epsilon}
\newcommand{\vep}{\varepsilon}

\newcommand{\mass}{x}

\newcommand{\Llm}{L_{\lambda,\underline{x},\mu}(x;\underline{p},\underline{q})}
\newcommand{\Llmr}{L_{\lambda,\underline{x},\mu}(x;\underline{p},\underline{q})}
\newcommand{\Lm}{L_{\mu}(x;\underline{p},\underline{q})}

\newcommand{\Bun}{\mathrm{Bun}}
\newcommand{\mHiggs}{\mathrm{mHiggs}}
\DeclareMathOperator{\tr}{tr}
\DeclareMathOperator{\End}{End}
\DeclareMathOperator{\diag}{diag}

\numberwithin{equation}{section}

\newcommand{\pt}{\hat{P}}

\newcommand{\parti}{
\begin{tikzpicture}
\foreach \a in {1,2,3,4,5} {
 \begin{scope}[shift={(0.5*\a,-3)}]
 \draw (0,0) rectangle (0.5,0.5);
 \end{scope}
 }
 \foreach \a in {1,2,3,4,5} {
 \begin{scope}[shift={(0.5*\a,-2.5)}]
 \draw (0,0) rectangle (0.5,0.5);
 \end{scope}
 }
 \foreach \a in {1,2} {
 \begin{scope}[shift={(0.5*\a,-2)}]
 \draw (0,0) rectangle (0.5,0.5);
 \end{scope}
 }
 \foreach \a in {1,2} {
 \begin{scope}[shift={(0.5*\a,-1.5)}]
 \draw (0,0) rectangle (0.5,0.5);
 \end{scope}
 }
 \foreach \a in {1} {
 \begin{scope}[shift={(0.5*\a,-1)}]
 \draw (0,0) rectangle (0.5,0.5);
 \end{scope}
 }
 \foreach \a in {1} {
 \begin{scope}[shift={(0.5*\a,-0.5)}]
 \draw (0,0) rectangle (0.5,0.5);
 \end{scope}
 }
 \foreach \a in {1} {
 \begin{scope}[shift={(0.5*\a,0)}]
 \draw (0,0) rectangle (0.5,0.5);
 \end{scope}
 }
 \node [below right] at (0.5,1) {$m_7=1$};
 \node [below right] at (1,-0.5) {$m_4=1$};
 \node [below right] at (1.5,-1.5) {$m_2=3$};
\end{tikzpicture}}

\newcommand{\plot}{
\begin{tikzpicture}[scale=1.3]
 \draw[step=0.5cm,lightgray,very thin] (-0.5,-2) grid (9,4);
 \draw [thick, ->] (-0.5,-2) -- (-0.5,4);
 \draw [thick, ->] (-0.5,-2) -- (9,-2);
 \node [above] at (-0.5,4) {j};
 \node [right] at (9,-2) {i};
 \node [below] at (3,-2) {$\nu^t_1 $};
 \node [below] at (8,-2) {$r=|\nu|$};
 \node [below left] at (-0.5,-2) {$0$};
\draw[black,very thick](-0.5,-2)--(0.5,2)--(1.5,3)--(3,3)--(8,-2);
 \node [above left] at (0.5,2) {$m_2$};
 \node [above left] at (1.5,3) {$m_4$};
 \node [above right] at (3,3) {$m_7$};

\foreach \a in {0,1,2,3,4,5,6,7,8,9,10,11,12,13} {
 \begin{scope}[shift={(0.5*\a,-0.5)}]
 \node at (0,-0.01) {\textcolor{black}{\lamdot}};
 \end{scope}
 }

\foreach \a in {0,1,2,3,4,5,6,7,8,9,10,11,12,13,14} {
 \begin{scope}[shift={(0.5*\a,-1)}]
 \node at (0,-0.01) {\textcolor{black}{\lamdot}};
 \end{scope}
 }

\foreach \a in {0,1,2,3,4,5,6,7,8,9,10,11,12,13,14,15} {
 \begin{scope}[shift={(0.5*\a,-1.5)}]
 \node at (0,-0.01) {\textcolor{black}{\lamdot}};
 \end{scope}
 }

\foreach \a in {-1,0,1,2,3,4,5,6,7,8,9,10,11,12,13,14,15,16} {
 \begin{scope}[shift={(0.5*\a,-2)}]
 \node at (0,-0.01) {\textcolor{black}{\lamdot}};
 \end{scope}
 }
\foreach \a in {0,1,2,3,4,5,6,7,8,9,10,11,12} {
 \begin{scope}[shift={(0.5*\a,0)}]
 \node at (0,-0.01) {\textcolor{black}{\lamdot}};
 \end{scope}
 }
 \foreach \a in {1,2,3,4,5,6,7,8,9,10,11} {
 \begin{scope}[shift={(0.5*\a,0.5)}]
 \node at (0,-0.01) {\textcolor{black}{\lamdot}};
 \end{scope}
 }
 \foreach \a in {1,2,3,4,5,6,7,8,9,10} {
 \begin{scope}[shift={(0.5*\a,1)}]
 \node at (0,-0.01) {\textcolor{black}{\lamdot}};
 \end{scope}
 }
 \foreach \a in {0,1,2,3,4,5,6,7,8} {
 \begin{scope}[shift={(0.5*\a+0.5,1.5)}]
 \node at (0,-0.01) {\textcolor{black}{\lamdot}};
 \end{scope}
 }

 \foreach \a in {0,1,2,3,4,5,6,7} {
 \begin{scope}[shift={(0.5*\a+0.5,2)}]
 \node at (0,-0.01) {\textcolor{black}{\lamdot}};
 \end{scope}
 }

 \foreach \a in {0,1,2,3,4,5} {
 \begin{scope}[shift={(0.5*\a+1,2.5)}]
 \node at (0,-0.01) {\textcolor{black}{\lamdot}};
 \end{scope}
 }
 \foreach \a in {-2,-1,0,1} {
 \begin{scope}[shift={(0.5*\a+2.5,3)}]
 \node at (0,-0.01) {\textcolor{black}{\lamdot}};
 \end{scope}
 }

\end{tikzpicture}
}
\newcommand{\quiver}{
\begin{tikzpicture}
\foreach \a in {1,2,3,4,5,6,7,8,9,10,11,12,13,14,15} {
 \begin{scope}[shift={(0.7*\a,0)}]
 \draw (0.3*\a,0) circle (0.3cm);
 \draw[black,thick] (0.3*\a+0.3,0)--(0.3*\a+0.7,0);
 \node [below] at (0.3*\a,-0.5) {$\a$};
 \end{scope}
 }
 \draw (16,0) circle (0.3cm);
 \node [below] at (16,-0.5) {$16$};
 \foreach \a in {10,9,8,7,6,5,4,3,2,1} {
 \begin{scope}[shift={(17-\a,0)}]
 \node at (0,0) {$\a$};
 \end{scope}
 }
 \node at (1,0) {$4$};
 \node at (2,0) {$8$};
 \node at (3,0) {$9$};
 \node at (4,0) {$10$};
 \node at (5,0) {$10$};
 \node at (6,0) {$10$};
 \draw[black,thick] (2,0.3) -- (2,0.7);
 \draw (1.7,0.7) rectangle (2.3,1.3);
 \node at (2,1) {$3$};
 \draw[black,thick] (4,0.3) -- (4,0.7);
 \draw (3.7,0.7) rectangle (4.3,1.3);
 \node at (4,1) {$1$};
 \draw[black,thick] (7,0.3) -- (7,0.7);
 \draw (6.7,0.7) rectangle (7.3,1.3);
 \node at (7,1) {$1$};
\end{tikzpicture}
}

\begin{document}

\allowdisplaybreaks

\newcommand{\arXivNumber}{1808.00799}

\renewcommand{\PaperNumber}{031}

\FirstPageHeading

\ShortArticleName{A Family of ${\rm GL}_r$ Multiplicative Higgs Bundles on Rational Base}

\ArticleName{A Family of $\boldsymbol{{\rm GL}_r}$ Multiplicative Higgs Bundles\\ on Rational Base}

\Author{Rouven FRASSEK and Vasily PESTUN}

\AuthorNameForHeading{R.~Frassek and V.~Pestun}

\Address{Institut des Hautes \'Etudes Scientifiques, Bures-sur-Yvette, France}
\Email{\href{mailto:frassek@ihes.fr}{frassek@ihes.fr}, \href{mailto:pestun@ihes.fr}{pestun@ihes.fr}}

\ArticleDates{Received September 09, 2018, in final form April 10, 2019; Published online April 25, 2019}

\Abstract{In this paper we study a restricted family of holomorphic symplectic leaves in the Poisson--Lie group ${\rm GL}_r(\mathcal{K}_{\mathbb{P}^1_x})$ with rational quadratic Sklyanin brackets induced by a~one-form with a single quadratic pole at $\infty \in \mathbb{P}_{1}$. The restriction of the family is that the matrix elements in the defining representation are linear functions of $x$. We study how the symplectic leaves in this family are obtained by the fusion of certain fundamental symplectic leaves. These symplectic leaves arise as minimal examples of (i) moduli spaces of multiplicative Higgs bundles on $\mathbb{P}^{1}$ with prescribed singularities, (ii) moduli spaces of $U(r)$ monopoles on $\mathbb{R}^2 \times S^1$ with Dirac singularities, (iii) Coulomb branches of the moduli space of vacua of 4d $\mathcal{N}=2$ supersymmetric $A_{r-1}$ quiver gauge theories compactified on a circle. While degree~1 symplectic leaves regular at $\infty \in \mathbb{P}^1$ (Coulomb branches of the superconformal quiver gauge theories) are isomorphic to co-adjoint orbits in $\mathfrak{gl}_{r}$ and their Darboux parametrization and quantization is well known, the case irregular at infinity (asymptotically free quiver gauge theories) is novel. We also explicitly quantize the algebra of functions on these moduli spaces by presenting the corresponding solutions to the quantum Yang--Baxter equation valued in Heisenberg algebra (free field realization).}


\Classification{16T25; 53D30; 81R12}

\section{Introduction}

Complex completely integrable Hamiltonian systems can be typically constructed starting from a locus $\mathcal{M}$ in the moduli space $\Bun_G(\Sigma)$ of holomorphic $G$-bundles or sheaves of certain type on a complex holomorphic symplectic surface $\Sigma$ with a structure of Lagrangian elliptic fibration $\Sigma \to X$, where the fibers $\Sigma_x$ are possibly degenerate elliptic curves, and $X$ is an algebraic curve typically called the base curve, see for example Section~0.3.6 in~\cite{Donagi:2000dr} and Section~3.8.2 in Donagi's lectures in~\cite{mason2003geometry} and~\cite{MR1397059,Donagi:1997dp,MR2042693,Donagi:2000dr} for more complete details.

Indeed, the symplectic structure on $\Sigma$ induces the symplectic structure on the space $\mathcal{M}$ which becomes the phase space of integrable system, the structure of Lagrangian fibration $\Sigma \to X$ induces the structure of Lagrangian fibration on $\mathcal{M}$, and the fact that the fibers $\Sigma_x$ are abelian varieties (possibly degenerate elliptic curves) induces the structure of abelian varieties on the Lagrangian fibers in~$\mathcal{M}$.

There are three cases to consider depending on whether the elliptic fibers are generically cusped elliptic, nodal elliptic or smooth elliptic.

\looseness=-1 (1) Fibers are cusped elliptic. If $X$ is an algebraic curve, and $\Sigma \to X$ is a cotangent bundle whose fibers are compactified to cusped elliptic curves, this construction produces algebraic integrable system called Hitchin system on the curve $X$ \cite{Donagi1995,hitchin1987,MR1300764}. Hitchin system is an example of an abstract Higgs bundle on~$X$ valued in an abelian group $K$ over $X$ for the case when the group~$K$ is the canonical line bundle on $X$ endowed with natural linear additive group structure in the fiber direction. The Higgs field $\phi(x)$ is a holomorphic 1-form valued in the Lie algebra adjoint bundle $\operatorname{ad} \mathfrak{g}$. The respective integrable system is of additive type in the fiber direction.

(2) Fibers are nodal elliptic. If $\Sigma \to X$ is a fibration whose fibers are nodal elliptic curves, then $\Bun_{G}(\Sigma)$ is equivalently described as a moduli space of multiplicative Higgs bundles $\mHiggs_G(X)$, that is moduli space of pairs $(P, g)$ where $P$ is a principal $G$-bundle on $X$, and Higgs field $g (x)$ is a section of Lie group adjoint bundle $\operatorname{ad} G$. The respective integrable system is of multiplicative type in vertical direction. In Donagi's lectures in \cite[Section 3.9]{MR2042693} one finds a remark on three types of integrable system in the fiber direction corresponding to the three types of connected 1-dimensional complex groups: an elliptic curve, the multiplicative group $G_m = \mathbb{C}^{\times}$ and the additive group $G_a = \mathbb{C}$, and that the latter two can be considered as groups of non-singular points in the elliptic case in the nodal and cuspidal limit respectively. He goes on to clarify that Hitchin systems are associated to the cuspidal type and principal bundles on smooth elliptic fibrations to smooth elliptic type, and then asks ``Is there an interesting geometric interpretation of the remaining ``trigonometric'' case, where the values are taken in multiplicative group $G_{m}$?'' We believe so and we refer to several geometrical perspectives on the multiplicative case further in the introduction. For the basic definitions see \cite{Cherkis:2001gm,Elliott2018,Frenkel2011,MR1974589,Nekrasov:2012xe}.

(3) Fibers are smooth elliptic. If $\Sigma \to X$ is an elliptically fibered complex surface with generically smooth fibers, the corresponding case was studied in \cite{Donagi:1997dp,Friedman:1997yq}. Using Loojienga description of moduli space of $G$-bundles on a smooth elliptic fiber as a space conjugacy classes in the affine Kac--Moody Lie group $\hat G$ \cite{Loojienga:1976}, we can also interpret $\Bun_{G}(\Sigma)$ as a moduli space $\mHiggs_{\hat G}(X)$ of multiplicative Higgs bundles for the affine Kac--Moody group $\hat G$. The respective integrable system is of elliptic type in vertical direction.

The case (1) of additive Higgs bundles (Hitchin systems) received large amount of attention in the mathematical literature in the context of geometrical Langlands correspondence and in the physical literature in the context of $6d$ $(2,0)$ superconformal self-dual tensor theory compactified on algebraic complex curve $X$ for $G$ of ADE type \cite{Alday:2009aq,Beilinson,Kapustin:2006pk,Nekrasov:2010ka}. Quantization of additive Higgs bundles on the curve $X$ relates to the theory of Kac--Moody current algebras on $X$, conformal blocks of $W$-algebra on $X$ with punctures, D-modules on $\Bun_{G}(X)$, and monodromy problems for various related differential equations.

The case (2) of multiplicative Higgs bundles on a complex curve $X$ appeared first in the context of current Poisson--Lie groups $G(x)$ with spectral parameter $x \in X$. A~Poisson--Lie group is a Lie group equipped with Poisson structure compatible with the group multiplication law. There is a standard way to equip $G(x)$ with Poisson structure called quadratic Skylanin bracket given a holomorphic no-where vanishing differential 1-form on $X$ (possibly with poles). Quantization of this Poisson structure leads to the theory of quantum groups \cite{Drinfeld:1986,MR1062425} which have been discovered in the context of the inverse scattering method, quantum integrable spin chains, Yang--Baxter equation and $R$-matrix with spectral parameter. The standard horizontal trichotomy of the rational, trigonometric or elliptic $R$-matrix corresponds to taking the base $X$ to be the $\mathbb{P}^{1}$ with 1-form with a single quadratic pole (rational type), the $\mathbb{P}^{1}$ with 1-form with two simple poles (trigonometric type), or smooth elliptic curve (elliptic type).

For the smooth elliptic base curve $X$ the multiplicative Higgs bundle was studied in \cite{MR1974589}, following \cite{MR1346215, MR1334607}. Independently, the definition of multiplicative Higgs bundles was given in~\cite{Frenkel2011} where they were called $G$-pairs. On another hand, multiplicative Higgs bundles on $X$ have been studied as periodic monopoles on real three-dimensional Riemannian manifold $X \times S^1$ via the monodromy map \cite{Charbonneau2008,Cherkis:2000cj,Cherkis:2001gm,Cherkis:2000ft,Gorsky:1997jq,Gorsky:1997mw,Nekrasov:2012xe}. The relation between quantization of the moduli space of monopoles on $\BR^3$ and Yangian has been proposed in \cite{Gerasimov:2005qz} and further work in this direction has been in \cite{MR3248988}. Recently a quantization of the holomorphic symplectic phase space of the moduli space of monopoles on $X \times S^1$ by a formal semi-holomorphic Chern--Simons functional on $X \times S^1 \times \BR_t$, where $\BR_{t}$ is the time direction, has been studied in \cite{Costello:2013zra, Costello:2017dso}. For simple Lie groups $G$ of the ADE type these moduli spaces appear as Coulomb branches of the moduli space of vacua of the $\mathcal{N}=2$ supersymmetric ADE quiver gauge theory on~\mbox{$\mathbb{R}^3 \times S^1$} \cite{Braverman:2016pwk,Braverman:2016wma,Nakajima:2015txa, Nekrasov:2012xe}. Some constructions from the world of additive Higgs bundles have their versions in the world of multiplicative Higgs bundles \cite{Elliott2018} leading to difference equations and their monodromy problems~\cite{birkhoff1913generalized, sauloy2004isomonodromy}, $q$-geometric Langlands correspondence \cite{Aganagic:2017smx}, $q{-}W$ algebras \cite{Kimura:2015rgi, Nekrasov:2015wsu,Nekrasov:2013xda}.

\looseness=-1 The goal of this paper is to present very concretely a Darboux coordinate system on a moduli space ${\rm GL}_{r}$ multiplicative Higgs bundles of degree 1 on the rational base $X = \mathbb{P}^1_x$. The base curve~$X$ is equipped with a holomorphic one-form ${\rm d} x$ that has the quadratic pole at $x_\infty = \infty$. The holomorphic one-form ${\rm d}x$ together with the Killing form on the Lie algebra induces the quadratic Sklyanin Poisson structure with the classical $\mathfrak{r}$-matrix of rational type in the spectral parameter~$x$. Equivalently, we are studying degree $1$ symplectic leaves in the rational Poisson--Lie group ${\rm GL}_{r}(\mathcal{K}_{\mathbb{P}^1})$, where $\mathcal{K}_{\mathbb{P}^1}$ denotes the field of rational functions on $\mathbb{P}^1$, and degree $1$ means that all matrix elements $(L_{ij}(x))_{1 \leq i,j \leq r}$ of the multiplicative Higgs field $g(x)$ in the defining representation of ${\rm GL}_{r}$ by $r \times r$ matrices~$L_{ij}(x)$ are degree~1 polynomials of~$x$, i.e., linear functions of~$x$.

\looseness=-1 By concrete presentation we mean introduction of explicit Darboux coordinates (canonically conjugated set of $(\underline p, \underline q) = \big(p_I, q^I\big)$ variables with $\big\{p_I, q^{J}\big\} = \delta_I^J$) and presentation of explicit formulae for the matrix elements $L_{ij}(x)$ in terms of $\big(p_I, q^J\big)$. The complete set of commuting Hamiltonian functions is obtained from the coefficients of the spectral determinant of $L(x)$. The matrix $L(x)$ valued in functions on the phase space is called Lax matrix and its matrix elements satisfy quadratic Sklyanin Poisson brackets, see in particular \cite{babelon:hal-00101459, FaddeevBook} but also the recent lecture notes~\cite{Torrielli:2016ufi}.

\looseness=-1 The quadratic Sklyanin Poisson brackets can be also defined as semi-classical limit of the quantum Yang--Baxter equation \cite{Sklyanin:121210,Sklyanin:1982tf}. In this paper we find all rational solutions of degree~$1$ in the spectral parameter $x$ associated to the classical Yang--Baxter equation defined by the rational $\mathfrak{gl}(r)$-invariant $\rr$-matrix, cf.~\cite{Belavin1982}. In another note we plan to consider the trigonometric case associated to the base curve being a punctured nodal elliptic curve $X =\mathbb{C}^{\times}_x$ equipped with the holomorphic one-form $\frac{{\rm d}x}{x}$ that has simple pole at $x = 0$ and $x = \infty$, a related work appears in~\cite{2017arXiv170801795F}.

The case of $G = {\rm GL}_2$ is well studied. Here the Sklyanin relation admits three different elementary types of non-trivial solutions with matrix elements linear in the spectral parameter~$x$ that yield integrable models. These solutions are called the $2 \times 2$ elementary Lax matrices for the Heisenberg magnet, the DST chain and the Toda chain. For an overview we refer the reader to lecture notes of Sklyanin~\cite{sklyanin00}.

For higher rank $r$, to the best knowledge of the authors the explicit presentation of all linear solutions is missing in the literature. The case regular at the infinity $x_\infty \in X$ has been described in \cite{ShapiroThesis} and many other places. Some partial cases of Toda like solutions for irregular case have been described in \cite{Gorsky:1997jq,Gorsky:1997mw,Meneghelli:thesis}. The classifying labels appeared in \cite{Haouzi:2016ohr}. In the quantum case some solutions to the Yang--Baxter equation were studied in connection to non-compact spin chains and Baxter $Q$-operators, in particular for the case of $\mathfrak{gl}_r$ we refer the reader to \cite{Bazhanov:2010jq, Derkachov:2006fw}. The solutions relevant for non-compact spin chains can be obtained by realising the quantum $R$-matrix in terms of an infinite-dimensional oscillators algebra which is also known as free-field realisation, see, e.g.,~\cite{DiFrancesco:1997nk}. The Lax matrices relevant for $Q$-operators are certain degenerate solutions in the sense that the term proportional to the spectral parameter is not the identity matrix but a matrix of lower rank. These Lax matrices can be obtained from the non-degenerate case through a limiting procedure as discussed in \cite{Bazhanov:2008yc} for $\mathfrak{gl}(2|1)$, \cite{Gorsky:1997jq,Gorsky:1997mw} for $\mathfrak{gl}_3$ or directly from the universal $R$-matrix as shown in~\cite{Boos:2010ss} for~$\mathfrak{gl}(3)$. Vice versa to the limiting procedure and as discussed in \cite{Bazhanov:2010jq}, one can also obtain the Lax matrices of non-compact spin chains by fusing the degenerate solutions relevant for $Q$-operators.

Here we follow the strategy of fusion in order to construct a family of ${\rm GL}_r$ Lax matrices $L(x)$ whose matrix elements are linear in spectral parameter~$x$.

The discrete data of labels in our family is specified by two partitions $\lambda$ and $\mu$ such that the total size is $|\lambda| + |\mu| = r$ and whose columns $\lambda_i^{t}$, $\mu_i^{t}$ are restricted by $r$. In addition to the discrete partition labels $(\lambda, \mu)$ we have a sequence of complex labels. There is a complex parameter $x_i$ assigned to each column $\lambda_i^{t}$ of the partition $\lambda$. Geometrically speaking, each pair $\big(\lambda_i^{t}, x_i\big)$ describes a type of singularity of the multiplicative Higgs field $g(x)$ at finite point $x_i \in \BC= \mathbb{P}^1 \setminus \{x_\infty\}$ given by the conjugacy class of $ (x - x_i)^{\check \omega_{\lambda_i^{t}}}$ where $\check \omega_k$ denotes $k$-th fundamental co-weight of ${\rm GL}_{r}$: that is the highest weight of the $k$-th antisymmetric power of the fundamental representation for the Langlands dual group ${\rm GL}_r$. Such highest weight is encoded by the column of height $\lambda_i^{t}$ in the partition~$\lambda$. Equivalently, in the neighborhood of the point $x_i$ in the spectrum of the $r \times r$ Lax matrix $L(x)$ there are exactly $\lambda_i^{t}$ eigenvalues which vanish linearly as~$x$ approaches~$x_i$, and the remaining $r - \lambda_i^{t}$ eigenvalues are regular non-zero at~$x_i$.

The partition $\mu$ specifies a dominant co-weight of singularity of the multiplicative Higgs field at the infinity point $x_{\infty} \in \mathbb{P}^1$, or equivalently the asymptotics of the eigenvalues of the Lax matrix
$L(x)$ as $x \to \infty$: given $r$ rows $(\mu_j)_{j \in [1, \dots, r]}$ of the partition $\mu$, the $j$-th eigenvalue of the Lax matrix $L(x)$ has asymptotics $(x^{-1})^{\mu_j-1}$ as $x \to \infty$.

We remark that the restriction on the total size of two partitions $|\mu| + \sum_{i} \lambda_i^{t} =r $ is a~consequence of the restriction of the present paper to consider only Lax matrices whose matrix elements are linear functions of~$x$. In the complete classification, if we allow higher degree of $x$ in the matrix elements, which is not in the scope of the present paper, the label of a singularity at any finite point $x_i$ is an arbitrary dominant ${\rm GL}_{r}$ co-weight described by an arbitrary partition~$\lambda_{i}$, so that if rows of partition $\lambda_i$ are denoted by $(\lambda_{ij})_{j \in [1, \dots, r]}$ then $j$-th eigenvalue of the Lax matrix $L(x)$ behaves as $ (x - x_i)^{\lambda_{ij}}$ as $ x \to x_i$. We leave for another note the presentation of explicit formulae for complete classification of the symplectic leaves of the degree $d$ whose matrix elements are degree $d$ polynomials of~$x$ for $|\mu| + \sum_{i} \lambda_i^{t} = d r$. (By looking at the determinant of $g(x)$ we see that the moduli space is non-empty only if the total size $|\mu| + \sum_{i} \lambda_i^{t}$ is integral multiple of rank~$r$, cf., e.g.,~\cite{MR1974589,Nekrasov:2012xe}. This condition means that the total dominant co-weight summed over all singularities $\check \omega_{\rm tot}$ belongs to the lattice of co-roots.\footnote{In the monopole picture, the topological degree of gauge bundle induced on a surface enclosing all singularities is trivial. The topological degree is an element in $\pi_1(G) \simeq \check \Lambda / \check Q$, where $\check \Lambda$ and $\check Q$ denote the lattice of co-weights and co-roots.} To summarize, near every singularity on $\mathbb{CP}^{1}$ in a local coordinate $w$ such that $w =0$ is a position of singularity, we have asymptotics $[g(w)] \sim w^{\omega^{\vee}}$ where $\omega^{\vee}\colon \mathbb{C} \to T_{G}$ is a co-weight (either $\lambda_i^{t}$ or $\mu$) that characterizes the singularity. Normally, because the total degree ($U(1)$-charge) vanishes, the sum of degrees of all co-weights $\omega$ must vanish. We have chosen to shift the notational representation of the singularity co-weight at infinity by adding $1$ to each row of the co-weight $\omega_{\infty}^{\vee}$ so that is described by a positive partition $\mu$. In consequence, the sum over all partitions $\lambda_i$'s and $\mu$ is $r$ is no longer zero but $r$, since there are $r$ rows in $\omega_{\infty}^{\vee}$, and each has been increased by $1$ in our notations: $\mu_j = \omega_{\infty,j} + 1$ for each row $j =1,\dots, r$.

In our solutions we can obtain higher (non-fundamental co-weight) singularities at finite point $x_{*}$ by collision of several fundamental singularities at $x_{i_1}, x_{i_2}, \dots , x_{i_k}$ which are associated to some columns $\lambda_{i_1}^{t}, \dots, \lambda_{i_k}^{t}$ of the partition $\lambda$ by sending all of them to the common point $ x_{*}$. In this case, generically, the multiplicative Higgs field $g(x)$ develops the singularity at point $x_{*}$ specified by a higher (non-fundamental) co-weight $\sum\limits_{j=1}^{k} \check \omega_{\lambda_{i_j}^{t}}$.

\looseness=-1 As we will see, all Lax matrices regular at the infinity $x_\infty$, that is $\mu=\varnothing$ in the current notations, and arbitrary $\lambda$ can be obtained by the fusion procedure of the elementary Lax matrices used in the $Q$-operator construction~\cite{Bazhanov:2010jq}. Also the case regular at infinity has been described in~\cite{ShapiroThesis}, where it was shown that degree~1 rational symplectic leaves for $G = {\rm GL}_{r}$ correspond to the co-adjoint orbits in the dual Lie algebra $\mathfrak{gl}_r^{*}$. The parametrization by Darboux coordinates of the holomorpic symplectic co-adjoint orbits in $\mathfrak{gl}_{r}^{*}$ identical to the present paper has been proposed in \cite{babich2016birational}.

Then we proceed to build Lax matrices irregular at infinity from the fusion of a certain set of elementary Lax matrices whose irregularity at infinity is of the simplest type.

Let us clarify the geometrical meaning of fusion. A Lax matrix $L_{\underline{\lambda}, \underline{x}, \mu} (x; \underline{p}, \underline{q})$ with a certain prescribed type of singularities at $\underline{x}, x_\infty$ parametrizes
by a system of Darboux coordinates $(\underline{p}, \underline{q})$ a
finite-dimensional symplectic leaf in the infinite-dimensional
Poisson--Lie group $\mathcal{G} = {\rm GL}_{r}(\mathcal{K}_{\mathbb{P}^1})$ where $\mathcal{K}_{\mathbb{P}^1}$ denotes the field of rational functions on $\mathbb{P}^1$. More geometrically, a Lax matrix $L(x, \underline{p}, \underline{q})$ is
a universal group valued (multiplicative)
Higgs field on a Darboux chart in the second factor of $\mathbb{C}_{x} \times \mathcal{M}_{_{\underline{\lambda}, \underline{x}, \mu} }$ represented in $r \times r$ matrices, where $\mathcal{M}_{{\underline{\lambda}, \underline{x}, \mu} }$ is a moduli
space of multiplicative Higgs fields of a certain type $({\underline{\lambda}, \underline{x}, \mu})$,
and complex spectral plane $\mathbb{C}_{x}$ is the domain of the Higgs field $g(x)$.
So for us a Lax matrix $L_{{\underline{\lambda}, \underline{x}, \mu}}$ is a composition of Darboux chart
parametrization
\begin{gather*}
 \mathbb{C}^{2 d_{\underline{\lambda}, \mu}} \to \mathcal{M}_{{\underline{\lambda}, \underline{x}, \mu}}
\end{gather*}
with a universal Higgs field map
\begin{gather*}
 \mathbb{C}_{x} \times \mathcal{M}_{\underline{\lambda}, \underline{x}, \mu} \to \mathrm{Mat}_{r \times r}.
\end{gather*}

Suppose we are given a symplectic leaf $\mathcal{M}_{\underline{\lambda}, \underline{x}, \mu} \subset \mathcal{G}$ described by a Lax matrix $L_{\underline{\lambda}, \underline{x}, \mu} (x; \underline{p}, \underline{q})$ and a symplectic leaf $\mathcal{M}'_{\underline{\lambda}',\underline{x}', \mu'} \subset \mathcal{G}$ described by a Lax matrix $L_{\underline{\lambda}', \underline{x}', \mu'} (x; \underline{p'}, \underline{q'})$.
By definition of Poisson--Lie group structure on $\mathcal{G}$ the group multiplication map
\begin{gather}\label{eq:groupm}
 m\colon \ \mathcal{G} \times \mathcal{G} \to \mathcal{G}
\end{gather}
is a Poisson map, i.e., the pushforward of the product Poisson structure on $\mathcal{G} \times \mathcal{G}$ coincides with the Poisson structure on $\mathcal{G}$. The symplectic leaves $\mathcal{M}$, $\mathcal{M}'$ are, in particular, co-isotropic submanifolds of $\mathcal{G}$, hence $ \mathcal{M} \times \mathcal{M}'$ is a co-isotropic submanifold of $\mathcal{G} \times \mathcal{G}$. Now, since the group multiplication map~$m$ in~\eqref{eq:groupm} is a Poisson map, and since the Poisson map preserves the co-isotropic property of the submanifolds, the image $m(\mathcal{M} \times \mathcal{M}') \subset \mathcal{G}$ is a co-isotropic subspace.

The $\mathcal{G}$-elements in the co-isotropic subspace $m(\mathcal{M} \times \mathcal{M}') \subset \mathcal{G}$ are represented by Lax matrices
\begin{gather}\label{eq:fusion1}
 L_{\underline{\lambda}, \underline{x}, \mu} (x; \underline{p}, \underline{q}) L_{\underline{\lambda}', \underline{x}', \mu'} (x; \underline{p'}, \underline{q'})
\end{gather}
and their type of singularities is typically a combination of the types of singularities of $(\underline \lambda, \underline x, \mu)$ and $(\underline{\lambda}', \underline x', \mu')$. However, $m(\mathcal{M} \times \mathcal{M}') \subset \mathcal{G}$ is not in general a symplectic leaf but a co-isotropic submanifold, and we can further slice it into symplectic leaves by determining the set of Casimir functions~$\tilde q'$ on $m(\mathcal{M} \times \mathcal{M}')$ and a set of new conjugated coordinates $\underline{\tilde p}$, $\underline{ \tilde q}$. We find that
\begin{gather*}
 L_{\underline{\lambda}, \underline{x}, \mu} (x; \underline{p}, \underline{q}) L_{\underline{\lambda}', \underline{x}', \mu'} (x; \underline{p'}, \underline{q'})
 = \tilde C(\underline{\tilde q}') \tilde{L}_{{\underline{\tilde{\lambda}}, \underline{\tilde{x}}, \tilde{\mu}}} (x; \underline{\tilde p}, \underline{\tilde q})
\end{gather*}
with the canonical transformation
\begin{gather*}
 {\rm d}\underline{p} \wedge {\rm d}\underline{q} + {\rm d}\underline{p}' \wedge {\rm d}\underline{q}' = {\rm d} \tilde {\underline{p}} \wedge {\rm d} \tilde {\underline{q}} + {\rm d} \tilde {\underline{p}} ' \wedge {\rm d} \tilde {\underline{q}'} .
\end{gather*}
Notice that the conjugate variables $\tilde p'$ to the Casimir functions $\tilde q'$ on $\tilde S$ do not appear on the right side of (\ref{eq:fusion1}).
The Lax matrices $ \tilde{L}_{{\underline{\tilde{\lambda}}, \underline{\tilde{x}}, \tilde{\mu}}} (x; \underline{\tilde p}, \underline{\tilde q})$ represent elements of $\mathcal{G}$ in a new symplectic leaf $\mathcal{M}_{{\underline{\tilde{\lambda}}, \underline{\tilde{x}}, \tilde{\mu}}}$ covered by Darboux coordinates $\underline{\tilde p}$, $\underline{\tilde q}$.

The symplectic leaves $\mathcal{M}_{\underline{\lambda}, \underline{x}, \mu} $ arise as moduli spaces of multiplicative Higgs bundles of certain type~\cite{Elliott2018}, and like additive Higgs bundles (Hitchin system), the
symplectic leaves $\mathcal{M}_{\underline{\lambda}, \underline{x}, \mu} $ support the structure of an algebraic completely integrable system. In fact, the moduli spaces $\mathcal{M}_{\underline{\lambda}, \underline{x}, \mu} $ can be also interpreted as moduli spaces of $U(r)$ monopoles on 3-dimensional Riemannian space $\mathbb{R}^2 \times S^1$ where $\BR^2 \simeq \BC = \mathbb{P}^1 \setminus \{x_\infty\}$, and consequently \cite{Cherkis:2000cj,Cherkis:2001gm,Cherkis:2000ft,Nekrasov:2012xe} as moduli spaces of vacua of certain $\mathcal{N}=2$ supersymmetric quiver gauge theories on~$\mathbb{R}^{3} \times S^1$ of quiver type $A_{r-1}$. The complex parameters $x_i \in \mathbb{C}$ which specify the position of singularities of the Lax matrix~$ L_{\underline{\lambda}, \underline{x}, \mu} $ play the role of the masses of the fundamental multiplets
attached to the quiver node~$\lambda_{i}^{t}$ in the $A_{r-1}$ quiver diagram (i.e., the node associated to a simple root dual to the fundamental co-weight $\lambda_{i}^{t}$), and at the same time they play the role of the
complex part of the coordinates of the positions of the Dirac singularities of the $U(r)$ monopoles on $\mathbb{R}^2 \times S^1$ under identification $\BR^2 \simeq \BC$. For polynomial Lax matrices that we consider in this paper the eigenvalues of $L(x)$ at the singularities $\underline{x}$ can have only zeros and no poles, thus the corresponding periodic monopoles can have only negatively charged Dirac singularities.

If the partition $\mu$ is empty, then the corresponding $A_{r-1}$ quiver gauge theory is $\mathcal{N}=2$ superconformal theory, and corresponding monopoles on $\mathbb{R}^2 \times S^1$ are regular at infinity. Non-empty partition $\mu$ corresponds to monopoles on $\mathbb{R}^2 \times S^1$ with non-trivial growth (or charge) at infinity controlled by~$\mu$, or to the Coulomb branches of asymptotically-free quiver gauge theories with $\beta$-function controlled by~$\mu$.

Consequently, the integrable system supported on a symplectic leaf $\mathcal{M}_{\underline{\lambda}, \underline{x}, \mu} $ is identical to Seiberg--Witten integrable system for a certain $A_{r-1}$ quiver gauge theory.

The complete set of commuting Hamiltonians functions $H_{ij}$ can be extracted from the spectral determinant of the associated Lax matrix
\begin{gather}\label{eq:spectral1}
 \det \big( y - g_{\infty} L_{\underline{\lambda}, \underline{x}, \mu} (x; \underline{p}, \underline{q})\big) = \sum_{i,j} H_{ij} x^i y^j
\end{gather}
by taking coefficients at the monomials $x^i y^j$ where the appearing pairs of indices $(i,j)$ can be described by certain profiles like Newton diagrams. The spectral curves~(\ref{eq:spectral1}) coincide with the spectral curves of the integrable systems studied in~\cite{Nekrasov:2012xe,Nekrasov:2013xda}. Equivalently, since the determinant can be expanded in terms of the characters $\tr_{R_k}$ of the $k$-th external powers of the fundamental representation, the commuting Hamiltonians are expressed as coefficients at powers of $x$ in the characters~$ \tr_{R^k} L(x)$.

We remark that by switching the role of variables $x \in \BC$ and $y \in \BC^{\times}$ (fiber-base duality) the spectral curve (\ref{eq:spectral1}) of multiplicative Higgs bundle on $X$ can be also interpreted as the spectral curve of additive Higgs bundle (Hitchin system) on $Y = \BC^{\times} = \mathbb{P}^{1}_{0, \infty}$. This is a peculiarity related to the fact that we are considering the rational case of the base $X = \mathbb{P}^{1}$ corresponding to the monopoles on $\BR^2 \times S^1$ and $4d$ quiver gauge theories rather than the trigonometric or elliptic base $X$ corresponding to the monopoles on $\BR \times S^1 \times S^1$ or $S^1 \times S^1 \times S^1$ that relate to~$5d$ or~$6d$ quiver gauge theories compactified on $S^1$ or $S^1 \times S^1$, and also that we take the gauge group to be of type ${\rm GL}_{r}$. In this situation, the moduli space of $U(r)$ monopoles on $\BR^2 \times S^1$ with several singularities has alternative presentation (Nahm duality) as ${\rm GL}_n$ Hitchin moduli space on $\mathbb{C}^{\times}$ with $r$ singularities where $n$ depends on the number and type of the singularities of the multiplicative Higgs bundle on~$X$ \cite{Cherkis:2000cj, Cherkis:2001gm,Cherkis:2000ft,Nekrasov:2012xe}.

\looseness=-1 Anyways, the fusion method of this paper allows us to analyze the multiplicative Higgs bundles in more general cases, which we leave for a future work, when Nahm duality of multiplicative Higgs bundle to a Hitchin system is not known. In particular, in the future one can study classification of symplectic leaves with matrix elements of higher degree in~$x$, one can analyze trigonometric case with the base curve $X$ is $\mathbb{C}^{\times} = \mathbb{P}^{1} \setminus \{0, \infty\}$ or elliptic case when the base curve~$X$ is a smooth elliptic curve like \cite{MR1974589} and consider arbitrary complex reductive Lie groups~$G$.

The article is organised as follows. In Section~\ref{sec:main} we remind and set notations about Poisson Lie groups, Sklyanin brackets and Lax matrices. In Section~\ref{sec:lambda} we build the Lax matrices for arbitrary partitions $\underline{\lambda}$ and empty $\mu=\varnothing$ from certain elementary building blocks by fusion. Similarly, in Section~\ref{sec:mupart} we build Lax matrices for arbitrary partitions $\mu$ with $\underline{\lambda}=\varnothing$ again employing certain elementary solutions using a slightly modified fusion procedure. In Section~\ref{sec:lmpart} we combine the solutions of Sections~\ref{sec:lambda} and \ref{sec:mupart} to write down the Lax matrices for arbitrary $\underline{\lambda}$ and $\underline{\mu}$. In Section~\ref{sec:specdet} we study the spectral determinant of the derived Lax matrices and compare our results with~\cite{Nekrasov:2012xe}. In Section~\ref{sec:higherdegree} we say a few words on higher degree symplectic leaves. In Section~\ref{sec:quantum} we consider the quantization of the algebra of functions and the integrable system.

\section{Rational Poisson--Lie group and Sklyanin brackets}\label{sec:main}

Let $X =\mathbb{P}^1$ be the base curve equipped with the differential holomorphic volume form $dx$ that has a single quadratic pole at $x_{\infty} \in \mathbb{P}^{1}$. Fix a Killing form $\tr$ on $\mathfrak{g}$. Then the residue pairing
\begin{gather*}
 \tr \oint_{x =0} f(x) g(x) {\rm d}x
\end{gather*}
induces the metric on $\mathfrak{g}_{D} = \mathfrak{g}((x))$ with respect to which $\mathfrak{g}[[x]]$ and $x^{-1} \mathfrak{g}\big[x^{-1}\big]$ are isotropic subspaces and we have $\mathfrak{g}_{D} = \mathfrak{g}_{+} \oplus \mathfrak{g}_{-}$. This splitting induces the structure of the Lie bi-algebra on $\mathfrak{g}_{+}$, which means that the space of functions on $\mathfrak{g}_{+}$ is equipped with the Poisson bracket (induced from the Lie bracket on $\mathfrak{g}_{-}$). The data $(\mathfrak{g}_{D}, \mathfrak{g}_{+}, \mathfrak{g}_{-})$ is called Manin triple. The Poisson bracket on the functions on $\mathfrak{g}_{+}$ can be extended to the Poisson bracket on the functions on the Lie group $G_{+}$ with the Lie algebra $\mathfrak{g}_{+}$, and the resulting bracket is called Sklyanin quadratic bracket with the rational $\mathfrak{r}$-matrix.

The space of rational multiplicative Higgs fields on $X = \mathbb{P}^{1}$ with a fixed framing of the gauge bundle at $x_\infty$ forms a Poisson--Lie group~\cite{Elliott2018}.

In the following we consider gauge group $G = {\rm GL}_{r}$ and for a Higgs field $g(x)$ we call $L(x)$ the representation of $g(x)$ by $r \times r$ matrix valued functions $L(x)$ called Lax matrices. The space of Lax matrices $L(x)$ carries the quadratic Poisson bracket of rational Sklyanin type
\begin{gather}\label{eq:skl}
 \{L(x)\otimes \ID ,\ID \otimes L(y)\}=[L(x)\otimes L(y),\rr(x-y)] ,
\end{gather}
the quantization of which gives quantum Yang--Baxter equation \cite{Sklyanin:1982tf}. Here the $\ID $ denotes the $r\times r$ identity matrix, and the classical rational $\rr$-matrix of $\mathfrak{gl}(r)$ is
\begin{gather}\label{eq:perm}
\rr(x)=x^{-1}\mathbb{P} ,\qquad\text{with}\quad \mathbb{P}=\sum_{a,b=1}^r e_{ab}\otimes e_{ba} .
\end{gather}
The bracket on the right-hand-side of \eqref{eq:skl} denotes the commutator $[X,Y]=XY-YX$. In a~system of Darboux coordinates $\big(\underline{p},\underline{q}\big)=\big(p_{I}, q^{I}\big)$, the Poisson bracket is
\begin{gather*}
\{X,Y\}=\sum_{I}\left(\frac{\partial X}{\partial p_{I}}\frac{\partial Y}{\partial q^{I}}-\frac{\partial X}{\partial q^{I}}\frac{\partial Y}{\partial p_{I}}\right),
\end{gather*}
where we sum over all conjugate variables $(\underline{p},\underline{q})$ in the Lax matrices $L$. In index notations, the Poisson bracket of matrix elements~(\ref{eq:perm}) reads as follows
\begin{gather*}%\label{eq:permclass}
 \{ L_{ij}(x), L_{kl}(y) \} = - \frac{1} {x - y} (L_{kj}(x) L_{il}(y) - L_{kj}(y) L_{il}(x)) .
\end{gather*}

The solutions to the Sklyanin relation \eqref{eq:skl} that appear in this paper are labelled by two partitions
\begin{gather*}
 \lambda=(\lambda_1,\lambda_2,\ldots,\lambda_r) ,\qquad \mu=(\mu_1,\mu_2,\ldots,\mu_r) ,
\end{gather*}
with $\lambda_1\geq \lambda_2\geq \cdots\geq \lambda_r$ and $\mu_1\geq \mu_2\geq \cdots\geq \mu_r$ where $\lambda_i,\mu_i\in \mathbb{Z}_{\geq 0}$. The total number $|\lambda|$ of elements in the partition $\lambda$ combined with total number $|\mu|$ of elements in the partition $\mu$ is equal to~$r$. We study solutions $\Llm$ whose matrix elements are no higher than of degree 1 in the spectral parameter $x$. We can assume that
\begin{gather*}
\Llm=x\times \diag(\underbrace{0,\ldots,0}_{\mu_1^t},\underbrace{1,\ldots,1}_{r-\mu^t_1})+M_{\lambda,\underline{x},\mu}(\underline{p},\underline{q}) .
\end{gather*}
Here $\mu_1^t$ denotes the first column, i.e., the first element in the transposed partition $\mu^t=\big(\mu_1^t,\mu_2^t,\ldots,\mu_r^t\big)$ and $M_{\lambda,\underline{x},\mu}(\underline{p},\underline{q})$ denotes an $r\times r$ matrix which is independent of the spectral parameter $x$. In total the matrix $M_{\lambda,\underline{x},\mu}(\underline{p},\underline{q})$ contains
\begin{gather}\label{eq:dimform}
 d_{\lambda,\mu}=\frac{1}{2}\left( r^2-\sum_{i=1}^{\lambda_1}\big(\lambda_i^t\big)^2-\sum_{i=1}^{\mu_1}\big(\mu_i^t\big)^2\right) ,
\end{gather}
pairs of variables $\big(p_I,q^I\big)$, i.e., $I=1,2,\ldots, d_{\lambda,\mu}$. Again the transposed partition is denoted as $\lambda^t=\big(\lambda_1^t,\lambda_2^t,\ldots,\lambda_r^t\big)$, and $\lambda_{i}^{t}$ are called columns. The dimension of the corresponding symplectic leaf or the moduli space of multiplicative Higgs bundles will be given by
\begin{gather*} %\label{eq:dimleaf}
 \dim_{\mathbb{C}} \mathcal{M}_{\lambda,\underline{x},\mu} = 2 d_{\underline{\lambda}, \mu} .
\end{gather*}

We fix the singularity of the $L(x)$ at points $x \to x_i$ to be of the form $[g(x)] \sim ( x - x_{i})^{\check \omega_{\lambda_{i}^{t}}}$, up to a~regular factor, where $\check \omega_{k}$ is the $k$-th fundamental co-weight associated in Young notations to a~column of height $k$, and at infinity $x \to x_\infty$ we take the singularity to be $[g(x)] \sim (x^{-1})^{\sum_i \check \omega_{\mu^{t}_i}- \check \omega_{r}} $. Here $\check \omega_{r}$
is a co-weight associated to the column of height $r$ and denoting the diagonal homomorphism ${\rm GL}_1 \to T_{{\rm GL}_{r}}$ where $T$ stands for the maximal torus, that is a co-weight dual to the weight of the determinant line representation.

The determinant of $L(x)$ determined by the partition $\lambda$ is a polynomial of degree $|\lambda|$ with roots $x_i$ of degeneracy $\lambda_i^t$:
\begin{gather}\label{eq:detlax}
 \det \Llm=\prod_{i=1}^{\lambda_1}(x-x_i)^{\lambda_i^t} .
\end{gather}
The explicit form of the matrices $\Llm$ is given in Section~\ref{sec:lambda} for $\mu=\varnothing$, in Section~\ref{sec:mupart} for $\lambda=\varnothing$ and for arbitrary $\lambda$ and $\mu$ with $|\lambda| + |\mu| = r$ in Section~\ref{sec:lmpart}.

As explained in the introduction, we can allow parameters $x_i$ to collide in which case the dominant co-weight $\check \omega_{*}$ of the singularity at the collision point $x_{*}$ is represented by a partition composed of several columns from $\lambda$, and is equal to the sum of the fundamental co-weights associated to each individual column in $\lambda$. In this way we get a symplectic leaf $\mathcal{M}_{\underline{\lambda}, \underline{x}, \mu}$ whose singularity type at $x_{*} \in \underline{x}$ is described by a partition $\lambda_{*} \in \underline{\lambda}$. Bearing this in mind, in the following we assume that parameters $x_i$ are assigned to individual
columns $\lambda_i^{t}$ of a single partition~$\lambda$.

\section[Degree 1 symplectic leaves regular with fundamental singularity at infinity]{Degree 1 symplectic leaves regular\\ with fundamental singularity at infinity}\label{sec:lambda}

In this section we focus on the ${\rm GL}_r$ Lax matrices that correspond to arbitrary partitions $\lambda$ of size $|\lambda|$ and a single column $\mu$-partition, $\mu=1^{[r-|\lambda|]}$. In particular if $|\lambda| = r$ then $\mu$ is empty.

Since $\mu$ is a single column, the singularity at infinity is specified by a fundamental co-weight. The associated $A_{r-1}$ quiver gauge theory with fundamental hypermultiplets~\cite{Nekrasov:2012xe} differs from the
conformal class by absence of a~single fundamental multiplet in the node~$\mu_1^{t}$.

We will assume that each element of the transposed partition $\lambda^t$, i.e., each column $\lambda^{t}_i$ of the partition $\lambda$ specifies a singularity of the Lax matrix~$\Llmr$ at point $ x = x_i$ of the type~$\check{\omega}_{\lambda_i^{t}}$. Here $\check{\omega}_{k}$ denotes a fundamental co-weight of ${\rm GL}_r$ of the form $( \underbrace{1,\ldots,1}_{k},\underbrace{0, \ldots, 0}_{r-k})$ in the basis dual to the standard basis of weights of the defining representation.

This type of ${\rm GL}_r$ Lax matrices can be obtained by fusion of the fundamental solutions associated to a single column $\lambda = 1^{[|\lambda|]}$ and a single column $\mu = 1^{[r-|\lambda|]}$. The fundamental solutions are given in Section~\ref{sec:eleml}, and the fusion is described in Section~\ref{sec:fusel2}. The Lax matrices for arbitrary partitions $\lambda$ are given in Section~\ref{sec:fusel}. We closely follow~\cite{Bazhanov:2010jq} where the elementary building blocks were derived, the factorisation was discussed on the quantum level and a closed formula for the Lax matrices was obtained for the case $\lambda=(r)$, see also~\cite{Derkachov:2006fw}.

\subsection[Fundamental $(\lambda,\mu)$ orbits]{Fundamental $\boldsymbol{(\lambda,\mu)}$ orbits}\label{sec:eleml}
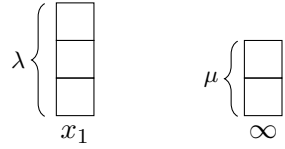
\begin{figure}[t] \centering

\begin{tikzpicture}
 \foreach \a in {1} {
 \begin{scope}[shift={(0.5*\a-0.1,1)}]
 \draw (0,0) rectangle (0.5,0.5);
 \end{scope}
 }
 \foreach \a in {1} {
 \begin{scope}[shift={(0.5*\a-0.1,0.5)}]
 \draw (0,0) rectangle (0.5,0.5);
 \end{scope}
 }
 \foreach \a in {1} {
 \begin{scope}[shift={(0.5*\a-0.1,0)}]
 \draw (0,0) rectangle (0.5,0.5);
 \end{scope}
 }

 \foreach \a in {6} {
 \begin{scope}[shift={(0.5*\a-0.1,0.5)}]
 \draw (0,0) rectangle (0.5,0.5);
 \end{scope}
 }
 \foreach \a in {6} {
 \begin{scope}[shift={(0.5*\a-0.1,0)}]
 \draw (0,0) rectangle (0.5,0.5);
 \end{scope}
 }

 \node [below ] at (0.65,0) {$x_1$};
 \node [below ] at (3.15,0) {$\infty$};

\draw [decorate,decoration={brace,amplitude=5pt},xshift=80pt,yshift=-14.5pt]
(0,0.5) -- (0,1.5)node [black,midway,xshift=-10pt] {\footnotesize
$\mu$};
\draw [decorate,decoration={brace,amplitude=5pt},xshift=7pt,yshift=-14.5pt]
(0,0.5) -- (0,2)node [black,midway,xshift=-10pt] {\footnotesize
$\lambda$};
\end{tikzpicture}
\caption{Single column partition for $r=5$ with $\lambda=1^{[3]}$ and $\mu=1^{[2]}$.}\label{fig:singcol}
\end{figure}

The fundamental building blocks are $r\times r$ matrices that correspond to the partition
\begin{gather*}
\mu=(\underbrace{1,\ldots,1}_{|\mu|}) ,\qquad \lambda=(\underbrace{1,\ldots,1}_{|\lambda|}) ,
\end{gather*}
with $r=|\lambda|+|\mu|$, see Fig.~\ref{fig:singcol}. They contain $|\lambda|\cdot|\mu|$ pairs of conjugate variables $(p_{ij},q_{ji})$ where $1\leq i\leq |\mu|$ and $|\mu|< j \leq r$ and can be written as
\begin{gather}\label{eq:elax}
\Llmr= \left(
 \begin{BMAT}[5pt]{c:c}{c:c}\ID&-P_{\mu,\lambda}\\
Q_{\lambda,\mu}& (x-x_1)\ID-Q_{\lambda,\mu}P_{\mu,\lambda}
 \end{BMAT}\right) .
\end{gather}
Here the upper diagonal block is of the size $|\mu|\times |\mu|$ and the lower one of size $|\lambda|\times |\lambda|$. The block matrices on the off-diagonal are parametrized as follows
\begin{gather*}
 (P_{\mu,\lambda})_{i,j} = p_{i, |\mu|+j}, \qquad (Q_{\lambda,\mu})_{i,j} = q_{|\mu|+i, j}.
\end{gather*}
The letter $\ID$ denotes the identity matrix of appropriate size. In particular we have $L_{1^{[r]},\underline{x},\varnothing}(x)=(x-x_1)\ID$ and $L_{\varnothing,\varnothing,1^{[r]}}=\ID$.

The matrices $\Llmr$ satisfy the Sklyanin relation \eqref{eq:skl} as can be verified by a direct computation using
\begin{gather*}
 \{(P_{\mu,\lambda})_{i,j},(Q_{\lambda,\mu})_{k,l}\}=\delta_{i,l}\delta_{k,j} .
\end{gather*}
Consequently one finds
\begin{gather*}
\{(Q_{\lambda,\mu}P_{\mu,\lambda})_{i,j},(Q_{\lambda,\mu})_{k,l}\}=+(Q_{\lambda,\mu})_{i,l}\delta_{k,j} ,\qquad \{(Q_{\lambda,\mu}P_{\mu,\lambda})_{i,j},(P_{\mu,\lambda})_{k,l}\}=-(P_{\mu,\lambda})_{k,j}\delta_{i,l} ,
\end{gather*}
and
\begin{gather*}
\{(Q_{\lambda,\mu}P_{\mu,\lambda})_{i,j},(Q_{\lambda,\mu}P_{\mu,\lambda})_{k,l}\}=\delta_{k,j}(Q_{\lambda,\mu}P_{\mu,\lambda})_{i,l}-\delta_{i,l}(Q_{\lambda,\mu}P_{\mu,\lambda})_{k,j} ,
\end{gather*}
which is sufficient in order to check the Sklyanin Poisson bracket. It is instructive to see that $\Llmr$ is factorized into a product of upper diagonal, diagonal and lower diagonal matrices:
\begin{gather*}
\Llmr=
 \left(
 \begin{BMAT}[5pt]{c:c}{c:c}
\ID& 0 \\
Q_{\lambda,\mu} & \ID
 \end{BMAT}\right)
\left( \begin{BMAT}[5pt]{c:c}{c:c}
\ID& 0 \\
0 & (x-x_1)\ID
 \end{BMAT}\right)
\left(
 \begin{BMAT}[5pt]{c:c}{c:c}
\ID& -P_{\mu,\lambda} \\
0 & \ID
 \end{BMAT}
\right) .
\end{gather*}
The determinant is
\begin{gather*}
 \det \Llmr=(x-x_1)^{|\lambda|} .
\end{gather*}

\subsection{Canonical coordinates on regular orbits}\label{sec:fusel}
In this section we will construct solutions $\Llmr$ for arbitrary partitions with $\lambda$ composed of columns $\lambda_i^{t}$ and a single column partition $\mu^t = \big(\mu_1^{t}\big)$.

The columns $\lambda_i^{t}$ are associated to fundamental singularities at $x = x_i$ of type $\lambda_i^{t}$, which means that the singularity of $\Llmr$ is in the conjugacy class of
\begin{gather*}
\operatorname{diag}(\underbrace{ (x - x_i), \dots, (x-x_i)}_{\lambda_i^{t}}, 1, \dots, 1), \qquad i=1,\ldots,\lambda_1,
\end{gather*}
i.e., distinct $\lambda_i^{t}$ eigenvalues of $\Llmr$ are vanishing linearly at $x = x_i$.

The column $\mu_1^{t}$ describes a fundamental singularity at $x = \infty$ which means that the singularity of $\Llmr$ at $ x \to \infty$ is in conjugacy class of
\begin{gather*}
x \operatorname{diag}\big(\underbrace{ x^{-1}, \dots, x^{-1}}_{\mu_1^{t}}, 1, \dots, 1\big) .
\end{gather*}
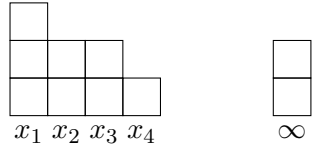
\begin{figure}\centering
\begin{tikzpicture}
\foreach \a in {1,2,3,4} {
 \begin{scope}[shift={(0.5*\a,0)}]
 \draw (0,0) rectangle (0.5,0.5);
 \end{scope}
 }
 \foreach \a in {1,2,3} {
 \begin{scope}[shift={(0.5*\a,0.5)}]
 \draw (0,0) rectangle (0.5,0.5);
 \end{scope}
 }
 \foreach \a in {1} {
 \begin{scope}[shift={(0.5*\a,1)}]
 \draw (0,0) rectangle (0.5,0.5);
 \end{scope}
 }
 \node [below ] at (0.75,0) {$x_1$};
 \node [below ] at (1.25,0) {$x_2$};
 \node [below ] at (1.75,0) {$x_3$};
 \node [below ] at (2.25,0) {$x_4$};

 \foreach \a in {0,1} {
 \begin{scope}[shift={(4,0.5*\a)}]
 \draw (0,0) rectangle (0.5,0.5);
 \end{scope}
 }
 \node [below] at (4.25,0) {$\infty$};
 \end{tikzpicture}
 \caption{Regular partition with $\lambda=(4,3,1)$, $\mu=(1,1)$ and $r=10$.}
 \end{figure}

We will prove recursively that regular Lax matrices $\Llmr$ can be parametrized as a~block matrix
\begin{gather}\label{eq:laxregular}
\Llmr =
\left(\begin{BMAT}[5pt]{c:c}{c:c}
 \ID& - P_{\mu,\lambda} \\
 Q_{\lambda, \mu} &x\ID- J_{\lambda, \lambda} - Q_{\lambda, \mu} P_{\mu, \lambda}
 \end{BMAT}\right) ,
\end{gather}
where the upper-left block is of size $\mu_1^{t} \times \mu_1^{t}$ and bottom-right block is of size $|\lambda| \times |\lambda|$. The matrix elements of block $P_{\mu, \lambda}$ and block $Q_{\lambda, \mu}$ are canonically conjugated variables with
\begin{gather*}
 \{ (P_{\mu,\lambda})_{ij}, (Q_{\lambda, \mu})_{kl} \} = \delta_{il} \delta_{jk}
\end{gather*}
and the matrix elements of $J_{\lambda,\lambda}$ satisfy the algebra of $\lambda \times \lambda$-matrices with respect to the Poisson brackets{\samepage
\begin{gather}\label{eq:gln}
 \{\J_{ij},\J_{kl}\}=\delta_{il} \J_{kj}-\delta_{kj} \J_{il} ,
\end{gather}
while Poisson commuting with matrix elements of $P_{\mu, \lambda}$ and $Q_{\lambda, \mu}$.}

The matrix elements of the $|\lambda| \times |\lambda|$ matrix $J_{\lambda,\lambda}$ have an explicit parametrization in terms of the canonically conjugated coordinates as follows
\begin{gather}\label{eq:Jmatrix}
 J_{\lambda,\lambda} = Q_{\lambda, \lambda} ( X_{\lambda} + [P_{\lambda, \lambda} Q_{\lambda, \lambda}]_{+})Q_{\lambda, \lambda}^{-1} ,
\end{gather}
cf.~\cite{babich2016birational, ShapiroThesis} and Appendix~\ref{sec:twisted-flag}. Here $X_\lambda$ denotes the diagonal matrix
\begin{gather}\label{eq:Xlambda}
 X_{\lambda}=\diag (\underbrace{x_1,\ldots,x_1}_{\lambda_1^t},\underbrace{x_2,\ldots,x_2}_{\lambda_2^t},\ldots,\underbrace{x_{\lambda_1},\ldots,x_{\lambda_1}}_{\lambda_{\lambda_1}^t}) .
\end{gather}
The corresponding blocks on the diagonal are of the size $\lambda^t_1,\ldots, \lambda_{\lambda_1}^t$. The matrix $[P_{\lambda, \lambda} Q_{\lambda, \lambda}]_+$ is strictly upper block triangular and reads
\begin{gather*}%\label{eq:dmatrix}
[P_{\lambda, \lambda} Q_{\lambda, \lambda}]_+=\left(
 \begin{BMAT}[5pt]{c:c:c:c:c}{c:c:c:c:c}
 0& \pt_{1,2}& \pt_{1,3}&\cdots& \pt_{1,\lambda_1}\\
0&0& \pt_{2,3}&\cdots& \pt_{2,\lambda_1}\\
 0&0&0&\ddots&\vdots\\
0&0&0&0&\pt_{\lambda_1-1,\lambda_1}\\
0&0&0&0&0\\
 \end{BMAT}
\right) .
\end{gather*}
Here the matrices $\pt_{ij}$ are of the size $\lambda_{i}^t\times \lambda_{j}^t$ and explicitly given by
\begin{gather}\label{eq:pt}
 \pt_{ij}=(P_{\lambda, \lambda})_{ij}+\sum_{k=j+1}^{\lambda_1}(P_{\lambda, \lambda})_{ik}(Q_{\lambda, \lambda})_{kj} .
\end{gather}
The matrix $Q_{\lambda, \lambda}$ is lower triangular and only depends on the variables $\underline{q}$ while $P_{\lambda,\lambda}$ is upper triangular and only depends on the variables $\underline{p}$. They read
\begin{gather}
 P_{\lambda,\lambda}=\left(
 \begin{BMAT}[5pt]{c:c:c:c:c}{c:c:c:c:c}
0&P_{1,2}&P_{1,3}&\cdots&P_{1,\lambda_1}\\
 0&0&P_{2,3}&\cdots&P_{2,\lambda_1}\\
 0&0&0&\ddots&\vdots\\
 0&0&0&\ddots&P_{\lambda_1-1,\lambda_1}\\
 0&0&0&0&0\\
 \end{BMAT}
\right) ,\nonumber\\
Q_{\lambda,\lambda}=\left(
 \begin{BMAT}[5pt]{c:c:c:c:c}{c:c:c:c:c}
 \ID &0&0&0&0\\
 Q_{2,1}&\ID &0&0&0\\
 Q_{3,1}&Q_{3,2}&\ID &0&0\\
 \vdots&\vdots&\ddots&\ddots&0\\
 Q_{\lambda_1,1}&Q_{\lambda_1,2}&\cdots&Q_{\lambda_1,\lambda_1-1}&\ID \\
 \end{BMAT},
\right)\label{eq:Umatrix}
\end{gather}
where $Q_{ij}$ and $P_{ij}$ denote $\lambda_i^t\times\lambda_j^t$ block matrices explicitly given by
\begin{alignat*}{3}
& (Q_{ij})_{kl} = q_{\ell(i)+k,\ell(j)+l}, \qquad && k \in [1,\lambda_i^{t}], \quad l \in [1, \lambda_j^{t}], & \\
& (P_{ij})_{kl}=p_{\ell(i) + k , \ell(j) + l}, \qquad && k \in [1,\lambda_i^{t}], \quad l \in [1, \lambda_j^{t}] .&
\end{alignat*}
Here we defined $\ell(i)=|\mu|+\sum\limits_{k=1}^{i-1}\lambda_k^t$.

The realization (\ref{eq:Jmatrix}) of the $\mathfrak{gl}(|\lambda|)$ algebra, also known as free field representation, can be constructed as algebra of twisted differential operators on the flag variety $G/P_{\lambda, +}$.
Here $G = {\rm GL}(|\lambda|)$ and $P_{\lambda, +}$ denotes a parabolic subgroup of ${\rm GL}(|\lambda|)$ whose Levi is $\prod_{i} {\rm GL}(\lambda_i^{t})$. The big cell of the flag variety $G/P_{\lambda,+}$ is identified with the $\lambda^{t}$-blocks unipotent subgroup $N_{\lambda, -}$ whose elements are represented by matrices $Q_{\lambda, \lambda}$ as in~(\ref{eq:Umatrix}). In the classical limit twisted differential operators in $J_{\lambda, \lambda}$ form a co-adjoint orbit $O_{X_{\lambda}}$ in the dual Lie algebra $\mathfrak{g}^{*}$ for $\mathfrak{g} = \mathfrak{gl}(|\lambda|)$ of the semi-simple element $X_{\lambda}$ (\ref{eq:Xlambda}). See details in Appendix~\ref{sec:twisted-flag}.

The number of pairs of conjugate variables in the Lax matrix \eqref{eq:laxregular} agrees with~\eqref{eq:dimform}. There are $\mu_1^t\times |\lambda|$ pairs in $P_{\mu,\lambda}$, $Q_{\lambda,\mu}$ and $\sum\limits_{i<j}\lambda_i^t\lambda_j^t$ in~$J_{\lambda,\lambda}$. Further we verify that the determinant of~\eqref{eq:laxregular} agrees with~\eqref{eq:detlax}.

\subsection{Regular orbits from fusion of fundamental orbits} \label{sec:fusel2}

\newcommand{\blockA}{{\tilde\mu}}
\newcommand{\blockB}{\lambda'}
\newcommand{\blockC}{\lambda}
\newcommand{\kblockA}{{|\tilde\mu|}}
\newcommand{\kblockB}{|\lambda'|}
\newcommand{\kblockC}{{|\lambda|}}

We will construct the solution in the form of (\ref{eq:laxregular}) associated to regular $(\tilde \lambda, \tilde \mu)$ by fusion of two solutions associated to $(\lambda, \mu)$ and $(\lambda', \mu')$. Here $(\tilde \lambda, \tilde \mu)$ is defined such that
\begin{gather*}
 \tilde \lambda^t = \big({\lambda'}^t, {\lambda^{t}}\big), \qquad |\tilde \lambda| = |\lambda| + |\lambda'| ,
 \end{gather*}
 where $({\lambda'}^t, {\lambda^{t}})$ denotes the partition given by the union of ${\lambda'}^t$ and ${\lambda^{t}}$. The partitions~$\mu$,~$\mu'$ and~$\tilde \mu$ are single columns
\begin{gather*}
 {\mu}^{t} = ( r - |\lambda|),\qquad {\mu'}^{t} = ( r - |\lambda'|),\qquad {\tilde \mu}^{t} = \big( r - |\tilde \lambda|\big) .
 \end{gather*}

Then, by assumption of the recursion we represent $\Llm$ in the form
\begin{gather}\label{eq:lax1}
\Llm=\left(\begin{BMAT}[5pt]{c:c:c}{c:c:c}
 \ID&0&-P_{\blockA,{\blockC}}\\
 0&\ID&-P_{{\blockB},{\blockC}}\\
 Q_{{\blockC},\blockA}&Q_{{\blockC},{\blockB}}&x\ID-\J_{{\blockC},{\blockC}}-Q_{{\blockC},\blockA}P_{\blockA,{\blockC}}-Q_{{\blockC},{\blockB}}P_{{\blockB},{\blockC}}
 \end{BMAT}
\right) .
\end{gather}
The blocks on the diagonal are of the size $\kblockA$, $\kblockB$ and $\kblockC$ respectively, with $\kblockA+\kblockB+\kblockC=r$. The matrix $\Llm$ explicitly depends on $\kblockC(r-\kblockC)$ pairs of conjugate variables arranged in the matrices $P_{\blockA,{\blockC}}$, $P_{{\blockB},{\blockC}}$ and $Q_{{\blockC},\blockA}$, $Q_{{\blockC},{\blockB}}$ defined as
\begin{alignat*}{3}%\label{eq:qmat1}
& (P_{\blockA,{\blockC}})_{ij} = p_{i,\kblockA +\kblockB+ j} ,\qquad &&(P_{{\blockB},{\blockC}})_{ij}= p_{\kblockA+i, \kblockA + \kblockB + j} ,&\\
& (Q_{{\blockC},\blockA})_{ij} = q_{\kblockA+\kblockB+i, j} ,\qquad&& (Q_{{\blockC},{\blockB}})_{ij}= q_{\kblockA + \kblockB + i,\kblockA+ j} ,&
\end{alignat*}
and the matrix $J_{{\blockC},{\blockC}}$ of the size $\kblockC \times \kblockC$ defined in~(\ref{eq:Jmatrix}).

Similarly, we consider another Lax matrix
\begin{gather}\label{eq:lax2}
 L'_{\lambda',\underline{x}',\mu'}(x;\underline{p}',\underline{q}')=\left(\begin{BMAT}[5pt]{c:c:c}{c:c:c}
 \ID& -P_{\blockA,{\blockB}}'&0\\
 Q_{{\blockB},\blockA}' &x\ID- \J_{{\blockB},{\blockB}}'- Q_{{\blockB},\blockA}' P_{\blockA,{\blockB}}'+ P_{{\blockB},{\blockC}}' Q_{{\blockC},{\blockB}}'&- P_{{\blockB},{\blockC}}'\\
 0&-Q_{{\blockC},{\blockB}}'&\ID
 \end{BMAT}
\right) ,
\end{gather}
with the same block structure as in \eqref{eq:lax1}. This matrix $ L'_{\lambda',\underline{x}',\mu'}(x;\underline{p}',\underline{q}')$ explicitly depends on $\kblockB(r-\kblockB)$ pairs of conjugate variables
\begin{alignat*}{3}%\label{eq:qmat2}
& (Q'_{{\blockB},\blockA})_{ij}= q'_{\kblockA + i, j}, \qquad&& (Q_{{\blockC},{\blockB}}')_{ij} = q_{\kblockA + \kblockB + i, \kblockA+j}',& \\
& (P'_{\blockA,{\blockB}})_{ij} = p_{i, \kblockA + j}', \qquad &&(P_{{\blockB},{\blockC}}')_{ij} = p_{\kblockA + i, \kblockA + \kblockB + j}',&
 \end{alignat*}
and another set of variables appearing in the expression for $\J'_{\blockB,\blockB}$ like in~(\ref{eq:Jmatrix}). The matrix $ L'_{\lambda',\underline{x}',\mu'}(x;\underline{p}',\underline{q}')$ that appears in~(\ref{eq:lax2}) is obtained from the canonical form~(\ref{eq:laxregular}) by permutation, that is a conjugation by an element of the Weyl group of ${\rm GL}_r$, and a canonical transformation in the variables~$\underline{p}$ and~$\underline{q}$.

\looseness=-1 In the next step we multiply the matrices \eqref{eq:lax1} and \eqref{eq:lax2}. It was pointed out in \cite{Bazhanov:2010jq} for the corresponding solutions of the quantum Yang--Baxter equation that the product can be written as
\begin{gather}\label{eq:prodll}
\Llm L'_{\lambda',\underline{x}',\mu'}(x;\underline{p}',\underline{q}')=\tilde Q' \tilde L_{\tilde \lambda,\tilde{\underline{x}},\tilde\mu}(x;\underline{\tilde p},\underline{\tilde q}) .
\end{gather}
Here $\tilde L_{\tilde \lambda,\tilde{\underline{x}},\tilde\mu}(x;\underline{\tilde p},\underline{\tilde q})$ denotes a~spectral parameter dependent Lax matrix and Casimir $\tilde Q'$ is a~lower triangular matrix. They are of the form
\begin{gather*} \tilde L_{\tilde \lambda,\tilde{\underline{x}},\tilde\mu}(x;\underline{\tilde p},\underline{\tilde q})=WU\!\left(\!\begin{BMAT}[5pt]{c:c:c}{c:c:c}
 \ID&0 &0\\
 0&x\ID-\J'_{\blockB,\blockB}&-\tilde P_{{\blockB},{\blockC}}\\
 0&0&x\ID-\J_{\blockC,\blockC}\\
 \end{BMAT}\!
\right)\!U^{-1}V^{-1} ,\qquad
 \tilde Q'=\left(\!\begin{BMAT}[5pt]{c:c:c}{c:c:c}
 \ID&0&0\\
 0&\ID&0\\
 0&\tilde Q_{{\blockC},{\blockB}}'&\ID
 \end{BMAT}\!
\right) ,
\end{gather*}
where
\begin{gather*}
 W=\left(\begin{BMAT}[5pt]{c:c:c}{c:c:c}
 \ID&0&0\\
 \tilde Q_{{\blockB},\blockA}&\ID&0\\
 \tilde Q_{{\blockC},\blockA}&0&\ID
 \end{BMAT}
\right) ,\qquad
 U=\left(\begin{BMAT}[5pt]{c:c:c}{c:c:c}
 \ID&0&0\\
 0&\ID&0\\
 0& \tilde Q_{{\blockC},{\blockB}}&\ID
 \end{BMAT}
\right) ,\qquad V=\left(\begin{BMAT}[5pt]{c:c:c}{c:c:c}
 \ID&\tilde P_{\blockA,{\blockB}}&\tilde P_{\blockA,{\blockC}}\\
 0&\ID&0\\
 0&0&\ID
 \end{BMAT}
 \right)
\end{gather*}
expressed in terms of the new variables
\begin{alignat}{3}
& \tilde P_{{\blockB}{\blockC}}=P_{{\blockB}{\blockC}}'+P_{{\blockB}{\blockC}}-Q'_{{\blockB}\blockA}P_{\blockA{\blockC}}, \qquad && \tilde Q_{{\blockC}{\blockB}}=Q_{{\blockC}{\blockB}}',& \nonumber\\
&\tilde P_{\blockA{\blockB}}=P_{\blockA{\blockB}}'-P_{\blockA{\blockC}}Q'_{{\blockC}{\blockB}}, \qquad && \tilde Q_{{\blockB}\blockA}=Q_{{\blockB}\blockA}', &\nonumber\\
&\tilde P_{\blockA{\blockC}}=P_{\blockA{\blockC}}, \qquad &&\tilde Q_{{\blockC}\blockA}=Q_{{\blockC}\blockA}+Q'_{{\blockC}{\blockB}}Q'_{{\blockB}\blockA},& \nonumber\\
&\tilde P'_{{\blockB}{\blockC}}= P_{{\blockB}{\blockC}}, \qquad && \tilde Q'_{{\blockC}{\blockB}}=Q_{{\blockC}{\blockB}}-Q'_{{\blockC}{\blockB}} .&\label{eq:trans1}
 \end{alignat}

The polynomial change of variables (\ref{eq:trans1}) is a symplectomorphism (i.e., canonical transformation) as we can directly verify. Indeed, computing the differentials we find
\begin{gather*}
 {\rm d} \tilde P_{{\blockB}{\blockC}} \wedge {\rm d} \tilde Q_{{\blockC}{\blockB}} = ({\rm d} P_{{\blockB}{\blockC}} ' + {\rm d}P_{{\blockB}{\blockC}} - {\rm d}Q_{{\blockB}\blockA}' P_{\blockA{\blockC}} - Q_{{\blockB}\blockA}' {\rm d} P_{\blockA{\blockC}}) \wedge {\rm d} Q_{{\blockC}{\blockB}}', \\
 {\rm d} \tilde P_{\blockA{\blockB}} \wedge {\rm d} \tilde Q_{{\blockB}\blockA} = ({\rm d} P_{\blockA{\blockB}}' - {\rm d}P_{\blockA{\blockC}} Q_{{\blockC}{\blockB}}' - P_{\blockA{\blockC}} {\rm d} Q_{{\blockC}{\blockB}}') \wedge {\rm d} Q_{{\blockB}\blockA}', \\
 {\rm d} \tilde P_{\blockA{\blockC}} \wedge {\rm d} \tilde Q_{{\blockC}\blockA} = {\rm d} P_{\blockA{\blockC}} \wedge ( {\rm d}Q_{{\blockC}\blockA} + {\rm d}Q_{{\blockC}{\blockB}}' Q_{{\blockB}\blockA}' + Q_{{\blockC}{\blockB}}' {\rm d} Q_{{\blockB}\blockA}')\\
 {\rm d} \tilde P_{{\blockB}{\blockC}}' \wedge {\rm d} \tilde Q_{{\blockC}{\blockB}}' = {\rm d}P_{{\blockB}{\blockC}} \wedge ( {\rm d}Q_{{\blockC}{\blockB}} - {\rm d}Q_{{\blockC}{\blockB}}'),
\end{gather*}
and hence, after cancellations, we find that the canonical symplectic form is invariant
\begin{gather*}
 \sum_{I \in \{\blockA{\blockB}, \blockA{\blockC}, {\blockB}{\blockC} \} } {\rm d} \tilde P_{I} \wedge {\rm d} \tilde Q_{I^t} + \sum_{I \in \{\blockB \blockC \} } {\rm d}\tilde P_{I}' \wedge {\rm d} \tilde Q_{i}' \\
 \qquad {} = \sum_{I \in \{\blockA{\blockC}, {\blockB}{\blockC}\}} {\rm d} P_{I} \wedge {\rm d} Q_{I^t} + \sum_{I \in \{\blockA{\blockB}, {\blockB}{\blockC}\}} {\rm d} P'_{I} \wedge {\rm d}Q'_{I^{t}} .
\end{gather*}

In analogy to the Yang--Baxter equation, the product of two solutions to the Sklyanin relation~\eqref{eq:skl} with different sets of conjugate variables $(\underline{p},\underline{q})$ is again a solution to the Sklyanin relation~\eqref{eq:skl}.

Therefore the matrix in \eqref{eq:prodll} satisfies the Sklyanin bracket when taking the Poisson bracket with respect to the variables $(\underline{\tilde p},\underline{\tilde q})$ which denote the elements of the matrices defined in~\eqref{eq:trans1}.

Finally, we note that the result is independent of $\tilde P_{{\blockB},{\blockC}}'$ which allows us to strip off the matrix~$\tilde Q'$ from~\eqref{eq:prodll}. Thus we conclude that
\begin{gather*}%\label{eq:lax3}
\tilde L_{\tilde \lambda,\tilde{\underline{x}},\tilde\mu}(x;\underline{\tilde p},\underline{\tilde q})=
\left(\begin{BMAT}[5pt]{c:c}{c:c}
 \ID&-\tilde P_{\blockA,\tilde\lambda}\\
 \tilde Q_{\tilde\lambda,\blockA} &x\ID-\tilde \J_{\tilde \lambda, \tilde\lambda} - \tilde Q_{\tilde\lambda,\blockA}\tilde P_{\blockA,\tilde\lambda}
 \end{BMAT}
\right) ,
\end{gather*}
with
\begin{gather*}%\label{eq:qmat3}
\big(\tilde P_{\blockA,\tilde\lambda}\big)_{ij} = \tilde p_{i, \kblockA +j}, \qquad \big( \tilde Q_{\tilde\lambda,\blockA} \big)_{ij} = \tilde q _{\kblockA + i, j}
\end{gather*}
is a solution of the Sklyanin relation. Here the generators $\tilde J_{\tilde \lambda, \tilde \lambda}$ of the $\mathfrak{gl}(\kblockB+\kblockC)$ subalgebra are realised as
\begin{gather*}%\label{eq:Jtlambda}
 \tilde \J_{\tilde \lambda ,\tilde \lambda}=\left(\begin{BMAT}[5pt]{c:c}{c:c}
 \ID&0\\
 \tilde Q_{\lambda,\lambda'}&\ID
 \end{BMAT}
\right)\cdot\left(\begin{BMAT}[5pt]{c:c}{c:c}
 \J_{\lambda', \lambda'}'& \tilde P_{\lambda',\lambda}\\
 0&\J_{\lambda, \lambda}
 \end{BMAT}
\right)\cdot\left(\begin{BMAT}[5pt]{c:c}{c:c}
 \ID&0\\
 -\tilde Q_{\lambda,\lambda'}&\ID
 \end{BMAT}
\right) ,
\end{gather*}
where
\begin{gather*}%\label{eq:qmat33}
\big(\tilde P_{{\blockB},{\blockC}}\big)_{ij} = \tilde p_{\kblockA+i, \kblockA+\kblockB + j}, \qquad \big(\tilde Q_{{\blockC},{\blockB}}\big)_{ij}= \tilde q _{\kblockA+\kblockB + i,\kblockA+ j} .
\end{gather*}
Let us remark that here we have chosen a certain order of fusion, but depending on the order we would get different parametrization related by a polynomial choice of variables, see, e.g., Appendix~\ref{sec:clusterstructures}. It would be interesting to explore the resulting cluster structure in more details.

\subsubsection{Linear fusion}
Now to demonstrate the particular parametrization~(\ref{eq:Jmatrix}) for $\tilde J_{\tilde \lambda, \tilde \lambda}$ it is sufficient to assume that~$\lambda'$ is a single column partition ${\lambda'}^{t} = ({\lambda'}^{t}_1)$ while $\lambda$ is an arbitrary collection of columns. In this case
\begin{gather*}
 J'_{\lambda', \lambda'} = X_{\lambda'} ,\qquad
 J_{\lambda, \lambda} = Q_{\lambda, \lambda} (X_{\lambda} + [P_{\lambda, \lambda} Q_{\lambda, \lambda}]_{+}) Q_{\lambda, \lambda}^{-1} .
\end{gather*}

Then we find that $\tilde J_{\tilde \lambda, \tilde \lambda}$ can be again represented in the form
\begin{gather}
 \tilde \J_{\tilde \lambda \tilde \lambda}=\left(\begin{BMAT}[5pt]{c:c}{c:c}
 \ID&0\\
 \tilde Q_{\lambda\lambda'}&\ID
 \end{BMAT}
\right)\left(\begin{BMAT}[5pt]{c:c}{c:c}
 \ID&0\\
 0 & Q_{\lambda\lambda}
 \end{BMAT}
\right)
\left(\begin{BMAT}[5pt]{c:c}{c:c}
 X_{\lambda'}& \tilde P_{\lambda'\lambda} Q_{\lambda, \lambda}\\
 0& X_{\lambda} + [P_{\lambda, \lambda} Q_{\lambda, \lambda}]_{+}
 \end{BMAT}
\right)\nonumber\\
\hphantom{\tilde \J_{\tilde \lambda \tilde \lambda}=}{}\times
\left(\begin{BMAT}[5pt]{c:c}{c:c}
 \ID&0\\
 0 & Q_{\lambda, \lambda}
 \end{BMAT}
\right)^{-1}
\left(\begin{BMAT}[5pt]{c:c}{c:c}
 \ID&0\\
 \tilde Q_{\lambda\lambda'}&\ID
 \end{BMAT}
\right)^{-1}\label{eq:Jfactor}
\end{gather}
or
 \begin{gather*}
 \tilde J_{\tilde\lambda, \tilde\lambda} = Q_{\tilde\lambda, \tilde\lambda} (X_{\tilde\lambda} + [P_{\tilde\lambda, \tilde\lambda} Q_{\tilde\lambda, \tilde\lambda}]_{+}) Q_{\tilde\lambda, \tilde\lambda}^{-1}
 \end{gather*}
 with
 \begin{gather*}
 Q_{\tilde \lambda, \tilde \lambda} =
 \left(\begin{BMAT}[5pt]{c:c}{c:c}
 \ID&0\\
 \tilde Q_{\lambda\lambda'}& Q_{\lambda, \lambda}
 \end{BMAT}
\right), \qquad P_{\tilde \lambda, \tilde \lambda} =
 \left(\begin{BMAT}[5pt]{c:c}{c:c}
 0 & \tilde P_{\lambda' \lambda} \\
 0 & P_{\lambda, \lambda}
 \end{BMAT}
\right) .
 \end{gather*}
As a consequence it follows that the Lax matrices \eqref{eq:laxregular} satisfy the Sklyanin relation \eqref{eq:skl}.

\section{Degree 1 symplectic leaves singular only at infinity}\label{sec:mupart}

\newcommand{\ia}{\alpha}
\newcommand{\ib}{\beta}
\newcommand{\ic}{\gamma}

In the following section we focus on the Lax matrices that correspond to $\lambda=\varnothing$ and arbitrary partition~$\mu$. Similar to the case labelled by pure $\lambda$ partitions in Section~\ref{sec:lambda} the present case can be obtained from fusion of the basic building blocks. These basic building blocks are generalisations of the well-known Lax matrix of the Toda chain~\cite{FaddeevBook} corresponding to the partition $\mu=(2)$. They are introduced in Sections~\ref{sec:elmu} and~\ref{sec:elmuu}. The Lax matrices for arbitrary partitions $\mu$ are presented in Section~\ref{sec:fullmupart}. As discussed in Section~\ref{sec:fac} we can apply a similar fusion procedure as in Section~\ref{sec:fusel} to derive the general form of the Lax matrices.

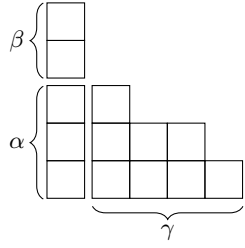
\begin{figure}\centering
\begin{tikzpicture}
\foreach \a in {2,3,4,5} {
 \begin{scope}[shift={(0.5*\a,0)}]
 \draw (0,0) rectangle (0.5,0.5);
 \end{scope}
 }
 \foreach \a in {2,3,4} {
 \begin{scope}[shift={(0.5*\a,0.5)}]
 \draw (0,0) rectangle (0.5,0.5);
 \end{scope}
 }
 \foreach \a in {2} {
 \begin{scope}[shift={(0.5*\a,1)}]
 \draw (0,0) rectangle (0.5,0.5);
 \end{scope}
 }
 \foreach \a in {1} {
 \begin{scope}[shift={(0.5*\a-0.1,1.5+0.1)}]
 \draw (0,0) rectangle (0.5,0.5);
 \end{scope}
 }
 \foreach \a in {1} {
 \begin{scope}[shift={(0.5*\a-0.1,2+0.1)}]
 \draw (0,0) rectangle (0.5,0.5);
 \end{scope}
 }
 \foreach \a in {1} {
 \begin{scope}[shift={(0.5*\a-0.1,1)}]
 \draw (0,0) rectangle (0.5,0.5);
 \end{scope}
 }
 \foreach \a in {1} {
 \begin{scope}[shift={(0.5*\a-0.1,0.5)}]
 \draw (0,0) rectangle (0.5,0.5);
 \end{scope}
 }
 \foreach \a in {1} {
 \begin{scope}[shift={(0.5*\a-0.1,0)}]
 \draw (0,0) rectangle (0.5,0.5);
 \end{scope}
 }

\draw [decorate,decoration={brace,amplitude=5pt},xshift=10pt,yshift=17pt]
(0,1) -- (0,2)node [black,midway,xshift=-10pt] {\footnotesize
$\beta$};

\draw [decorate,decoration={brace,amplitude=5pt},xshift=10pt,yshift=-14.5pt]
(0,0.5) -- (0,2)node [black,midway,xshift=-10pt] {\footnotesize
$\alpha$};

\draw [decorate,decoration={brace,amplitude=5pt},xshift=85.5pt,yshift=-3pt
,rotate=90]
(0,0) -- (0,2)node [black,midway,xshift=0pt,yshift=-10pt] {\footnotesize
$\gamma$};
\end{tikzpicture}
 \caption{Example of the decomposition in \eqref{eq:mudec} for $\mu=(5,4,2,1,1)$. We have $\alpha=(1,1,1)$, $\beta=(1,1)$ and $\gamma=(4,3,1)$.} \label{fig:decompose}
\end{figure}

To describe the Lax matrices it is convenient to introduce the partitions
\begin{gather}\label{eq:mudec}
 \alpha=\big(\underbrace{1,\ldots,1}_{\mu_2^t}\big) ,\qquad \beta=\big(\underbrace{1,\ldots,1}_{\mu_1^t-\mu_2^t}\big) ,\qquad \gamma=\big(\mu_2^t,\ldots,\mu_{\mu_1}^t\big)^t ,
\end{gather}
as shown in Fig.~\ref{fig:decompose}. The partition $\mu$ is then written as $ \mu^t=\big(|\ia|+|\ib|,\ic_1^t,\ldots,\ic_{\ic_1}^t\big)$. For simplicity we are only considering partitions with $\mu_i\geq \mu_j$ for $1\leq i<j\leq \mu^t_1$.

\subsection[Elementary $\mu$ partitions: $\ia=\ic$ and $\ib=0$]{Elementary $\boldsymbol{\mu}$ partitions: $\boldsymbol{\ia=\ic}$ and $\boldsymbol{\ib=0}$}\label{sec:elmu}

First we introduce the Lax matrices that correspond to the partitions $\lambda=\varnothing$ and $\mu=2^{[\frac{r}{2}]}$ with
\begin{gather*}
 \alpha=(\underbrace{1,\ldots,1}_{\mu_1^t}) ,\qquad \beta=0 ,\qquad \gamma=(\underbrace{1,\ldots,1}_{\mu_2^t}) ,
\end{gather*}
where $\mu_1^t=\mu_2^t=\frac{r}{2}$ and $r\in 2\mathbb{N}$. The Lax matrices $ \Lm=L_{\varnothing,\varnothing,\mu}(x;\underline{p},\underline{q})$ are $r\times r$ matrices with $|\ia|+|\ic|=r$ whose determinant evaluates to unity. They contain $\big(\frac{r}{2}\big)^2$ pairs of conjugate variables $\big(p_I,q^I\big)$ and can be written in the form
\begin{gather}\label{eq:lax22alt}
 \Lm=\left(\begin{BMAT}[5pt]{c:c}{c:c}
0&K_{\ia,\ic}\\ \bar K_{\ic,\ia}& x\ID-F_{\ic,\ic}
 \end{BMAT}\right) .
\end{gather}
For later purposes we labeled the upper block by $\ia$ and the lower block by $\ic$ such that the block on the diagonal are of equal size $|\ia|\times |\ia|$ and $|\ic|\times |\ic|$ respectively. Further we introduced the matrices
\begin{gather}\label{eq:G}
 F_{\ic,\ic}=Q_-GQ_-^{-1} ,\qquad\text{with}\quad G= P_0+[P_+Q_-]_++ Q_0[Q_+P_-]_-Q_0^{-1} ,
\end{gather}
where $[\phantom{x}]_{\pm}$ denotes the projection on the upper and lower diagonal part respectively and
\begin{gather}\label{eq:abc}
\bar K_{\ic,\ia}=Q_-Q_0Q_+ ,\qquad K_{\ia,\ic}=-Q_+^{-1}Q_0^{-1}Q_-^{-1}=-\bar K_{\ic,\ia}^{-1} .
\end{gather}
The matrices $Q_{\pm,0}$ are parametrized in terms of the conjugate variables $(\underline{p},\underline{q})$ as follows
\begin{gather*}
Q_-=\ID + \sum_{|\mu|\geq i>j> \mu_1^t }q_{ij} e_{ij} ,\qquad Q_+=\ID +\sum_{\mu_1^t< i<j\leq |\mu|}q_{ij}e_{ij} ,\qquad Q_0=\sum_{i=\mu_1^t+1}^{|\mu|} {\rm e}^{q_{ii}}e_{ii} ,
\end{gather*}
and
\begin{gather*}
P_-=\sum_{|\mu|\geq i>j>\mu^t_1 }p_{ij} e_{ij} ,\qquad P_+=\sum_{\mu_1^t< i<j\leq |\mu|}q_{ij}e_{ij} ,\qquad P_0=\sum_{i=\mu_1+1}^{|\mu|} p_{ii}e_{ii} .
\end{gather*}
We note that $Q_+$ is an upper triangular matrix containing variables $q_{ij}$ with $i>j$, while $Q_-$ is lower triangular containing the variables $q_{ij}$ with $i<j$. The diagonal matrix $Q_0$ only contains the exponential function of~$q_{ii}$. All variables~$p_{ij}$ are contained in $G$ which is decomposed as the sum of a diagonal, a lower diagonal and an upper diagonal matrix.

The Sklyanin relation is equivalent to the commutators
\begin{gather}\label{eq:FK}
 \{F_{ij},K_{kl}\}=-K_{kj}\delta_{il}, \qquad \{F_{ij},\Kb_{kl}\}=\Kb_{il}\delta_{kj}, \qquad \{\K_{ij},\Kb_{kl}\}=0,\\
 \label{eq:FF}
\{F_{ij},F_{kl}\}=\delta_{kj} F_{il}- \delta_{il} F_{kj} .
\end{gather}
Here the latter can be identified with commutators of the $\mathfrak{gl}(\frac{r}{2})$ algebra, while the parametrization of $K_{\alpha, \gamma}$ is given in terms of a Gauss decomposition of ${\rm GL}(\frac{r}{2})$. These relations are verified explicitly in Appendix~\ref{app:proof}.

For $\mu=(2)$, i.e., $|\ia|=|\ic|=1$, the Lax matrix in \eqref{eq:lax22alt} reproduces the well known Lax matrix for the Toda chain
\begin{gather*}
 L_{(2)}(x;p,q)=\left(\begin{matrix}
 0&-{\rm e}^{-q}\\
 {\rm e}^q&x-p
 \end{matrix}
\right) .
\end{gather*}

\subsection[Elementary $\mu$ partitions: $\ia=\ic$ and $\ib\neq 0$]{Elementary $\boldsymbol{\mu}$ partitions: $\boldsymbol{\ia=\ic}$ and $\boldsymbol{\ib\neq 0}$}\label{sec:elmuu}
We can extend the elementary Lax matrices to the case $\ib\neq 0$ which can be used to obtain the Lax matrices for arbitrary partitions $\mu$. They correspond to the partitions
\begin{gather*}
 \mu=(\underbrace{2,\ldots,2}_{|\ia|=|\ic|},\underbrace{1,\ldots,1}_{|\beta|}) ,
\end{gather*}
with $|\ia|+|\ib|+|\ic|=r$ and contain $|\ic|(|\ia|+|\ib|)$ pairs of conjugate variables. The Lax matrices can be defined from $\Lm$ with $\ia=\ic$ and $\ib=0$ given in \eqref{eq:lax22alt} as
\begin{gather}\label{eq:lax222111}
\Lm=\left(
 \begin{BMAT}[5pt]{c:c:c}{c:c:c}
 0&0&K_{\ia,\ic}\\
 0&\ID&-P_{\ib,\ic}\\
 \bar K_{\ic,\ia}&Q_{\ic,\ib}&x\ID-F_{\ic,\ic}-Q_{\ic,\ib}P_{\ib,\ic}\\
 \end{BMAT}
\right) .
\end{gather}
Here $F_{\ic,\ic}$, $\K_{\ia,\ic}$ and $\Kb_{\ic,\ia}$ are defined in \eqref{eq:abc} and do not depend on $\ib$. The diagonal block containing the identity matrix $\ID$ is of the size $|\ib|$. The matrices $P_{\ib,\ic}$ and $Q_{\ic,\ib}$ read
\begin{gather}\label{eq:Pbggg}
(P_{\ib,\ic})_{i,j}=p_{|\ia|+i,|\ia|+|\ib|+j} ,\qquad ( Q_{\ic,\ib})_{i,j}=q_{|\ia|+|\ib|+i,|\ia|+j} .
\end{gather}
The proof of Sklyanin relation is straightforward combining the proofs in Sections~\ref{sec:eleml} and \ref{sec:elmu}. The determinant can be obtained using \eqref{eq:detabcd} and yields unity. From here one may build all other Lax matrices corresponding to arbitrary~$\mu$ partitions by factorisation. The result is presented in the next subsection.

\subsection[Lax matrices for $\mu$ partitions]{Lax matrices for $\boldsymbol{\mu}$ partitions}\label{sec:fullmupart}
The Lax matrix for arbitrary $\mu$ partitions can be written in the form
\begin{gather}\label{eq:mulax}
\Lm=\left(\begin{BMAT}[5pt]{c:c:c}{c:c:c}
 0&0&K_{\ia,\ic}\\
 0&\ID&-P_{\ib,\ic}\\
 \bar{K}_{\ic,\ia}&Q_{{\ic},\ib}& x\ID -F_{\ic,{\ic}}-Q_{{\ic},\ib}P_{\ib,{\ic}}
 \end{BMAT}
\right) .
\end{gather}
The blocks on the diagonal of this Lax matrix are of the size $|\ia|$, $|\ib|$ and $|\ic|$, respectively, with $|\ia|+|\ib|+|\ic|=r$. The matrices $P_{\ib,\ic}$ and $Q_{\ic,\ib}$ are defined as in \eqref{eq:Pbggg} and contain $|\ic|\cdot|\ib|$ pairs of conjugate variables. The remaining matrix elements can then be expressed in terms of $\gamma_1$ copies of the matrices defined in \eqref{eq:G} and \eqref{eq:abc}. We have
\begin{gather}\label{eq:bigmat}
F_{\ic,\ic}=Q_{\ic,\ic}\left(
 \begin{BMAT}[5pt]{c:c:c:c:c}{c:c:c:c:c}
 F_{1,1}&\pt_{1,2}&\pt_{1,3}&\cdots&\pt_{1,\ic_1}\\
 W_{2,1}&F_{2,2}&\pt_{2,3}&\cdots&\pt_{2,\ic_1}\\
 W_{3,1}&W_{3,2}&F_{3,3}&\ddots&\vdots\\
 \vdots&\vdots&\ddots&\ddots&\pt_{\ic_1-1,\ic_1}\\
 W_{\ic_1,1}&W_{\ic_1,2}&\cdots&W_{\ic_1,\ic_1-1}&F_{\ic_1,\ic_1}\\
 \end{BMAT}
\right)Q^{-1}_{\ic,\ic} ,\\ \label{eq:bigK}
 {K}_{\ia,{\ic}}=\left(\begin{BMAT}[5pt]{c:c:c:c}{c}
 {D}^{[\ic_1]}_{\ia,\ia}\cdots {D}^{[2]} _{\ia,\ia} {K}_{\ia,1}&\cdots
& {D}^{[\ic_1]}_{\ia,\ia} {K}_{\ia,\ic_1-1}& {K}_{\ia,\ic_1}
 \end{BMAT}
\right)Q_{\ic,\ic}^{-1}
\end{gather}
and
\begin{gather}\label{eq:bigKb}
 \bar{ {K}}_{{\ic},\ia}=Q_{\ic,\ic}\left(\begin{BMAT}[5pt]{c}{c:c:c:c}
\bar{ {K}}_{1,\ia}\\
\bar{ {K}}_{2,\ia} {D}^{[1]}_{\ia,\ia}
\\
\cdots
\\\bar{ {K}}_{\ic_1,\ia} {D}^{[\ic_1-1]}_{\ia,\ia}\cdots {D}^{[1]}_{\ia,\ia}
 \end{BMAT}
\right) .
\end{gather}

Each block $(i,j)$ in \eqref{eq:bigmat} is of the size $\ic_i^t\times\ic_j^t$. The matrices $\pt_{ij}$ are defined as in \eqref{eq:pt} with
\begin{gather*}
 \pt_{ij}=(P_{\ic,\ic})_{ij}+\sum_{k=j+1}^{\ic_1}(P_{\ic,\ic})_{ik}(Q_{\ic,\ic})_{kj} .
\end{gather*}
The corresponding matrices $Q_{\ic,\ic}$ and $P_{\ic,\ic}$ as defined for the partition $\lambda$ in \eqref{eq:Umatrix} read
\begin{gather*}%\label{eq:Umatrixmu}
 Q_{\ic,\ic}=\left(\!
 \begin{BMAT}[5pt]{c:c:c:c:c}{c:c:c:c:c}
\ID &0&0&0&0\\
 Q_{2,1}&\ID &0&0&0\\
 Q_{3,1}&Q_{3,2}&\ID &0&0\\
 \vdots&\vdots&\ddots&\ddots&0\\
 Q_{\ic_1,1}&Q_{\ic_1,2}&\cdots&Q_{\ic_1,\ic_1-1}&\ID \\
 \end{BMAT}\!
\right) \!,\quad
 P_{\ic,\ic}=\left(\!
 \begin{BMAT}[5pt]{c:c:c:c:c}{c:c:c:c:c}
0&P_{1,2}&P_{1,3}&\cdots&P_{1,\ic_1}\\
 0&0 &P_{2,3}&\cdots&P_{2,\ic_1}\\
 0&0&0&\ddots&\vdots\\
 0&0&0&\ddots&P_{\ic_1-1,\ic_1}\\
 0&0&0&0&0 \\
 \end{BMAT}\!
\right) \!,
\end{gather*}
where $Q_{ij}$ and $P_{ij}$ denote block matrices explicitly given by
\begin{alignat*}{3}
&(Q_{ij})_{kl}= q_{\ell(s)+k,\ell(t)+l}, \qquad && k \in \big[1,\ic^t_i\big], \quad l \in \big[1, \ic^t_j\big] ,\\
& (P_{ij})_{kl}=p_{\ell(s) + k , \ell(t) + l}, \qquad && k \in \big[1,\ic^t_i\big], \quad l \in \big[1,\ic^t_j\big] .
 \end{alignat*}
Here we defined $\ell(i)=|\ia|+|\ib|+\sum\limits_{l=1}^{i-1}\ic_l^t$.

The elements on the lower diagonal of the middle part of $F_{\ic,\ic}$ in~\eqref{eq:bigmat} are defined as the product
\begin{gather*}
 W_{ij}=-\bar{ {K}}_{i,\ia} {D}^{[i-1]}_{\ia,\ia}\cdots{D}^{[j+1]}_{\ia,\ia} {K}_{\ia,j} ,
\end{gather*}
which in particular yields $W_{i+1,i}=-\bar{ {K}}_{i+1,\ia} {K}_{\ia,i}$. The remaining matrices are parametrized in terms of the matrices defined in~\eqref{eq:G} and~\eqref{eq:abc} as
\begin{gather}
 F_{k,k} =\big(Q_-GQ_-^{-1}\big)_{\ic^t_{k},\ic^t_{k}}+Q_{\ic^t_{k},|\ia|-\ic^t_k}P_{|\ia|-\ic^t_k,\ic^t_{k}} , \qquad \K_{\ia,k} =-\left(\begin{BMAT}[5pt]{c}{c:c}
 \left(Q_-Q_0Q_+\right)^{-1}_{\ic_{k}^t,\ic_{k}^t}\\
 P_{|\ia|-\ic_{k}^t,\ic_{k}^t}\\
 \end{BMAT}
\right) , \nonumber\\
 \Kb_{k,\ia}=\left(\begin{BMAT}[5pt]{c:c}{c}
 \left(Q_-Q_0Q_+\right)_{\ic_{k}^t,\ic_{k}^t}&Q_{ \ic_{k}^t,|\ia|-\ic_{k}^t}
 \end{BMAT} \right), \hspace{14mm}
 D^{[k]}_{\ia,\ia}=\diag(\underbrace{0,\ldots,0}_{\ic_{k}^t},\underbrace{1,\ldots,1}_{|\ia|-\ic_{k}^t}) . \label{eq:elemelem}
 \end{gather}
Here the matrices $(Q_-Q_0Q_+)_{\ic^t_{k},\ic^t_{k}}$, $(Q_-Q_0Q_+)_{\ic^t_{k},\ic^t_{k}}^{-1}$ and $\big(Q_-GQ_-^{-1}\big)_{\ic^t_{k},\ic^t_{k}}$ are built from the variables $q_{ij}$ and $p_{ji}$ where $\ell(k)<i,j\leq \ell(k+1)$ with $\ell(k)=|\ia|+|\ib|+\sum\limits_{l=1}^{k-1}\ic_l^t$. Further the matrices $Q_{\ic_{k}^t,|\ia|-|\ic|_{k}^t}$ and $P_{|\ia|-\ic_{k}^t,|\ic|_{k}^t}$ are of the form
\begin{alignat*}{3}
& \big(Q_{\ic_{k}^t,|\ia|-|\ic|_{k}^t}\big)_{ij} = q_{\ell(k)+i,\ic_{k}^t+j}, \qquad && i \in \big[1, \ic_{k}^{t}\big],\quad j \in \big[1,|\ia|-\ic_{k}^{t}\big] ,& \\
& \big(P_{|\ia|-\ic_{k}^t,|\ic|_{k}^t}\big)_{ij} =p_{\ic_{k}^t+ i , \ell(k) + j}, \qquad && i \in \big[1,|\ia|-\ic_k^t\big], \quad j \in \big[1, \ic_{k}^{t}\big] .&
 \end{alignat*}

The total number of pairs $(p_I,q_I)$ in the Lax matrix for general $\mu$ partitions is $\frac{1}{2}\Big(r^2-\sum\limits_{i=1}^{\mu_1}\big(\mu_i^t\big)^2\Big)$. Here $|\ib|\cdot|\ic|$ pairs come from the elements $P$ and $Q$ in~\eqref{eq:mulax}, the matrices~$P_{\ic,\ic}$ and~$Q_{\ic,\ic}$ contain $\sum\limits_{i<j}\ic_i^t\ic_j^t$ pairs of conjugate variables and the matrices~$F_{k,k}$, $\K_{\ia,k}$ and $\Kb_{k,\ia}$ in~\eqref{eq:elemelem} contain for $k=1,\ldots,\ic_1$ in total $|\ia|\cdot|\ic|$ pairs of variables.

The expression for the Lax matrix $\Lm$ in \eqref{eq:mulax} is in principle valid for any ordering of columns where $|\ia|+|\ib|$ denotes the height of the biggest columns and $\gamma$ the partition that remains after removing that column. If the partition is ordered, i.e., $\lambda_i\geq\lambda_j$ for $i<j$, we have that $D_{\ia,\ia}^{[i]}D_{\ia,\ia}^{[j]}=D_{\ia,\ia}^{[j]}D_{\ia,\ia}^{[i]}=D_{\ia,\ia}^{[i]}$ for $i<j$ and $D_{\ia,\ia}^{[1]}=0$ which simplifies the expressions above.
\subsection[Fusion procedure for $\mu$ partitions]{Fusion procedure for $\boldsymbol{\mu}$ partitions}\label{sec:fac}

\newcommand{\ba}{\tilde\ia}
\newcommand{\bb}{\tilde\ib}
\newcommand{\bc}{\ic'}
\newcommand{\bd}{\ic}
The formula for the Lax matrices of the $\mu$ partitions can be shown in analogy to Section~\ref{sec:fusel}. We define three partitions $\mu$, $\mu'$ and $\tilde\mu$ with $|\mu|=|\mu'|=|\tilde\mu|=r$. They are related by fusion via
\begin{gather}\label{eq:mufuse}
 |\tilde\ia| =\max(|\ia|,|\ia'|) ,\qquad |\tilde \ib|=\min(|\ib|,|\ib'|) ,\qquad \tilde\gamma ^t=\big(\ic^t,{\ic'}^t\big).
\end{gather}

Here we consider a solution to the Sklyanin relation of the form~\eqref{eq:lax222111} written as a $4\times 4$ block matrix
\begin{gather}\label{eq:ll1}
\Lm=\left(\begin{BMAT}[5pt]{c:c:c:c}{c:c:c:c}
 {D_{\ba,\ba}}&0&0&\K_{\ba,\bd}\\
 0&\ID&0&-P_{\bb,\bd}\\
 0&0&\ID&-P_{\bc,\bd}\\
 \Kb_{\bd,\ba}&Q_{\bd,\bb}&Q_{\bd,\bc}&x\ID-\F_{\bd,\bd}-Q_{\bd,\bb}P_{\bb,\bd}-Q_{\bd,\bc}P_{\bc,\bd}
 \end{BMAT}
\right) .
\end{gather}
The blocks on the diagonal are of the size $|\tilde \ia|$, $|\tilde \ib|$, $|\ic'|$ and $|\ic|$, respectively with $|\tilde \ia|+|\tilde \ib|+|\ic'|+|\ic|=r$. The matrices $Q_{\bd,\bb}$, $Q_{\bd,\bc}$ and $P_{\bb,\bd}$, $P_{\bc,\bd}$ are explicitly given in terms of the conjugate variables. They read
\begin{alignat*}{3}
 & (P_{\bb,\bd})_{i,j} =p_{|\tilde \ia|+i,|\tilde \ia|+|\tilde \ib|+|\ic'|+j}, \qquad && (P_{\bc,\bd})_{i,j} =p_{|\tilde \ia|+|\tilde \ib|+i,|\tilde \ia|+|\tilde \ib|+|\ic'|+j},& \\
 &(Q_{\bd,\bb})_{i,j}=q_{|\tilde \ia|+|\tilde \ib|+|\ic'|+i,|\tilde \ia|+j} , \qquad && (Q_{\bd,\bc})_{i,j}=q_{|\tilde \ia|+|\tilde \ib|+|\ic'|+i,|\tilde \ia|+|\tilde \ib|+j} .&
\end{alignat*}
Furthermore we define a second Lax matrix, cf.~\eqref{eq:lax2}, which also is a solution of the Sklyanin relation. It has the same block structure as~\eqref{eq:ll1} and reads
\begin{gather*}%\label{eq:ll2}
 {L}'_{\mu'}(x;\underline{p}',\underline{q}')=\left(\begin{BMAT}[5pt]{c:c:c:c}{c:c:c:c}
 {D}'_{\ba,\ba}&0&{\K}'_{\ba,\bc}&0\\
 0&\ID& -P_{\bb,\bc}'&0\\
 \Kb'_{\bc,\ba}&Q_{\bc,\bb}'&x\ID- \F'_{\bc,\bc}-Q_{\bc,\bb}' P_{\bb,\bc}'+ P_{\bc,\bd}'Q_{\bd,\bc}'& -P_{\bc,\bd}'\\
 0& 0&- Q_{\bd,\bc}'& \ID
 \end{BMAT}
\right) .
\end{gather*}
We got
\begin{alignat*}{3}
& (P'_{\bb,\bc})_{i,j}=p_{|\tilde \ia|+i,|\tilde \ia|+|\tilde \ib|+j} , \qquad && (P_{\bc,\bd}')_{i,j}=p_{|\tilde \ia|+|\tilde \ib|+i,|\tilde \ia|+|\tilde \ib|+|\ic'|+j} ,&\\
& (Q'_{\bc,\bb})_{i,j} =q_{|\tilde \ia|+|\tilde \ib|+i,|\tilde \ia|+j} , \qquad && (Q_{\bd,\bc})_{i,j} =q_{|\tilde \ia|+|\tilde \ib|+|\ic'|+i,|\tilde \ia|+|\tilde \ib|+j} .&
\end{alignat*}
We proceed as in Section~\ref{sec:fusel} and multiply the two solutions of the Sklyanin relation. The product can again be written as
\begin{gather*}%\label{eq:fussm}
\Llm {L}'_{\mu'}(x;\underline{p}',\underline{q}')=\tilde Q'\tilde{L}_{\tilde\mu}(x,\underline{\tilde p},\underline{\tilde q}) ,
\end{gather*}
cf.~\eqref{eq:prodll}. The spectral parameter dependent matrix $\tilde{L}_{\tilde\mu}(x,\underline{\tilde p},\underline{\tilde q})$ and the matrix $\tilde Q'$ take the form
\begin{gather*}%\label{eq:llaxx}
\tilde{L}_{\tilde\mu}(x,\underline{\tilde p},\underline{\tilde q})=WU\left(\begin{BMAT}[5pt]{c:c:c:c}{c:c:c:c}
 {D}_{\ba,\ba}{D}'_{\ba,\ba}&0&{D_{\ba,\ba}}{\K_{\ba,\bc}}' & {\K}_{\ba,\bd} \\
 0 &\ID&0&0\\
 {\Kb}'_{\bc,\ba}&0&x\ID-\F'_{\bc,\bc}& -\tilde P_{\ic',\ic}\\
{\Kb}_{\bd,\ba} {D} ' _{\ba,\ba}&0&{{\Kb}}_{\bd\ba}{\K}'_{\ba,\bc}&x\ID-\F_{\bd,\bd}
 \end{BMAT}
\right)U^{-1}V^{-1} ,
\end{gather*}
and
\begin{gather*}
 \tilde Q'=\left(\begin{BMAT}[5pt]{c:c:c:c}{c:c:c:c}
 \ID&0&0&0\\
 0&\ID&0&0\\
 0&0&\ID&0\\
 0 &0&\tilde Q_{\ic,\ic'}'&\ID
 \end{BMAT}
\right) .
\end{gather*}
Here we have written $\tilde{L}_{\tilde\mu}(x,\underline{\tilde p},\underline{\tilde q})$ in a factorised form and introduced the matrices
\begin{gather*}%\label{eq:WV}
 W=\left(\begin{BMAT}[5pt]{c:c:c:c}{c:c:c:c}
 \ID&0&0&0\\
 0&\ID&0&0\\
 0&\tilde Q_{\ic',\tilde \ib}&\ID&0\\
 0 &\tilde Q_{\ic,\tilde \ib}&0&\ID
 \end{BMAT}
\right) ,\qquad
 U=\left(\begin{BMAT}[5pt]{c:c:c:c}{c:c:c:c}
 \ID&0&0&0\\
 0&\ID&0&0\\
 0&0&\ID&0\\
 0 &0&\tilde Q_{\ic,\ic'}&\ID
 \end{BMAT}
\right) ,\\ V^{-1}=\left(\begin{BMAT}[5pt]{c:c:c:c}{c:c:c:c}
 \ID&0&0&0\\
 0&\ID&\tilde P_{\tilde \ib,\ic'}&\tilde P_{\tilde \ib,\ic}\\
 0& 0&\ID&0\\
 0 &0&0&\ID
 \end{BMAT}
\right) .
\end{gather*}
They are expressed in terms of the new variables
\begin{alignat*}{3}
& \tilde P_{{\bc}{\bd}}=P_{{\bc}{\bd}}'+P_{{\bc}{\bd}}-Q'_{{\bc}\bb}P_{\bb{\bd}}, \qquad && \tilde Q_{{\bd}{\bc}} =Q_{{\bd}{\bc}}',&\\
&\tilde P_{\bb{\bc}}=P_{\bb{\bc}}'-P_{\bb{\bd}}Q'_{{\bd}{\bc}}, \qquad && \tilde Q_{{\bc}\bb} =Q_{{\bc}\bb}', & \\
&\tilde P_{\bb{\bd}}=P_{\bb{\bd}}, \qquad && \tilde Q_{{\bd}\bb}=Q_{{\bd}\bb}+Q'_{{\bd}{\bc}}Q'_{{\bc}\bb}, &\\
& \tilde P'_{{\bc}{\bd}}= P_{{\bc}{\bd}}, \qquad && \tilde Q'_{{\bd}{\bc}}=Q_{{\bd}{\bc}}-Q'_{{\bd}{\bc}} .&
 \end{alignat*}
This is the same change of variables as in \eqref{eq:trans1} and therefore it is canonical. Following the same logic as in Section~\ref{sec:fusel} we conclude that $\tilde{L}_{\tilde\mu}(x,\underline{\tilde p},\underline{\tilde q})$ is a solution of the Sklyanin relation. For convenience we write it in the same form as $\Lm$ such that
\begin{gather}\label{eq:laxz}
\tilde{L}_{\tilde\mu}(x,\underline{\tilde p},\underline{\tilde q})=\left(\begin{BMAT}[5pt]{c:c:c}{c:c:c}
 \tilde{D} _{\tilde\ia,\tilde\ia}&0& {\K}_{\ba,\tilde \ic}\\
 0&\ID&-\tilde P_{\bb,\tilde\ic}\\
 {{\Kb}}_{\tilde\ic,\ba}&\tilde Q_{\tilde\ic,\bb}&x\ID-\tilde F_{\tilde\ic,\tilde\ic}-\tilde Q_{\tilde\ic,\bb}\tilde P_{\bb,\tilde\ic}\\
 \end{BMAT}
\right) .
\end{gather}
The size of the block matrices on the diagonal is $|\tilde \ia|$, $|\tilde \ib|$ and $|\tilde \ic|$. We defined the matrices
\begin{gather*}%\label{eq:pqdp}
 \big(\tilde P_{\bb,\tilde\ic}\big)_{i,j}=\tilde p_{|\tilde \ia|+i,|\tilde \ia|+|\tilde \ib|+j} ,\qquad \big(\tilde Q_{\tilde\ic,\bb}\big)_{i,j}=\tilde p_{|\tilde \ia|+|\tilde \ib|+i,|\tilde \ia|+j},
\end{gather*}
while the remaining elements are given by
\begin{gather*}
 \tilde F_{\tilde\ic,\tilde\ic}=\tilde Q_-\left(\begin{BMAT}[5pt]{c:c}{c:c}
 \F'_{\bc,\bc}& \tilde P'_{\bc,\bd} \\
 -{{\Kb}_{\bd,\ba}}{K_{\ba,\bc}'} &\F_{\bd,\bd}
 \end{BMAT}
\right)\tilde Q_-^{-1} ,\qquad\text{with}\quad
\tilde Q_-=
 \left(\begin{BMAT}[5pt]{c:c}{c:c}
 \ID&0\\
 \tilde Q_{ \bd,\bc}'&\ID
 \end{BMAT}
\right) ,\\
 {{\Kb}_{\tilde\ic,\ba}}=\tilde Q_-\left(\begin{BMAT}[5pt]{c}{c:c}
 {{\Kb}_{\bc,\ba}}'\\
 {{\Kb}_{\bd,\ba}}{D_{\ba,\ba}'}
 \end{BMAT}
\right) ,\qquad {K}_{\ba,\tilde\ic}=\left(\begin{BMAT}[5pt]{c:c}{c}
{D}_{\ba,\ba}{K_{\ba,\bc}'}& {K}_{\ba,\bd}
 \end{BMAT}
\right)\tilde Q_-^{-1} ,
\end{gather*}
and $\tilde {D}_{\ba,\ba}={D}_{\ba,\ba}{D}_{\ba,\ba}'$.

\subsubsection{Recursion}
\newcommand{\tia}{\tilde \alpha}
\newcommand{\tib}{\tilde \beta}
\newcommand{\tic}{\tilde \gamma}
We specify the matrix elements in the fusion procedure to describe the fusion of one arbitrary partition $\mu$ as proposed in Section~\ref{sec:fullmupart} and an elementary matrix \eqref{eq:lax222111} corresponding to the partition $\mu'$ with the restriction $\ia'=\ic'$. The resulting partition $\tilde\mu$ is then written in terms of~$\mu$ and~$\mu'$ as in~\eqref{eq:mufuse}.

This can be seen as follows.
The primed letters correspond to elements of the Lax matrix corresponding to $\mu'$ and read
\newcommand{\spa}{}
\begin{gather*}
 {F'}_{\ic',\ic'}=\left(Q_-^{\spa}G^{\spa}Q_-^{-1}\right)_{\ic',\ic'}+Q^{\spa}_{\ic',\tilde\ia-\ia'}P^{\spa}_{\tilde\ia-\ia',\ic'} ,
\\
 {\Kb}'_{{\ic'},\tilde\ia}=\left(\begin{BMAT}[5pt]{c:c}{c}
 \left(Q_-^{\spa}Q_0^{\spa}Q_+^{\spa}\right)_{\ic',\ia'}&Q^{\spa}_{\ic',\tilde\ia-\ia'}
 \end{BMAT}
\right), \qquad
 {K}'_{\tilde\ia,\ic'}=-\left(\begin{BMAT}[5pt]{c}{c:c}
 \left(Q_-^{\spa}Q_0^{\spa}Q_+^{\spa}\right)^{-1}_{\ia',\ic'}\\
 P^{\spa}_{\tilde\ia-\ia',\ic'}\\
 \end{BMAT}
\right)
\end{gather*}
and
\begin{gather*}
{D}'_{\tilde\ia,\tilde\ia}={D}^{[\ic']}_{\tilde \ia,\tilde \ia}=\diag(\underbrace{0,\ldots,0}_{\ia'},\underbrace{1,\ldots,1}_{\tilde \ia-\ia'}) .
\end{gather*}
The unprimed letters correspond to the partition $\mu$ as given in \eqref{eq:bigmat}, \eqref{eq:bigK} and \eqref{eq:bigKb}. We find that
\begin{gather*}%\label{eq:bigmat2}
F_{\tilde \ic,\tilde \ic}=Q_{\tilde \ic,\tilde \ic}\left(
 \begin{BMAT}[5pt]{c:c}{c:c}
 F'_{\ic',\ic'}&\tilde P_{\ic',\ic}Q_{\ic,\ic}\\
 -Q_{\ic,\ic}^{-1}\Kb_{\ic,\tilde\alpha}\K'_{\tilde\alpha,\ic'}&
 \begin{BMAT}[5pt]{c:c:c:c}{c:c:c:c}
F_{1,1}&\pt_{1,2}&\cdots&\pt_{1,\ic_1}\\
W_{2,1}&F_{2,2}&\ddots&\vdots\\
\vdots&\ddots&\ddots&\pt_{\ic_1-1,\ic_1}\\
W_{\ic_1,1}&\cdots&W_{\ic_1,\ic_1-1}&F_{\ic_1,\ic_1}\\
 \end{BMAT}
 \end{BMAT}
\right)Q^{-1}_{\tilde \ic,\tilde \ic} ,
\end{gather*}
where similar as for the case of $\lambda$-partitions in \eqref{eq:Jfactor} we identify
\begin{gather*}
 \tilde P_{\ic',\ic}Q_{\ic,\ic}=\left(\begin{BMAT}[5pt]{c:c:c:c}{c}
\pt_{\ic',1}&\pt_{\ic',2}&\cdots&\pt_{\ic',\ic_1}
 \end{BMAT}\right) .
\end{gather*}
Furthermore we identify
\begin{gather*}
 -Q_{\ic,\ic}^{-1}\Kb_{\ic,\tilde\alpha}\K'_{\tilde\alpha,\ic'}=-\left(\begin{BMAT}[5pt]{c}{c:c:c:c}
\Kb_{1,\tilde \ia}\K_{\tilde \ia,\ic'}'\\
\Kb_{2,\tilde \ia}D_{\tilde \ia,\tilde \ia}^{[1]}\K_{\tilde \ia,\ic'}'\\
\vdots\\
\Kb_{\ic_1,\tilde \ia}D_{\tilde \ia,\tilde \ia}^{[\ic_1-1]}\cdots D_{\tilde \ia,\tilde \ia}^{[1]}\K_{\tilde \ia,\ic'}'
 \end{BMAT}\right)=\left(\begin{BMAT}[5pt]{c}{c:c:c:c}
W_{1,\ic'}\\
W_{2,\ic'}\\
\vdots\\
W_{\ic_1,\ic'}
 \end{BMAT}\right)
\end{gather*}
and obtain
\begin{gather*}
 \Kb_{\tilde\ic,\tilde\ia}=Q_{\tilde\ic,\tilde \ic}\left(\begin{BMAT}[5pt]{c}{c:c:c:c:c}
\bar{ {K}}_{\ic',\tilde \ia}'\\
\bar{ {K}}_{1,\tilde \ia} {D}^{[\ic']}_{\tilde \ia,\tilde \ia}\\
\bar{ {K}}_{2,\tilde \ia} {D}^{[1]}_{\tilde \ia,\tilde \ia}{D}^{[\ic']}_{\tilde \ia,\tilde \ia}
\\
\vdots
\\\bar{ {K}}_{\ic_1,\tilde \ia} {D}^{[\ic_1-1]}_{\tilde \ia,\tilde \ia}\cdots {D}^{[1]}_{\tilde \ia,\tilde \ia}{D}^{[\ic']}_{\tilde \ia,\tilde \ia}
 \end{BMAT}
\right)
\end{gather*}
and
\begin{gather*}
 {K}_{\tilde \ia,\tilde {\ic}}=\left(\begin{BMAT}[5pt]{c:c:c:c:c}{c}
 {D}^{[\ic_1]}_{\tia,\tia}\cdots {D}^{[1]} _{\tia,\tia} {K}_{\tia,\ic'}'& {D}^{[\ic_1]}_{\tia,\tia}\cdots {D}^{[2]} _{\tia,\tia} {K}_{\tia,1}&\cdots
& {D}^{[\ic_1]}_{\ia,\ia} {K}_{\ia,\ic_1-1}& {K}_{\ia,\ic_1}
 \end{BMAT}
\right)Q_{\tic,\tic}^{-1} .
\end{gather*}
Thus we conclude that \eqref{eq:mulax} satisfies Sklyanin's quadratic Poisson bracket.

\section{Generic degree 1 symplectic leaves}\label{sec:lmpart}

We will now define the Lax matrices $\Llm$ for arbitrary partitions $ \lambda$ and $ \mu$. They can be obtained by fusing the Lax matrix for regular partitions~\eqref{eq:laxregular} with the Lax matrix for~$\mu$ partitions~\eqref{eq:mulax}.

\subsection[Lax matrix for $\lambda$, $\mu$ partitions]{Lax matrix for $\boldsymbol{\lambda}$, $\boldsymbol{\mu}$ partitions}
The Lax matrix for arbitrary partitions $ \lambda$ and $ \mu$ can compactly be written as
\begin{gather}\label{eq:finallax}
 \Llm=\left(\begin{BMAT}[5pt]{c:c:c}{c:c:c}
 0&0&K_{\ia,\ic\lambda}\\
 0 &\ID &-P_{\ib,\ic\lambda}\\
 \bar{K}_{\ic\lambda,\ia}& Q_{\ic\lambda,\ib}&x\ID -{F}_{\ic\lambda,\ic\lambda}-Q_{\ic\lambda,\beta}P_{\beta,\ic\lambda}\\
 \end{BMAT}
\right) .
\end{gather}
The blocks on the diagonal are of the size $|\ia|$, $|\ib|$ and $|\ic|+|\lambda|$ respectively.
Here $Q_{\lambda,\ic}$ and $P_{\ic,\lambda}$ are defined as
\begin{gather*}
 (P_{\ic,\lambda})_{i,j}= p_{|\ia|+|\ib|+i,| \ia|+| \ib|+|\ic|+j} ,\qquad ( Q_{\lambda,\ic})_{i,j}=q_{| \ia|+| \ib|+|\ic|+i,| \ia|+|\ib|+j}.
\end{gather*}
The remaining matrix elements in \eqref{eq:finallax} are given in terms of the components of the Lax matrix for regular partitions~\eqref{eq:laxregular} and the Lax matrix for~$\mu$ partitions~\eqref{eq:mulax}. We have
\begin{gather*}
{F}_{\ic\lambda,\ic\lambda}=\left(\begin{BMAT}[5pt]{c:c}{c:c}
 \ID &0\\
 Q_{\lambda,\ic}&\ID
 \end{BMAT}
\right)\cdot\left(\begin{BMAT}[5pt]{c:c}{c:c}
 F_{\ic,\ic}&P_{\ic,\lambda}\\
 -Q_{\lambda,\ia} K_{\ia,\ic} & J_{\lambda,\lambda}+Q_{\lambda,\ia}P_{\ia,\lambda}
 \end{BMAT}
\right)\cdot\left(\begin{BMAT}[5pt]{c:c}{c:c}
 \ID &0\\
 -Q_{\lambda,\ic}&\ID
 \end{BMAT}
\right) ,
\end{gather*}
and
 \begin{alignat*}{3}
& \bar{{K}}_{\ic\lambda,\ia}=\left(\begin{BMAT}[5pt]{c}{c:c}
 \bar K_{\ic,\ia}\\
 Q_{\lambda,\ic} \bar K_{\ic,\ia}
 \end{BMAT}
\right) , \qquad && {{K}}_{\ia,\ic\lambda}=\left(\begin{BMAT}[5pt]{c:c}{c}
 K_{\ia,\ic}+P_{\ia,\lambda} Q_{\lambda,\ic}&-P_{\ia,\lambda}
 \end{BMAT}
\right) ,&\\
& Q_{\ic\lambda,\ib}=\left(\begin{BMAT}[5pt]{c}{c:c}
 Q_{\ic,\ib}\\
 Q_{\lambda,\ib}
 \end{BMAT}
\right) ,\qquad &&
 P_{\ib,\ic\lambda}=\left(\begin{BMAT}[5pt]{c:c}{c}
 P_{\ib,\ic}&P_{\ib,\lambda}
 \end{BMAT}
\right) .
 \end{alignat*}
Again we can check that the number of pairs of conjugate variables agrees with~\eqref{eq:dimform}. First we note that $F$ contains $\sum\limits_{i<j}\ic_i^t\ic_j^t+|\ia|\cdot|\ic|$ and $J$ contains $\sum\limits_{i<j}\lambda_i^t\lambda_j^t$ pairs. The remaining variables are contained in $P_{\ib,\ic\lambda}$, $Q_{\ic\lambda,\ib}$, $P_{\ia,\lambda}$, $Q_{\lambda,\ia}$ and $P_{\ic,\lambda}$, $Q_{\lambda,\ic}$. By construction, cf.~Section~\ref{sec:fuslm}, the determinant of the Lax matrix in~\eqref{eq:finallax} satisfies~\eqref{eq:detlax}.

The symplectic leaves that we found in the Poisson--Lie group $\mathcal{G}$ are orbits of certain representative elements under the dressing action of the dual Poisson--Lie group $\mathcal{G}^{*}$. These representative elements are easily seen as Lax matrices at $\underline p = \underline q = 0$. Here the Lax matrix \eqref{eq:finallax} reduces to a~block matrix of the form
\begin{gather*}%\label{eq:basepoint}
L_{{\lambda},\underline{x},\mu}(x;\varnothing,\varnothing)=\left(\begin{BMAT}[5pt]{c:c}{c:c}
x\ID_\mu-\Sigma_\mu &0\\
0&x\ID-X_\lambda
 \end{BMAT}
\right) ,
\end{gather*}
where
\begin{gather*}
I_\mu=\diag(\underbrace{0,\ldots,0}_{|\alpha|+|\beta|},\underbrace{1\ldots,1}_{|\gamma|})
\end{gather*}
and $X_\lambda$ denotes the diagonal matrix defined in \eqref{eq:Xlambda}. The matrix $\Sigma_\mu$ is a permutation matrix containing the elements~$\pm 1$. This matrix can be block diagonalized such that it contains $|\alpha|+|\beta|$ blocks of the size $\mu_i$, $i=1,\ldots,|\alpha|+|\beta|$, corresponding to the rows of the partitions $\mu$. The diagonal of each block $i$ reads $\diag(0,x,\ldots,x)$ and its remaining elements $\pm 1$ correspond to a~cyclic permutation of length $\mu_i$. For example for a row of $\mu_i=4$ we obtain
\begin{gather*}
 \left(\begin{matrix}
 0&-1&0&0\\
 0&x&-1&0\\
 0&0&x&-1\\
 1&0&0&x
 \end{matrix}
\right) .
\end{gather*}
For $\mu_i=1$ where $i=|\alpha|+1,\ldots,|\alpha|+|\beta|$, we obtain a $1\times 1$ block containing only the element~$1$.

\subsection{Fusion procedure}\label{sec:fuslm}

The Lax matrix \eqref{eq:finallax} can be derived using the factorisation formula in Section~\ref{sec:fac} when substituting the $\ic$ block for a $\lambda=\tilde\lambda$ block as follows
\begin{gather*}
{F}_{\ic,\ic}\rightarrow J_{\tilde \lambda,\tilde \lambda}+Q_{\tilde \lambda,\tilde \alpha}P_{\tilde \alpha,\tilde \lambda} ,\qquad \bar{{K}}_{\ic,\tilde \ia}\rightarrow Q_{\tilde \lambda,\tilde \ia} ,\qquad{K}_{\tilde \ia,\ic}\rightarrow-P_{\tilde \ia,\tilde\lambda} ,\qquad {D}_{\tilde \ia,\tilde \ia}\rightarrow\ID .
\end{gather*}
Here $\ID $ is the $|\tilde\ia|\times| \tilde \ia|$ identity matrix. The primed elements in the second Lax matrix $L'$ are taken to be as defined in \eqref{eq:mulax} as
\begin{gather*}%\label{eq:defbbb}
{F}'_{\ic',\ic'}=F_{\tilde\ic,\tilde\ic} ,\qquad
 \bar{{K}}'_{\ic',\tilde \ia}=\bar{K}_{\tilde \ic,\tilde \ia} ,\qquad {K}'_{\tilde \ia,\ic'}=K_{\tilde \ia,\tilde \ic} ,\qquad {D}'_{\tilde \ia,\tilde \ia}=0 .
\end{gather*}
Here ${D'}$ is equal to the $|\tilde \ia|\times|\tilde \ia|$ zero matrix.

This factorisation corresponds to the fusion of the partitions $\lambda$, $\mu$ and $\lambda'$, $\mu'$ expressed in terms of the resulting partition $\tilde\lambda$, $\tilde\mu$ via
\begin{gather*}\begin{split}&
 \ia=\varnothing ,\qquad \ib=(\underbrace{1,\ldots,1}_{|\tilde\ia|+|\tilde\ib|+|\tilde\ic|}) ,\qquad \ic=\varnothing ,\qquad \lambda=\tilde\lambda ,
\\
& \ia'=\tilde\ia ,\qquad \ib'=(\underbrace{1,\ldots,1}_{|\tilde\ib|+|\tilde\lambda|}) ,\qquad \ic'=\tilde \ic ,\qquad \lambda'=\varnothing .\end{split}
\end{gather*}
The final result of the factorisation can be directly read off from \eqref{eq:laxz}. We conclude that \eqref{eq:finallax} is a solution to Skyanin's relation \eqref{eq:skl}.

\section[Algebraic completely integrable systems and Coulomb branches of $A_{r-1}$~quiver gauge theory]{Algebraic completely integrable systems \\ and Coulomb branches of $\boldsymbol{A_{r-1}}$~quiver gauge theory}\label{sec:specdet}

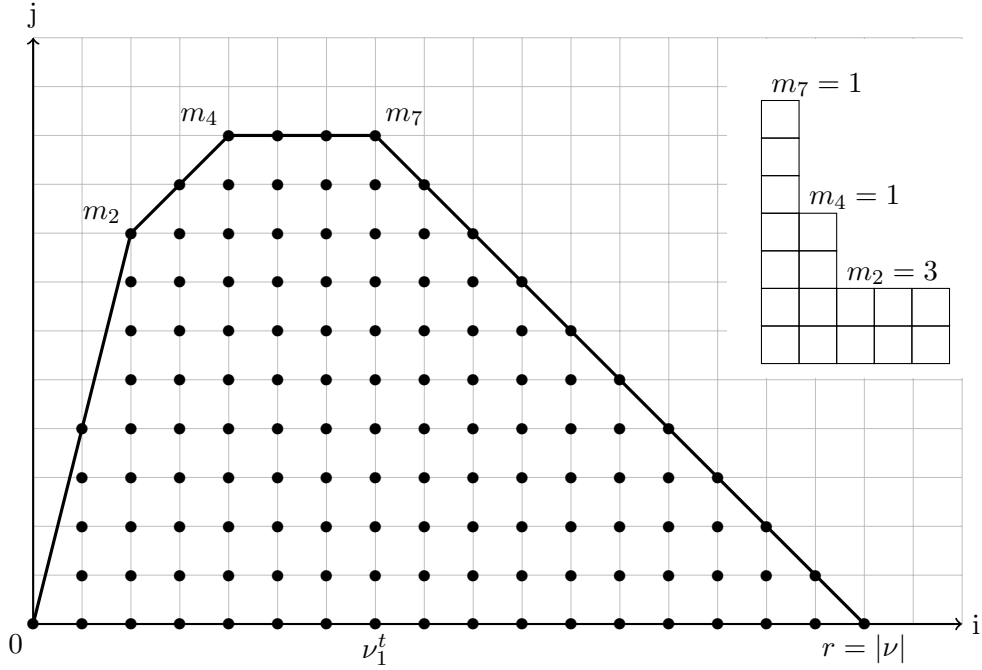
\begin{figure}[t]\centering
 \begin{tikzpicture}
 \node at (0,0) {\plot};
 \fill [white] (3.1,-0.58) rectangle (6.3,3.92);
 \node at (4.8,1.6) {\parti};
\end{tikzpicture}
\caption{An example of non-vanishing commuting Hamiltonians $\mathcal{X}_{i,j}^{[\lambda,\mu]}$ for a ${\rm GL}_{r}$ Lax matrix $\Llm$ corresponding to a
 partition $\nu=(5,5,2,2,1,1,1)$ with $r=17$ and $d_\nu=106$, with non-zero $m_2=3$ and $m_4=m_7=1$.
 The horizontal axis $i$ labels the nodes of the $A_{r-1}$ quiver diagram for $i \in [1, r-1]$,
 and the vertical coordinate $j$ of the enveloping profile denotes the color ranks $n_i$ in the quiver diagram.}\label{fig:newton}
\end{figure}

The symplectic leaf $\mathcal{M}_{\underline{\lambda}, \underline{x}, \mu}$, i.e., the moduli space of multiplicative Higgs bundles
with fixed singularities, supports fibration of an algebraic completely integrable system
\begin{gather} \label{eq:intsys}
 H\colon \ \mathcal{M}_{\underline{\lambda}, \underline{x}, \mu} \to \mathcal{U}_{\underline{\lambda}, \underline{x}, \mu} .
\end{gather}
Here $H$ denotes a complete set of independent commuting Hamiltonian functions (also known as conserved charges or action variable or integrals of motion of an integrable Hamilonian dynamical system) and $\mathcal{U}_{\underline{\lambda}, \underline{x}, \mu}$ denotes the space where the complete set of independent Hamiltonians takes value. The fibers $\mathcal{A}_{u} = H^{-1}(u), u \in \mathcal{U}_{\underline{\lambda}, \underline{x}, \mu}$ of the map \eqref{eq:intsys} are abelian varieties which are holomorphic Lagrangians with respect to the holomorphic symplectic structure on $\mathcal{M}_{\underline{\lambda}, \underline{x}, \mu}$,
so that
\begin{gather*}
 \dim_{\BC} \mathcal{U}_{\underline{\lambda}, \underline{x}, \mu} = \dim_{\BC} \mathcal{A}_{u} = \tfrac 1 2 \dim \mathcal{M}_{\underline{\lambda}, \underline{x}, \mu}
\end{gather*}
Let
\begin{gather*} %\label{eq:dimphas}
 d_{\underline{\lambda}, \underline{x}, \mu} = \tfrac 1 2 \dim \mathcal{M}_{\underline{\lambda}, \underline{x}, \mu}
 \end{gather*}
 denote the half-dimension of the symplectic leaf (phase space) $\mathcal{M}_{\underline{\lambda}, \underline{x}, \mu}$. In the context of Seiberg--Witten integrable systems \cite{Cherkis:2001gm, Cherkis:2000ft, Nekrasov:2012xe, Seiberg:1994rs} the holomorphic symplectic phase space $\mathcal{M}_{\underline{\lambda}, \underline{x}, \mu}$ is the Coulomb branch of the hyperk\"{a}hler moduli space of vacua of $\mathcal{N}=2$ supersymmetric quiver gauge theory on $\BR^{3} \times S^1$ viewed as a holomorphic symplectic manifold at a distinguished point $\zeta = 0$ on the twistor sphere of complex structures. The complex base space $\mathcal{U}_{\underline{\lambda}, \underline{x}, \mu}$ is the moduli space of vacua of the same $\mathcal{N}=2$ supersymmetric gauge theory on $\BR^{4}$ called $\mathcal{U}$-plane in the respective context. In terms of action-angle variables, the complex action variables parametrize the base $\mathcal{U}$-plane, and the complex angle variables parametrize the abelian fibers $\mathcal{A}_{u}$.

\looseness=-1 To realize the structure of an algebraic completely integrable system \eqref{eq:intsys} we need to construct~$d_{\underline{\lambda}, \underline{x}, \mu}$ independent Poisson commuting Hamiltonian functions on $\mathcal{M}_{\underline{\lambda}, \underline{x}, \mu}$. Like in the case of additive Higgs bundles (Hitchin system), the commuting Hamiltonian functions on multiplicative Higgs bundles (or more general abstract Higgs bundles) can be realized by the abstract cameral cover construction \cite{Donagi:2000dr}. In the case of additive Higgs bundles on $X$, the cameral cover construction generates Poisson commuting Hamiltonian functions as coefficients of $P(\phi(x))$ where the Higgs field $\phi(x)$ is a section of $\operatorname{ad} \mathfrak{g} \otimes K_{X}$ and $P$ is an adjoint invariant function on the Lie algebra $\mathfrak{g}$. Similarly, in the case of multiplicative Higgs bundles on $X$, the cameral cover construction generates Poisson commuting Hamiltonian functions as coefficients of $\chi(g (x))$ where $\chi$ is an adjoint invariant function on the group $G$ and multiplicative Higgs field $g(x)$ is a section of $\operatorname{Ad} G$ on~$X$. The complete set of independent Poisson commuting Hamiltonians for simple~$G$ is spanned by the characters $\chi_{R_i}$ of the fundamental irreducible highest weight modules $R_{k}$ whose highest weight is the fundamental weight $\omega_{k}$ for each $k$ in the set of nodes of the Dynkin diagram of~$g$.

If $G = {\rm GL}_{r}$, the irreducible highest weight module $R_{k}$ with highest weight $\omega_k$ associated to the $k$-th node of the $A_{r-1}$ Dynkin diagram of the simple factor ${\rm SL}_{r} \subset {\rm GL}_{r}$ is isomorphic to the $k$-th external power $R_{k} = \bigwedge^{k} R_{1}$, for $k = 1, \dots, r-1$, of the defining $r$-dimensional representation~$R_1$, and we set $R_{r} = \bigwedge^{r} R_{1}$ to be the determinant 1-dimensional representation. It is convenient to assemble the fundamental characters $\chi_{k} = \chi_{R_{k}}$ for $k = 1, \dots, r$ into the spectral polynomial
 \begin{gather*}
 \det ( y \ID_{r \times r} - L(x) )_{r \times r} = \sum_{k=0}^{r} (-1)^{k} y^{r-k} \chi_k (g(x)),
 \end{gather*}
where $\chi_k (g(x)) = \tr_{R_{k}} \rho_k (g(x))$ is a character for a fundamental representation $\rho_{k}\colon G \to \mathrm{End}(R_k)$ evaluated on Higgs field~$g(x)$.

Now we illustrate explicitly the construction of commuting Hamiltonians for the Lax matrices constructed in the previous sections that describe the symplectic leaves $ \mathcal{M}_{\underline{\lambda}, \underline{x}, \mu}$.

First, for any symplectic leaf $ \mathcal{M}_{\underline{\lambda}, \underline{x}, \mu}$ and its representing Higgs field $g_{\underline{\lambda}, \underline{x}, \mu}(x)$ we define its twisted version
 \begin{gather*} %\label{eq:twisted}
 g_{\underline{\lambda}, \underline{x}, \mu, g_L; g_R}(x) = g_L g_{\underline{\lambda}, \underline{x}, \mu}(x) g_R,
 \end{gather*}
which represents a symplectic leaf $\mathcal{M}_{\underline{\lambda}, \underline{x}, \mu; g_L, g_R}$. Here $g_L \in G, g_R \in G$ are arbitrary constant ($x$-independent) Higgs fields. We remark that the symplectic leaves $\mathcal{M}_{\underline{\lambda}, \underline{x}, \mu; g_L, g_R}$ are isomorphic for various $g_L$, $g_{R}$, and for certain relation between $g_{L}$ and $g_{R}$ they in fact coincide, in this sense the labeling by both $g_{L}$ and $g_{R}$ are redundant.\footnote{What is exactly the degree of redundancy? We can see that for the case regular at infinity $\mu = \varnothing$, when~$L(x)$ is an $x$-shifted co-adjoint orbit in $\mathfrak{g}$, the non-redundant label is the product $g_{R} g_{L}$. In any case, the resulting completely integrable system depends only on the product $g_R g_L$ as we see from the spectral polynomial~\eqref{eq:specdet}.} For the following, it is sufficient to take, $g_{L} \equiv g_{\infty}$, $g_{R} \equiv 1$, and define the Lax matrix
\begin{gather} \label{eq:twistedL}
 L_{\underline{\lambda}, \underline{x}, \mu, g_\infty}(x) = \rho_1 (g_\infty) L_{\underline{\lambda}, \underline{x}, \mu}(x),
 \end{gather}
where $\rho_1(g_\infty)$ is $r \times r$ matrix representing $g_\infty \in G$. Note that due to the symmetries of the $\mathfrak{r}$-matrix in \eqref{eq:skl} the product $\rho_1 (g_\infty) L_{\underline{\lambda}, \underline{x}, \mu}(x)$ is a solution to the Sklyanin relation if $L_{\underline{\lambda}, \underline{x}, \mu}$ is.

Now define the spectral determinant to be a polynomial of two variables $x$ and $y$
\begin{gather}\label{eq:specdet}
 W_{\underline{\lambda}, \underline{x}, \mu, g_\infty}(x,y)= \det(y - L_{\underline{\lambda}, \underline{x}, \mu, g_\infty}(x) ) = \sum_{k=0}^{r} (-1)^{k} y^{r-k} \chi_k (x) .
\end{gather}
The commuting Hamiltonians are coefficients of the monomials $x^j y^i $. With
\begin{gather} \label{eq:chik}
 \chi_k(x) = \mathcal{Q}^{[\lambda]}_{r-k} \mathcal{X}^{[\lambda, \mu]}_{k}(x),
\end{gather}
we find that the spectral determinant can be written as
\begin{gather}\label{eq:specdetQ}
 W_{\underline{\lambda}, \underline{x}, \mu, g_\infty}(x,y)=y^r+\sum_{i=1}^{r-1} (-1)^{r-i} \mathcal{Q}_i^{[\lambda]}(x) \mathcal{X}^{[\lambda,\mu]}_{r-i}(x) y^{i}+ (-1)^{r} \mathcal{Q}_0^{[\lambda]}(x) ,
\end{gather}
where $\mathcal{Q}_i^{[\lambda]}$ is a polynomial in $x$, cf.~\eqref{eq:appe}, which is independent of the conjugate variables $(p,q)$ of the Lax matrix and which takes the form
\begin{gather}\label{eq:Qpol}
 \mathcal{Q}_i^{[\lambda]}(x)=\prod_{j=1}^{\lambda_{i+1}}(x-x_j)^{\lambda_j^t-i} .
\end{gather}
All commuting Hamiltonians are thus contained in $\mathcal{X}_{i}^{[\lambda,\mu]}(x)$. More precisely $\mathcal{X}^{[\lambda,\mu]}_{r-k}(x)$ is a~polynomial in $x$ of degree
\begin{gather}\label{eq:ncolors}
 n_k^{[\lambda,\mu]}= \sum_{j=1}^{k} (\nu_j-1) \qquad\text{with}\quad \nu_i=\lambda_i+\mu_i ,\quad k \in [1, r-1] .
\end{gather}
We note that the number of independent commuting Hamiltonians only depends on the total dominant co-weight represented by the partition obtained by the union of columns of the partitions $\lambda$ and $\mu$ minus the shift by the diagonal co-representation (see below (\ref{eq:diagshift}). The charges are obtained as the coefficients of the expansion
\begin{gather*}
 \mathcal{X}_{r-i}^{[\lambda,\mu]}(x)=\sum_{j=0}^{n_i^{[\lambda,\mu]}}\mathcal{X}_{r-i,j}^{[\lambda,\mu]} x^j ,
\end{gather*}
cf. Fig.~\ref{fig:newton}. The highest coefficients do not depend on the conjugate variables $(p_I,q_I)$ but all other coefficients in the expansion do. The total number of linearly independent charges is equal to the number of conjugate pairs in the corresponding Lax matrix
\begin{gather}\label{eq:sumcharges}
 d_{\underline{\lambda},\mu}=\sum_{k=1}^{r-1} n_k^{\underline{\lambda},\mu} ,
\end{gather}
cf.~\eqref{eq:dimform}. This relation is shown using Frobenius-like coordinates for the partitions in Appendix~\ref{sec:proofcharges}.

For given partitions we can plot the non vanishing coefficients of the spectral determinant in a Newton diagram as done in Fig.~\ref{fig:newton}. Here we introduce the parameters{\samepage
\begin{gather}\label{eq:params}
 m_k^{[\lambda,\mu]}=\big(n_k^{[\lambda,\mu]}-n_{k-1}^{[\lambda,\mu]}\big)-\big(n_{k+1}^{[\lambda,\mu]}-n_{k}^{[\lambda,\mu]}\big)=\nu_k-\nu_{k+1} \qquad\text{for}\quad k\in [1, r-1]
\end{gather}
to label the partition and the corresponding Newton diagram.}

The representation theoretical meaning of the equations \eqref{eq:ncolors}, \eqref{eq:sumcharges} and \eqref{eq:params} is the following. The partitions $\underline{\lambda}$ and $\mu$ encode the ${\rm GL}_{r}$ co-weights of the respective singularity of the multiplicative Higgs field at finite points $\underline{x}$ and $x_\infty$. The encoding is in the $r$-dimensional basis of the dual to the weights of the defining representation that we call $\check e_k$ with $k = 1, \dots, r$. In terms of~$e_i$ define the simple co-roots $\alpha_i$ of ${\rm SL}_r$ to be
\begin{gather*}
 \check \alpha_k = \check e_k - \check e_{k-1}, \qquad k \in [1, r-1]
\end{gather*}
and define the fundamental weights to be
\begin{gather*}
 \check \omega_{k} = \sum_{j=1}^{k} \check e_k - \frac{k}{r} \sum_{j=1}^{r} \check e_j, \qquad k \in [1, r - 1].
\end{gather*}

The dominant co-weight associated to each singularity $x_{*} \in \underline{x}$ with associated partition $\lambda_{*} = \lambda_{*,1} \geq \lambda_{*, 2} \geq \dots \geq \lambda_{*, r} = 0$ is given by
\begin{gather*}
 \check \omega_{*} = \sum_{k=1}^{r} \lambda_{*, _i } \check e_i
\end{gather*}
and the dominant co-weight associated to the singularity at $x_{\infty} = \infty$ with associated partition~$\mu$ is given by
\begin{gather}\label{eq:diagshift}
 \check \omega_{\infty} = \sum_{k=1}^{r} (\mu_i - 1) \check e_i,
\end{gather}
so that the ${\rm GL}_r$ multiplicative Higgs field behaves up to a multiplication by a regular function as $(x - x_{*})^{\check \omega_{*}}$ as $ x \to x_{*}$ and as $ (1/x)^{\check \omega_{\infty}}$ as $ x \to \infty$.

Let $\rho$ be the Weyl vector
\begin{gather*}
 \rho = \frac{1}{2} \sum_{\alpha > 0} \alpha = \sum_{k=1}^{r-1} \omega_k = \sum_{k=1}^{r} \frac{ r - (2k+1)}{2} e_k
\end{gather*}
and let
\begin{gather*}
 \check \omega_{\rm tot} = \check \omega_{\infty} + \sum_{x_{*} \in \underline{x}} \check \omega_{x_*}
\end{gather*}
be the sum of the co-weights of all singularities in $\underline{x}$ and $x_\infty$. Then we see that the dimension formula \eqref{eq:sumcharges} is equivalent to
\begin{gather*} %\label{eq:dimensionformula}
 \dim \mathcal{M}_{\underline{\lambda}, \underline{x}, \mu; g_L, g_R} = 2 d_{\underline{\lambda}, \underline{x}, \mu} = 2 (\rho, \check \omega_{\rm tot})
\end{gather*}
in agreement with the general formula for the dimension of the moduli space of monopoles with Dirac singularities encoded by the total co-weight $\check \omega_{\rm tot}$.

The numbers $n_k$ and $m_k$ for $k \in [1, r-1]$ are the number of colors in the node $k$ and the number of fundamental flavours attached to the node $k$ of the Dynkin quiver~\cite{Nekrasov:2012xe}, including the ``deficit'' fundamental flavours multiplets of asymptotically free theory which would make it conformal if added. We have
\begin{gather*}
 \check \omega_{\rm tot} = \sum_{k=1}^{r-1} \check \alpha_k n_k = \sum_{k=1}^{r-1} \check \omega_k m_k
\end{gather*}
in agreement with \eqref{eq:ncolors} and \eqref{eq:params}. The position of each singularity $x_{*} \in \underline{x}$ to which we have associated a column $\lambda_{*}^{t}$ encoding a fundamental co-weight $\check \omega_{\lambda_{*}^t}$ is the mass of the fundamental flavour multiplet at the node~$\lambda_{*}^{t}$.

\begin{figure}[t]
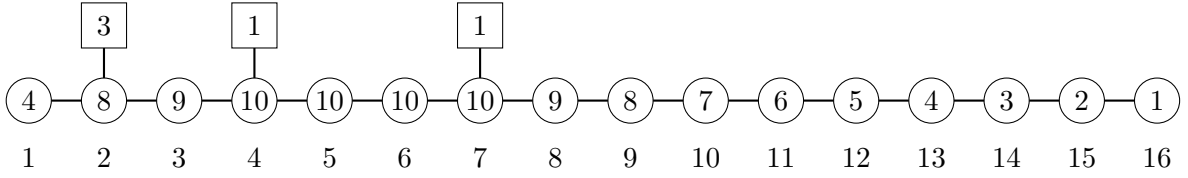
\centering
\quiver
\caption{Representation of the Newton polygon in Fig.~\ref{fig:newton} corresponding to the partition $\nu=(5,5,2,2,1,1,1)$ with $r=17$ as quiver diagram. Here the integers in the circles denote the number of charges for a given index $i$ indicated below. The parameters $m_i^{[\lambda,\mu]}$ are given in the squared boxes.}\label{fig:quiver}
\end{figure}

Now the spectral curve can be compared with \cite{Nekrasov:2012xe} where a slightly different notation is used. To do so we note that the polynomials $\mathcal{Q}_i^{[\lambda]}$ in \eqref{eq:Qpol} can be written in terms of the parameters~$m_i^{[\lambda,\varnothing]}$ introduced in~\eqref{eq:params} as
\begin{gather}\label{eq:QasP}
 \mathcal{Q}_k^{[\lambda]}(x)=\prod_{i=k+1}^{|\lambda|} \mathcal{P}_{r-i}^{ i-k}(x) ,\qquad\text{with}\quad \mathcal{P}_{r-i}(x)=\prod_{j=1}^{m_i^{[\lambda,\varnothing]}}(x-x_{\lambda_i-j+1}) .
\end{gather}
This relation is shown in Appendix~\ref{app:prf2}. Setting $m_i^{[\lambda,\varnothing]}=0$ for $i>|\lambda|$ we can now write the spectral determinant \eqref{eq:specdetQ} in the notation used in $(7.5)$ of \cite{Nekrasov:2012xe}. We find
\begin{gather*}
 W_{\underline{\lambda}, \underline{x}, \mu, g_\infty}(x,y)=y^r+\sum_{i=1}^{r-1}(-\zeta(x))^{ i} \prod_{j=1}^{i-1} \mathcal{P}_{j}^{ i-j}(x) \mathcal{X}_{i}(x) y^{r-i}+(-\zeta(x))^{ r}\prod_{j=1}^{r-1} \mathcal{P}_{j}^{ r-j}(x) ,
\end{gather*}
where we defined
\begin{gather*}
 \zeta(x)=\mathcal{P}_{0}(x) \qquad \mathcal{X}_{i}(x)=\mathcal{X}^{[\lambda,\mu]}_{i}(x) ,
\end{gather*}
with $i=1,\ldots,r-1$. We note that here the so-called matter polynomials $\mathcal{P}$ only depend on the partition $\lambda$ and not on $\mu$.

For singularities associated to the partition $\nu=(5,5,2,2,1,1,1)$ as plotted in Fig.~\ref{fig:newton} the quiver diagram is depicted in Fig.~\ref{fig:quiver}. Further examples are discussed in Appendix~\ref{app:examples}.

\section{Higher degree symplectic leaves}\label{sec:higherdegree}
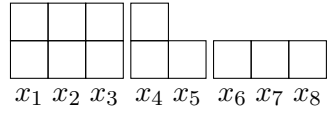
\begin{figure} \centering
\begin{tikzpicture}
\foreach \a in {0,1,2} {
 \begin{scope}[shift={(0.5*\a,0)}]
 \draw (0,0) rectangle (0.5,0.5);
 \end{scope}
 }
 \foreach \a in {0,1,2} {
 \begin{scope}[shift={(0.5*\a,0.5)}]
 \draw (0,0) rectangle (0.5,0.5);
 \end{scope}
 }
 \foreach \a in {3,4} {
 \begin{scope}[shift={(0.5*\a+0.1,0)}]
 \draw (0,0) rectangle (0.5,0.5);
 \end{scope}
 }
 \foreach \a in {3} {
 \begin{scope}[shift={(0.5*\a+0.1,0.5)}]
 \draw (0,0) rectangle (0.5,0.5);
 \end{scope}
 }
 \foreach \a in {5,6,7} {
 \begin{scope}[shift={(0.5*\a+0.2,0)}]
 \draw (0,0) rectangle (0.5,0.5);
 \end{scope}
 }
 \node [below ] at (0.25,0) {$x_1$};
 \node [below ] at (0.75,0) {$x_2$};
 \node [below ] at (1.25,0) {$x_3$};
 \node [below ] at (1.85,0) {$x_4$};
 \node [below ] at (2.35,0) {$x_5$};
 \node [below ] at (2.95,0) {$x_6$};
 \node [below ] at (3.45,0) {$x_7$};
 \node [below ] at (3.95,0) {$x_8$};
\end{tikzpicture}
\caption{Decomposition of the partition $\lambda=(8,4)$ for $r=3$ into three partitions $(3,3)$, $(2,1)$ and $(3)$.}\label{fig:ytdecc}
\end{figure}

In this section we discuss some symplectic leaves of higher degree $n$ in the spectral parameter $x$ corresponding to the partitions of the total size $ n r $ for $G = {\rm GL}_{r}$.

In the case $r=2$ we can factorize the higher degree Lax matrices for partitions $\lambda=(2n)$ as a product of degree~1 Lax matrices labelled by partitions $\lambda=(2)${\samepage
\begin{gather*}
 L_{(2n),\underline{x},\varnothing}(x;\underline{p},\underline{q})=L_{(2),(x_1,x_2),\varnothing}(x;p_1,q_1)\cdots L_{(2),(x_{2n-1},x_{2n}),\varnothing}(x;p_n,q_n),
\end{gather*}
cf.~\cite{ShapiroThesis}.}

For $r=3$ the partitions are of the form $\lambda=(\lambda_1,3n-\lambda_1)$ with $\lambda_1\leq 3n$. The Lax matrices for these partitions can be factorized as a product of Lax matrices of partitions $\lambda=(3)$, $\lambda=(2,1)$ and their conjugates. The conjugates correspond to the partitions $\lambda=(3,3)$ and $\lambda=(2,1)$ respectively and are obtained via
\begin{gather*}
 \bar{L} _{\lambda,\underline{x},\mu}(x;\underline{p},\underline{q})=\det \Llm L^{-1} _{\lambda,\underline{x},\mu}(-x;\underline{p},\underline{q}) .
\end{gather*}
This can be seen as follows. If $3n-\lambda_1=0$ we can build the partition from copies of $\lambda=(3)$ as for the case $r=2$. Any such partition is extended to the case where $3n-\lambda_1=1$ by adding a~partition $\lambda=(2,1)$ which again extends to $3n-\lambda_1=2$ by adding another partition $\lambda=(2,1)$. Now we note that any $\lambda=(\lambda_1,3n-\lambda_1)$ can be reduced to the cases discussed when stripping off multiples of the partition $\lambda=(3,3)$. An example is shown in Fig.~\ref{fig:ytdecc}.

A similar factorization of higher degree leaves to the product of the degree 1 leaves applies to the case of ${\rm GL}_4$. However in the case of ${\rm GL}_5$ and higher rank such factorization fails, first time for the $n=2$ and the partition $\lambda=(4,3,3)$, i.e., $\lambda^{t} = (3,3,3,1)$, of the total size $|\lambda|=10$. The Lax matrix associated to this partition is not factorized into a product of degree 1 Lax matrices. However, we can compute $L_{\lambda =(4,3,3)}$ using the fusion method.

\subsection{Fusion of degree 2}\label{app:quadnotfull}
In this subsection we present the degree 2 fusion of two elementary Lax matrices
\begin{gather*}
{L}(x)=\left(\begin{BMAT}[5pt]{c:c:c}{c:c:c}
x-x_1+P_{12}Q_{21}&-P_{12}&P_{12}P_{23}\\
 -Q_{21}&\ID&-P_{23}\\
 -Q_{32}Q_{21}&Q_{32}&x-x_1-Q_{32}P_{23}\\
 \end{BMAT}
\right)
\end{gather*}
and
\begin{gather*}
{L}'(x)=\left(\begin{BMAT}[5pt]{c:c:c}{c:c:c}
 x-x_2+P'_{13}Q'_{31}&P'_{13}Q'_{32}&-P'_{13}\\
 P'_{23}Q'_{31}&x-x_2+P'_{23}Q'_{32}&-P'_{23}\\
 -Q'_{31}&-Q'_{32}&\ID\\
 \end{BMAT}
\right)
\end{gather*}
as introduced in \eqref{eq:elax}. Here the blocks on the diagonal are of the size $k_1\times k_1$, $k_2\times k_2$ and $k_3\times k_3$ and the Lax matrices contain $k_2(k_1+k_3)$ and $k_3(k_1+k_2)$ pairs of conjugate variables respectively.

We find that their product can be decomposed as
\begin{gather*}
 L(x)L'(x)=\tilde Q\tilde L(x),
\end{gather*}
where
\begin{gather*}
\tilde Q=\left(\begin{BMAT}[5pt]{c:c:c}{c:c:c}
\ID&0&0\\
 0&\ID&0\\
 0&\tilde Q_{32}&\ID\\
 \end{BMAT}
\right)
\end{gather*}
and
\begin{gather}\label{eq:laxquad}
\tilde L(x)=\left(\begin{BMAT}[5pt]{c:c}{c:c}
 (x-x_1)(x-x_2)\ID+\tilde P_{1\tilde 2}(x\ID-\tilde\J)\tilde Q_{\tilde 2 1} &-\tilde P_{1\tilde 2}(x\ID-\tilde\J)\\
 -(x\ID-\tilde \J) \tilde Q_{\tilde 2 1}&x\ID-\tilde \J\\
 \end{BMAT}
\right) .
\end{gather}
Here we defined
\begin{gather*}
 \tilde P_{1\tilde 2}=\left(\begin{BMAT}[5pt]{c:c}{c}
 \tilde P_{12}& \tilde P'_{13}\\
 \end{BMAT}
\right) ,\qquad
 \tilde Q_{\tilde 21}=\left(\begin{BMAT}[5pt]{c}{c:c}
 \tilde Q_{21}\\ \tilde Q'_{31}\\
 \end{BMAT}
\right) ,
\end{gather*}
and
 \begin{gather*}
\tilde \J=\left(\begin{BMAT}[5pt]{c:c}{c:c}
\ID&0\\
 \tilde Q'_{32}&\ID\\
 \end{BMAT}
\right)\left(\begin{BMAT}[5pt]{c:c}{c:c}
x_2\ID&\tilde P'_{23}\\
 0&x_1\ID\\
 \end{BMAT}
\right)\left(\begin{BMAT}[5pt]{c:c}{c:c}
\ID&0\\
 -\tilde Q'_{32}&\ID\\
 \end{BMAT}
\right) .
\end{gather*}
As in the linear fusion we introduced the new canonical variables
\begin{alignat*}{3}
&\tilde P_{12}=P_{12}-P'_{13}Q'_{32}, \qquad && \tilde Q_{21}=Q_{21},&\\
&\tilde P'_{13}=P'_{13}, \qquad && \tilde Q'_{31}=Q'_{31}+Q'_{32}Q_{21},&\\
&\tilde P'_{23}=P'_{23}+P_{23}-Q_{21}P'_{13},\qquad && \tilde Q'_{32}=Q'_{32}, &\\
& \tilde P_{23} =P_{23}, \qquad && \tilde Q_{32}=Q_{32}-Q'_{32} .&
 \end{alignat*}
The final Lax matrix has $k_1k_2+k_1k_3+k_2k_3$ pairs $(\underline{p},\underline{q})$ and $\det \tilde L(x)=(x-x_1)^{k_1+k_3}(x-x_2)^{k_1+k_2}$. It corresponds to the partition $\tilde\lambda^t=(k_1+k_3,k_1+k_2)$. This can be seen when setting all~$p$ and~$q$ equal to zero in~\eqref{eq:laxquad}. One obtains
\begin{gather*}
 \left. \tilde L(x)\right|_{\underline{p},\underline{q}=\varnothing}=\diag \left(\begin{BMAT}[5pt]{c:c:c}{c}
 (x-x_1)(x-x_2)\ID_{k_1\times k_1} & (x-x_2)\ID_{k_2\times k_2} & (x-x_1)\ID_{k_3\times k_3}
 \end{BMAT}
\right) .
\end{gather*}

\subsection{Full fusion of degree 2}\label{app:fullquad}
We can further multiply the resulting Lax matrix in \eqref{eq:laxquad}
\begin{gather*}
L'(x)=\left(\begin{BMAT}[5pt]{c:c}{c:c}
 (x-x_1)(x-x_2)\ID+ P'_{1 \tilde 2}(x\ID-\J')Q'_{\tilde 2 1} &- P'_{1\tilde 2}(x\ID-\J')\\
 -(x\ID-\J') Q'_{\tilde 2 1}&x\ID- \J'\\
 \end{BMAT}
\right) ,
\end{gather*}
with
\begin{gather*}
L(x)=\left(\begin{BMAT}[5pt]{c:c}{c:c}
\ID&-P_{1\tilde 2}\\
 Q_{\tilde 2 1}&x\ID-J-Q_{\tilde 2 1}P_{1\tilde 2}\\
 \end{BMAT}
\right) ,
\end{gather*}
which corresponds to a general regular partition \eqref{eq:laxregular} with arbitrary $\lambda$ and $\mu=1^{[k_1]}$. The blocks on the diagonal are of the size $k_1\times k_1$ and $\tilde k_2=k_2+k_3$ as defined in Section~\ref{app:quadnotfull}. One finds
\begin{gather*}
 L(x)L'(x)=\tilde Q\tilde L(x)
\end{gather*}
with
\begin{gather*}
 \tilde Q=\left(\begin{BMAT}[5pt]{c:c}{c:c}
\ID&0\\
 \tilde Q_{\tilde 21}&\ID\\
 \end{BMAT}
\right) ,
\end{gather*}
and
 \begin{gather*}
\tilde L(x)=\left(\begin{BMAT}[5pt]{c:c}{c:c}
\ID&0\\
 \tilde Q'_{\tilde 21}&\ID\\
 \end{BMAT}
\right)\left(\begin{BMAT}[5pt]{c:c}{c:c}
(x-x_1)(x-x_2)\ID&-\tilde P'_{1\tilde 2}(x\ID-J')\\
 0&(x\ID-J)(x\ID-J')\\
 \end{BMAT}
\right)\left(\begin{BMAT}[5pt]{c:c}{c:c}
\ID&0\\
 -\tilde Q'_{\tilde 21}&\ID\\
 \end{BMAT}
\right) .
\end{gather*}
Here we defined
\begin{gather*}
\tilde P'_{1\tilde 2} =P'_{1\tilde 2}+P_{1\tilde 2}, \qquad \tilde P_{1\tilde 2}=P_{1\tilde 2}, \qquad \tilde Q'_{\tilde 21}=Q'_{\tilde 21} ,\qquad \tilde Q_{\tilde 21}=Q_{\tilde 21}-Q'_{\tilde 21} .
\end{gather*}
The final Lax matrix contains $k_1\tilde k_2+k_J+k_{J'}$ pairs of conjugate variables where $k_J$ and $k_{J'}$ denote the number of pairs in $J$ and $J'$ respectively. It corresponds to the partition $\tilde \lambda^t=(k_1+k_3,k_1+k_2,\lambda^t)$. Setting $p,q=0$ yields the block matrix
\begin{gather}\label{eq:qbas}
 \left. \tilde L(x)\right|_{\underline{p},\underline{q}=\varnothing}(x)=\diag \left(\begin{BMAT}[5pt]{c:c}{c}
 (x-x_1)(x-x_2)\ID_{k_1\times k_1} & (x-X_{\lambda})(x-X'_{\lambda'})\ID_{\tilde k_2\times\tilde k_2}
 \end{BMAT}
\right) .
\end{gather}
Here $X_\lambda$ is defined in \eqref{eq:Xlambda} corresponding to an arbitrary partition $\lambda$ and $X'_{\lambda'}$ follows from \eqref{eq:laxquad} and reads
\begin{gather*}
 X_{\lambda'}'=\diag \left(\begin{BMAT}[5pt]{c:c}{c}
 ((x-x_2)\ID_{k_2\times k_2} & (x-x_1)\ID_{k_3\times k_3} \end{BMAT}\right) .
\end{gather*}

\subsection[Example $\tilde \lambda=(4,3,3)$]{Example $\boldsymbol{\tilde \lambda=(4,3,3)}$}
The case ${\rm GL}_5$ and $\tilde \lambda=(4,3,3)$ we discussed at the beginning of this section corresponds to setting $k_1=1$ and $\tilde k_2=4$ in Section~\ref{app:fullquad} while setting $\lambda=(2,1,1)$ such that $n_J=3$. The Lax matrix $L'$ is obtained in the case $k_1=1$, $k_2=2$ and $k_3=2$ from the fusion in Section~\ref{app:quadnotfull} and thus yields $n_{J'}=4$. The half-dimension is $\frac 1 2 \dim_{\BC}\mathcal{M}_{4,3,3} = 11$. For $p,q=0$ the diagonal of the Lax matrix follows from~\eqref{eq:qbas} when taking $X_{\lambda}=\diag(x_3,x_3,x_3,x_4)$.

\section{Quantization}\label{sec:quantum}

Notice that the classical Yang--Baxter equation (\ref{eq:skl}) is a limit of quantum Yang--Baxter equation
\begin{gather*}
 R_{12}(x-y) \hat L_{13}(x) \hat L_{23}(y) = \hat L_{23}(y) \hat L_{13}(x) R_{12}( x- y)
\end{gather*}
in $\End(V) \otimes \End(V) \otimes A$ where $V \simeq \BC^{r}$ is the fundamental representation of $\mathfrak{gl}_{r}$ and $A$ is the quantized algebra of functions on the classical phase space parametrized locally by $\big(p_I, q^I\big)$. Here the quantum $R$-matrix is $R \in \End(V) \otimes \End(V)$ and the quantum L-operator is $\hat L \in \End(V) \otimes A$, that is an $r \times r$ matrix valued in operators in $A$. The quantum $R$-matrix is
\begin{gather*}
 R(x) = 1 + \frac{\ep \mathbb{P}}{x},
\end{gather*}
where $\ep = - i \hbar$ is the quantization parameter and $\mathbb{P}$ is the permutation operator (\ref{eq:perm}).

In terms of matrix elements $\hat L_{ij}$ we have the quantum Yang Baxter equation in $A$
\begin{gather*}
 \big[ \hat L_{ij}(x), \hat L_{kl}(y)\big] = -\frac{\ep}{x - y} \big(\hat L_{kj}(x) \hat L_{il}(y) - \hat L_{kj}(y) \hat L_{il}(x)\big)
\end{gather*}
and its classical limit is
\begin{gather*}
 \{ L_{ij}(x), L_{kl}(y) \} = - \frac{1} {x - y} (L_{kj}(x) L_{il}(y) - L_{kj}(y) L_{il}(x))
\end{gather*}
with the standard convention
\begin{gather*}
 \big[\hat \phi , \hat \psi\big] = \ep \{ \phi, \psi \} + O\big(\ep^2\big) , \qquad \ep \to 0,
\end{gather*}
where $ \big[\hat \phi , \hat \psi\big] $ denotes the commutator of the elements $\hat \phi$, $\hat \psi$ of the algebra $A$ that correspond to the quantization of the functions $\phi$, $\psi$ on the classical phase space with Poisson brackets $\{\phi, \psi\}$. In particular the canonical coordinates $ p_I$, $q^I$ have Poisson bracket
\begin{gather*}
 \big\{ p_I, q^J \big\} = \delta_{I}^J
\end{gather*}
and the respective operators have commutation relations
\begin{gather*}
 \big[ \hat p_I, \hat q^J \big] = \ep \delta_{I}^J
\end{gather*}
that can be represented in the algebra of differential operators acting on Hilbert space of states represented by function of $q^I$ as
\begin{gather*}
 \hat q^I \mapsto q^I, \qquad
 \hat p_I \mapsto \ep \frac{ \partial}{ \partial q^I}.
\end{gather*}

For a polynomial function $f(\underline{\hat q},\underline{ \hat p})$ the normal ordering notation ${:}f(\underline{\hat q},\underline{\hat p}){:}$ means placing all operators $\hat p_I$ to the right of the operators $\hat q^I$ in each monomial.

The quantum version $\hat L _{\underline{\lambda}, \underline{x}, \mu}(x)$ of all our classical solutions $L_{\underline{\lambda}, \underline{x}, \mu}(x)$ is obtained by replacing all variables $(\underline{p},\underline{q})$ by the operators $\underline{\hat p}$, $\underline{ \hat q}$ and assuming normal ordering convention. One can check that such operator valued matrix $\hat L_{\underline{\lambda}, \underline{x}, \mu}(x)$ satisfies quantum Yang--Baxter equation. The commuting Hamiltonians are obtained from the expansion of the quantum spectral determinant (quantum spectral curve) as in~\cite{Chervov:2006xk}
\begin{gather}
\hat W_{x,y} =\tr A_r \big(y-{\rm e}^{\epsilon \partial_x}\hat L_1'(x)\big)\big(y-{\rm e}^{\epsilon \partial_x}\hat L_2'(x)\big)\cdots\big(y- {\rm e}^{\epsilon \partial_x}\hat L_r'(x)\big)\nonumber\\
\hphantom{\hat W_{x,y}}{} =\sum_{k=0}^r (-1)^k y^{r-k} \hat \chi_k(x+\epsilon){\rm e}^{\epsilon k \partial_x},\label{eq:qspecdet}
\end{gather}
where $\hat L'(x)=\rho_1(g_\infty)\hat L _{\underline{\lambda}, \underline{x}, \mu}(x)$, cf.~\eqref{eq:twistedL}, and $A_k$ is the normalised antisymmetrizer acting on the $k$-fold tensor product of $\mathbb{C}^r$.
The quantum characters whose coefficients generate the algebra of quantum commuting Hamiltonians (Bethe subalgebra) are
\begin{gather}\label{eq:qchar}
\hat \chi_k(x)=\tr A_k \hat L_{1}'(x)\cdots \hat L_{k}'(x+\epsilon (k-1))
\end{gather}
see also \cite{molevbook}. The definition of the quantum spectral determinant~(\ref{eq:qspecdet}) is a quantum version of the classical spectral curve (\ref{eq:specdet}), and there is a quantum version of the
factorization (\ref{eq:chik})
\begin{gather*}%\label{eq:quantumchik}
 \hat \chi_k (x) = \hat {\mathcal{Q}}^{[\lambda]}_{r-k}(x+\ep|\mu|) \hat {\mathcal{X}}^{[\lambda, \mu]}_{k}(x),
\end{gather*}
where the $c$-valued polynomials are
\begin{gather*}%\label{eq:Qpolquantum}
\hat{ \mathcal{Q}}_i^{[\lambda]}(x)=\prod_{j=1}^{\lambda_{i+1}}\prod_{k=1}^{\lambda_{j}^t-i}\left(x-x_j+\ep\left(\sum_{l=1}^{j-1}\lambda_l^t+k-1\right)\right) .
\end{gather*}

The quantization of the corresponding integrable systems in the context of the $\mathcal{N}=2$ supersymmetric quiver gauge theories has been considered in~\cite{Nekrasov:2013xda}, in particular the $q$-character functions appearing in \cite{Nekrasov:2013xda} after \cite{Frenkel:1998} stand for the eigenvalue of the quantum commuting Hamiltonians~(\ref{eq:qchar}).

The quantized symplectic leaves $\hat{\mathcal{M}}_{\underline{\lambda}, \underline{x}, \mu}$ are modules, typically infinite-dimensional, for the dual Yangian algebra $\mathbf{Y}(\mathfrak{gl}_{r})^{*}$ which is a quantum deformation algebra of the space of functions on the Poisson--Lie group ${\rm GL}_{r}(\mathcal{K}_{\mathbb{P}^{1}_x})$. This representation theory relates to the `pre-fundamental' modules of Hernandez--Jimbo~\cite{hernandez2012asymptotic} associated to the individual singularities at points $x_i$ labeled by a fundamental co-weight $\check \omega_{\lambda_{i}^{t}}$.

\appendix

\section{Twisted cotangent bundles of generalized flag varieties}\label{sec:twisted-flag}
Let $\fg$ be a reductive Lie algebra and let $\fg = \fn_{-} \oplus \fh \oplus \fn_{+}$ be a decomposition of $\fg$ into the Cartan subalgebra $\fh$, the negative nilpotent subspace $\fn_{-} = \oplus_{\alpha < 0} \fg_{\alpha}$
and positive nilpotent subspace $\fn_{+} = \oplus_{\alpha> 0} \fg_{\alpha}$. Here $\alpha$ denote a root of $\fg$ and $\fg_{\alpha}$ the $\alpha$-root subspace of $\fg$. Let $\fb_{+} = \fh + \fn_{+}$ and $\fb_{-} = \fh + \fn_{-}$ be the respective Borel subalgebras. If $\fg = \fgl_r$ then $\fb_{+}$ (or $\fb_{-}$) is represented by upper (or lower) triangular matrices including the diagonal, and $\fn_{+}$ (or $\fn_{-}$) is represented by strictly upper (or lower) triangular matrices excluding the diagonal.

Let $G$, $H$, $N_{\pm}$, $B_{\pm}$ the respective Lie groups with Lie algebras $\fg$, $\fh$, $\fn_{\pm}$, $\fb_{\pm}$, and let $\mass \in \fh^{*}$ be a~weight. Here we record explicit formulas for representation of $U\fg$ in $\mass$-twisted differential operators on the complete flag manifold $G/B_{+}$ following the approach of Harish-Chandra, Springer, Kostant, Beilinson--Bernstein. We identify the big cell of $G/B_{+}$ with $N_{-}$ and denote elements of $N_{-}$ by $Q$.

We compute the vector field $L_{X}$ associated to the action of Lie algebra element $X$ on $G/B_{+}$ from the left. Let $\vep$ be infinitesimal parameter, and let $ 1 + \vep X$ be a group element corresponding to Lie algebra element $X \in \fg$. Let $\tilde Q = Q + \vep \delta_{X} Q$ denotes a coset representative in $G/B_{+}$ obtained from the action of $1 + \vep X$ on $Q$ from the left:
\begin{gather}\label{eq:var}
 (1 + \vep X) Q = (Q + \vep \delta_{X} Q)(1 + \vep n_{+} + \vep h), \qquad n_{+} \in \fn_{+}, \quad h \in \fh,
\end{gather}
where $(1 + \vep n_{+} + \vep h)$ is an element of $B_{+}$ that gauges the deformation of $Q$. We find
\begin{gather*}
 X Q = Q ( n_{+} + h) + \delta_{X} Q, \qquad n_{+} \in \fn_{+},\quad h \in \fh
\end{gather*}
and thus
\begin{gather*}
 \delta_{X} Q = Q \big[Q^{-1} X Q\big]_{-},
\end{gather*}
where $[\phantom{x}]_{-}$ denotes the projection $ \fg \to \fn_{-}$. The corresponding vector field and the differential operator on scalar functions on $N_{-}$ is $ L_{X} = - \delta_{X} Q \frac {\partial} {\partial Q} $ that is
\begin{gather*}
 L_{X} = - Q \big[Q^{-1} X Q\big]_{-} \frac {\partial}{\partial Q} ,
\end{gather*}
where the minus sign comes from the standard convention of defining the vector fields associated to the group actions on manifolds in such a way as to preserve the Lie algebra bracket.

We are actually interested in a more general situation, when the differential operator $L_{X}$ acts not on functions on $G/B_{+}$ but on sections of line bundle induced from the $H$-bundle $G/N_{+} \to G/B_{+}$ by a semi-simple co-weight $\mass \in \mathfrak{h}^{*}$, e.g.,
\begin{gather*}
 \mass =
 \begin{pmatrix}
 \mass_1 & &\\
 & \mass_2 & &\\
 & & \mass_3
 \end{pmatrix} .
\end{gather*}
The additional connection term is $-\mass(h_X)$ for the diagonal variation $h_X$ in the coset computation~(\ref{eq:var})
\begin{gather*}
 L_{X, \mass} = L_{X} - \mass(h)
\end{gather*}
and since
\begin{gather*}
 h_X = \big[Q^{-1} X Q\big]_{0}
\end{gather*}
where $[\phantom{x}]_{0}$ denotes the projection to the diagonal part $ \fg \to \fh$, we find the differential operator
\begin{gather*}
 L_{X, \mass} = - Q \big[Q^{-1} X Q\big]_{-} \frac {\partial}{\partial Q} - \big\langle \mass, \big[Q^{-1} X Q\big]_{0} \big\rangle .
\end{gather*}
acting on sections of the line bundle on $G/B_{+}$.

Now we fix $\fg = \fgl_{r}$. Let $(e_{ij})_{i,j \in [1, r]}$ denote the standard basis elements of $\fgl_{r}$ represented by matrices whose $(i,j)$-entry is equal to 1 and the rest is 0. The upper-triangular Borel subgroup~$B_{+}$ preserves the standard full flag
\begin{gather*}
 0 \subset \BC e_1 \subset \BC e_1 \oplus \BC e_2 \subset \dots \subset \BC e_1 \oplus \BC e_2 \dots \oplus \BC e_{r}.
\end{gather*}
Further we define the coordinates $(q_{i,j})$ with $1 \leq j<i \leq r$ on $N_{-}$ taking the matrix elements of $Q \in N_{-}$ in the defining representation of $\fgl_r$
\begin{gather*}
 q_{ij} := Q_{ij}, \qquad 1 \leq j<i \leq r
\end{gather*}
for example, for $\fgl_3$ we have
\begin{gather*}
 Q =
 \begin{pmatrix}
 1 & & \\
 q_{2,1} & 1 & \\
 q_{3,1} & q_{3,2} & 1
 \end{pmatrix} .
\end{gather*}

We evaluate in coordinates $(q)_{ij}$ the differential operator $L_{x, \mass}$ associated to each basis element $X = e_{ij}$ in $\fg$, and we assemble $r \times r$ matrix $\hat L$ valued in twisted differential operators
\begin{gather*}
 \hat L_{ij, \mass} = L_{e_{ji}, \mass} .
\end{gather*}
Let us denote
\begin{gather*}
 p_{ij} = \frac{\partial}{\partial q_{ij}} , \qquad i > j
\end{gather*}
with
\begin{gather*}
 [p_{ij}, q_{kl}] = \delta_{ik} \delta_{jl},
\end{gather*}
and assemble the upper triangular matrix with only non-zero entries $(P)_{ij} = p_{ji}$ for $ i > j $
\begin{gather*}
 P =
 \begin{pmatrix}
 0 & p_{21} & p_{31} \\
 0 & 0 & p_{32} \\
 0 & 0 & 0
 \end{pmatrix} .
\end{gather*}
Then
\begin{gather*}
 L_{e_{ij}, \mass} = - \tr Q \big[Q^{-1} e_{ij} Q\big]_{-} P - \tr Q^{-1} e_{ij} Q \mass
 = -{:}\tr e_{ij} Q (\mass + [PQ]_{+}) Q^{-1}{:} \\
 \hphantom{L_{e_{ij}, \mass}}{} = -{:}\big( Q (\mass + [PQ]_{+}) Q^{-1}\big)_{ji}{:}
\end{gather*}
where normal ordering notation $:\phantom{x}$: means that all symbols of the operators $p_{ij}$ are kept to the right,
and consequently we find
\begin{gather*}
 \hat L_{ij} = - {:}\big( Q (\mass + [PQ]_{+}) Q^{-1}\big)_{ij}{:} .
\end{gather*}

\section{Determinant formula}
The determinant of a block matrix can be written as
\begin{gather}\label{eq:detabcd}
\left|
 \begin{matrix}
 A&B\\C&D
 \end{matrix}
\right|=\det\big(A-BD^{-1}C\big)\det D .
\end{gather}
\section[Sklyanin relation for elementary $\mu$ partitions]{Sklyanin relation for elementary $\boldsymbol{\mu}$ partitions}\label{app:proof}

In order to show that the Lax matrix \eqref{eq:lax22alt} satisfies the Sklyanin relation \eqref{eq:skl} we verify~\eqref{eq:FK} and~\eqref{eq:FF} in the following.

Starting with \eqref{eq:FF} we first note that $F_{\ic,\ic}$ in \eqref{eq:G} can be written as
\begin{gather*}
 F_{\ic,\ic}=J_{\ic,\ic}+Q_-G'Q_-^{-1} \qquad\text{with}\quad G'=P_0+Q_0[Q_+P_-]_-Q_0^{-1},
\end{gather*}
where $J_{\ic,\ic}$ has been defined in~\eqref{eq:Jmatrix} and satisfies the $\mathfrak{gl}\big(\frac{r}{2}\big)$ commutation relations~\eqref{eq:gln}.

It follows that \eqref{eq:FF} is equivalent to
\begin{gather}
 \big[\big\{([P_+Q_-]_+)_2,\big(Q_-^{-1}\big)_1\big\}(Q_-)_1,G'_1\big] -\big[\big\{([P_+Q_-]_+)_1,\big(Q_-^{-1}\big)_2\big\}(Q_-)_2,G'_2\big]\nonumber \\
 \qquad{} +\{G'_1,G'_2\}=(G'_1-G'_2)\mathbb{P} ,\label{eq:conditions1}
\end{gather}
where $[X,Y]=XY-YX$ denotes the anticommutator. Further we use the notation $X_1=X\otimes\ID $ and $X_2=\ID \otimes X$ and $\mathbb{P}$ act as a permutation such that $\mathbb{P}X_1=X_2\mathbb{P}$.
It is convenient to consider different cases. Writing~\eqref{eq:conditions1} in components and taking into account that $G'$ and $Q_-$ are lower diagonal while $[P_+Q_-]_+$ is upper diagonal results in the conditions
\begin{gather*}
 \{G'_{ab},G'_{cd}\} =\delta_{cb}G'_{ad}-\delta_{ad}G'_{cb}\qquad \text{for} \quad a\geq b \wedge c\geq d ,\\
\big[\big\{([P_+Q_-]_+)_{cd},Q_-^{-1}\big\}Q_-,G'\big]_{ab} =\delta_{cb}G'_{ad}-\delta_{ad}G'_{cb}\qquad \text{for} \quad c<d .
\end{gather*}
The first relation can be verified straightforwardly. The second relation follows when noting that $\big\{([P_+Q_-]_+)_{ab},Q_-^{-1}\big\}Q_-=-e_{ba}$ for $a<b$.

To show \eqref{eq:FK} we note that $ \Kb_{\ic,\ia}$ and $\K_{\ia,\ic}$ are independent of the variables $p_{ij}$. Thus the brackets reduces to
\begin{gather}\label{eq:skl1}
 \sum_{s,t=1}^{\frac{r}{2}} \frac{\partial F_{ij}}{\partial p_{st}}\frac{\partial \bar K_{kl}}{\partial q_{ts}}=+\bar K_{il}\delta_{kj} ,\qquad \sum_{s,t=1}^{\frac{r}{2}} \frac{\partial F_{ij}}{\partial p_{st}}\frac{\partial K_{kl}}{\partial q_{ts}}=-K_{kj}\delta_{il} .
\end{gather}
Here we suppressed the subindeces $\ia$ and $\ic$. The Latin indices take values $i,j,k,l=1,\ldots,\frac{r}{2}$.

Using the relation $\partial_q\K=\K\big(\partial_q\Kb\big)\K$ which follows from \eqref{eq:abc}, one finds that the two equations in \eqref{eq:skl1} are equivalent. They can be written as
\begin{gather}\label{eq:equation}
 \sum_{s,t} \frac{\partial G_{ij}}{\partial p_{st}}\frac{\partial \Kb_{kl}}{\partial q_{ts}}=(Q_0Q_+)_{il}(Q_-)_{kj} .
\end{gather}
In order to show this relation it is again convenient to consider different cases and take into account the dependence of $G$ on $p$. For a particular choice of the indices $i$ and $j$ we see that~\eqref{eq:equation} is equivalent to:\\
Case $i<j$
\begin{gather*}
 \sum_{s<t} \frac{\partial (P_+Q_-)_{ij}}{\partial p_{st}}\frac{\partial (Q_-)_{kl}}{\partial q_{ts}}=\delta_{il}(Q_-)_{kj} .
\end{gather*}
Case $i>j$
\begin{gather*}
 \sum_{s>t} \frac{\partial (Q_+P_-)_{ij}}{\partial p_{st}}\frac{\partial (Q_+)_{kl}}{\partial q_{ts}}=(Q_+)_{il}\delta_{kj} .
\end{gather*}
Case $i=j$
\begin{gather*}
 \sum_{s=t} \frac{\partial (P_0)_{ij}}{\partial p_{st}}\frac{\partial (Q_0)_{kl}}{\partial q_{ts}}=(Q_0)_{il}\delta_{kj} .
\end{gather*}
These equations can be checked explicitly. The derivatives with respect to $q_{ij}$ for $i\neq j$ essentially yield delta functions while the derivatives with respect to~$p_{ij}$ give the $q$-dependence.
\section{The number of independent commuting Hamiltonians}\label{sec:proofcharges}
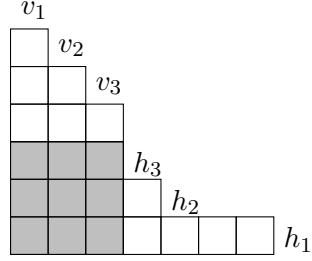
\begin{figure}[t]\centering
\begin{tikzpicture}
 \draw [fill=lightgray] (0.5,-1) rectangle (2,-2.5);
\foreach \a in {1,2,3,4,5,6,7} {
 \begin{scope}[shift={(0.5*\a,-2.5)}]
 \draw (0,0) rectangle (0.5,0.5);
 \end{scope}
 }
 \foreach \a in {1,2,3,4} {
 \begin{scope}[shift={(0.5*\a,-2)}]
 \draw (0,0) rectangle (0.5,0.5);
 \end{scope}
 }
 \foreach \a in {1,2,3} {
 \begin{scope}[shift={(0.5*\a,-1.5)}]
 \draw (0,0) rectangle (0.5,0.5);
 \end{scope}
 }
 \foreach \a in {1,2,3} {
 \begin{scope}[shift={(0.5*\a,-1)}]
 \draw (0,0) rectangle (0.5,0.5);
 \end{scope}
 }
 \foreach \a in {1,2} {
 \begin{scope}[shift={(0.5*\a,-0.5)}]
 \draw (0,0) rectangle (0.5,0.5);
 \end{scope}
 }
 \foreach \a in {1} {
 \begin{scope}[shift={(0.5*\a,0)}]
 \draw (0,0) rectangle (0.5,0.5);
 \end{scope}
 }
 \node [right] at (4,-2.3) {$h_1$};
 \node [right] at (2.5,-1.8) {$h_2$};
 \node [right] at (2,-1.3) {$h_3$};
 \node [above right] at (0.5,0.5) {$v_1$};
 \node [above right] at (1,0) {$v_2$};
 \node [above right] at (1.5,-0.5) {$v_3$};
\end{tikzpicture}
\caption{Frobenius-like coordinates of the partition $(7,4,3,3,2,1)$: $f=3$, $h_1=4$, $h_2=1$, $h_3=0$, $v_1=3$, $v_2=2$, $v_3=1$.}\label{fig:yt}
\end{figure}

In order to show the relation \eqref{eq:sumcharges} it is convenient to introduce Frobenius-like coordinates to label the partitions, compare, e.g.,~\cite{macdonald}.

Let us focus on the case $\mu=\varnothing$ which is sufficient as we will argue at the end of this section. Using Frobenius-like coordinates any partition $\lambda$ can be written as
\begin{gather*}
 \lambda=(h_1+f,\ldots,h_f+f,\underbrace{f,\ldots,f}_{v_f},\underbrace{f-1,\ldots,f-1}_{v_{f-1}-v_f},\ldots,\underbrace{1,\ldots,1}_{v_1-v_2}) .
\end{gather*}
Here $f$ denotes the length of the sides of the maximal square which fits into the lower left corner of the corresponding Young diagram as shown in Fig.~\ref{fig:yt}. the variables $h_i$ denote the number of boxes on the right of the square in the $i$th row while the variables $v_k$ denote the number of boxes above the square in the $k$th row. The coordinates introduced in this way have the advantage that the transpose can be obtained by interchanging $h_i$ and $v_i$ where $i=1,\ldots,f$ such that
\begin{gather*}
 \lambda^t=\lambda\big|_{v_i\leftrightarrow h_i}=(v_1+f,\ldots,v_f+f,\underbrace{f,\ldots,f}_{h_f},\underbrace{f-1,\ldots,f-1}_{h_{f-1}-h_f},\ldots,\underbrace{1,\ldots,1}_{h_1-h_2}) .
\end{gather*}
To introduce Frobenius-like coordinates in \eqref{eq:sumcharges} we decompose the sum over the elements of the partition $\lambda$ as
\begin{align}
 \sum_{k=0}^rn_k^{[\lambda,\varnothing]}&=\sum_{k=1}^r\sum_{j=1}^k\lambda_j-\frac{r(r+1)}{2}\nonumber\\
 &= \sum_{k=1}^f(r-k+1)\lambda_k+\sum_{k=1}^{r-f}(r-f-k+1)\lambda_{k+f}-\frac{r(r+1)}{2} .\label{eq:app1}
 \end{align}
Now the first term on the right-hand-side of \eqref{eq:app1} can be written in terms of the variables $h_i$ as
\begin{gather*}
 \sum_{k=1}^f(r-k+1)\lambda_k=\sum_{k=1}^f(r-k+1)(h_k+f) ,
\end{gather*}
while the second one can be written in terms of the variables $v_i$ as
\begin{align*}
 \sum_{k=1}^{r-f}(r-f-k+1)\lambda_{k+f}&=f\sum_{k=1}^{v_f}(r-f-k+1)+\sum_{l=1}^{f-1} l\sum_{k=v_{l+1}+1}^{v_l}(r-f-k+1)\\
 &=-\frac{1}{2}\left(\sum_{l=1}^fv_l^2+(2f-2r-1)\sum_{l=1}^fv_l\right) .
 \end{align*}
Finally, using the relation $ r=\sum\limits_{i=1}^fh_i+f^2+\sum\limits_{i=1}^fv_i$ we find that
\begin{gather*}
 \sum_{k=0}^rn_k^{[\lambda,\varnothing]} =\frac{1}{2}\left(r^2-\sum_{l=1}^f(v_l+f)^2-\sum_{l=1}^f h_{l}(2l-1)\right) =\frac{1}{2}\left(r^2-\sum_{i=1}^{\lambda_1}\left(\lambda_i^t\right)^2\right)=d_{[\lambda,\varnothing]} .
 \end{gather*}
As $\lambda$ is arbitrary it follows that the same relation holds for partitions $\mu$ with $\lambda=\varnothing$. Combining the two relations we find that
\begin{align*}
 \sum_{k=0}^r n_k^{[\lambda,\mu]}&= \sum_{k=0}^r\big( n_k^{[\lambda,\varnothing]}+n_k^{[\varnothing,\mu]}\big)+\frac{r(r+1)-|\lambda|(|\lambda|+1)-|\mu|(|\mu|+1)}{2}\\
 &=d_{[\lambda,\varnothing]}+d_{[\varnothing,\mu]}+\frac{r^2-|\lambda|^2-|\mu|^2}{2}=d_{[\lambda,\mu]} ,
 \end{align*}
where we used that $r=|\lambda|+|\mu|$. Another way to show \eqref{eq:sumcharges} is to use the relation \eqref{eq:transp}.

\section{Rewritten polynomials}\label{app:prf2}
The relation in \eqref{eq:QasP} can be shown using the relation
\begin{gather}\label{eq:transp}
\lambda^t=(\underbrace{r,\ldots,r}_{\lambda_r},\underbrace{r-1,\ldots,r-1}_{\lambda_{r-1}-\lambda_r},,\ldots,\underbrace{1,\ldots,1}_{\lambda_1-\lambda_2}) .
\end{gather}
First it follows that $\mathcal{Q}_i^{[\lambda]}$ in \eqref{eq:Qpol} can be written as
\begin{gather}\label{eq:appe}
 \mathcal{Q}_i^{[\lambda]}(x)=\prod_{k=1}^{\lambda_{i+1}} (x-x_k)^{\lambda^t_k-i}=\prod_{l=i+1}^{r} \prod_{k=\lambda_{l+1}+1}^{\lambda_{l}}(x-x_k)^{\lambda_{k}^t-i}=\prod_{l=i+1}^{r} \prod_{k=\lambda_{l+1}+1}^{\lambda_{l}}(x-x_k)^{l-i} ,
\end{gather}
with $\lambda_{r+1}=0$ and using that $\lambda_{k}^t=l$ for $k=\lambda_{l+1}+1,\ldots,\lambda_l$. Note that the polynomiality of~$\mathcal{Q}_i^{[\lambda]}$ is now manifest. Further we find
\begin{gather*}
 \mathcal{Q}_i^{[\lambda]}(x)=\prod_{l=i+1}^{r} \prod_{k=\lambda_{l+1}+1}^{\lambda_{l}}(x-x_k)^{l-i}=\prod_{l=i+1}^{r} \prod_{k=1}^{\lambda_{l}-\lambda_{l+1}}(x-x_{\lambda_{l}-k+1})^{l-i} .
\end{gather*}
Identifying $m_l=\lambda_l-\lambda_{l+1}$ we recover \eqref{eq:QasP}. We remark that we can write the polynomial simply as
\begin{gather*}
 \mathcal{Q}_i^{[\lambda]}(x)=\prod_{l=i+1}^{r} \prod_{k=\lambda_{l+1}+1}^{\lambda_{l}}(x-x_k)^{l-i}= \prod_{l=i+1}^{r} \prod_{k=1}^{\lambda_{l}}(x-x_k) ,
\end{gather*}
after rearranging the product.

\section{Examples}\label{app:examples}
\newcommand{\quz}{
\begin{tikzpicture}
\foreach \a in {1} {
 \begin{scope}[shift={(0.7*\a,0)}]
 \draw (0.3*\a,0) circle (0.3cm);
 \draw[black,thick] (0.3*\a+0.3,0)--(0.3*\a+0.7,0);
 \node [below] at (0.3*\a,-0.5) {$\a$};
 \end{scope}
 }
 \draw (2,0) circle (0.3cm);
 \node [below] at (2,-0.5) {$2$};
 \node at (1,0) {$1$};
 \node at (2,0) {$1$};

 \draw[black,thick] (2,0.3) -- (2,0.7);
 \draw (1.7,0.7) rectangle (2.3,1.3);
 \node at (2,1) {$1$};
 \draw[black,thick] (1,0.3) -- (1,0.7);
 \draw (0.7,0.7) rectangle (1.3,1.3);
 \node at (1,1) {$1$};
\end{tikzpicture}
}

\newcommand{\nuz}{
\begin{tikzpicture}
 \draw[step=0.5cm,lightgray,very thin] (-0.5,-2) grid (1.5,-0.5);
 \draw [thick, ->] (-0.5,-2) -- (-0.5,-0.5);
 \draw [thick, ->] (-0.5,-2) -- (1.5,-2);
 \node [above] at (-0.5,-0.5) {j};
 \node [right] at (1.5,-2) {i};
 \node [below] at (1,-2) {$r$};
 \node [below left] at (-0.5,-2) {$0$};
\draw[black,very thick](-0.5,-2)--(0,-1.5)--(0.5,-1.5)--(1,-2);
 \node [above] at (0,-1.5) {$m_1$};
 \node [above right] at (0.5,-1.5) {$m_2$};
\foreach \a in {0,1} {
 \begin{scope}[shift={(0.5*\a,-1.5)}]
 \node at (0,-0.01) {\textcolor{black}{\lamdot}};
 \end{scope}
 }
\foreach \a in {-1,0,1,2} {
 \begin{scope}[shift={(0.5*\a,-2)}]
 \node at (0,-0.01) {\textcolor{black}{\lamdot}};
 \end{scope}
 }
\end{tikzpicture}
}

\newcommand{\nuuz}{
\begin{tikzpicture}
 \draw[step=0.5cm,lightgray,very thin] (-0.5,-2) grid (2,0);
 \draw [thick, ->] (-0.5,-2) -- (-0.5,-0);
 \draw [thick, ->] (-0.5,-2) -- (2,-2);
 \node [above] at (-0.5,-0) {j};
 \node [right] at (2,-2) {i};
 \node [below] at (1.5,-2) {$r$};
 \node [below left] at (-0.5,-2) {$0$};
\draw[black,very thick](-0.5,-2)--(0.5,-1)--(1.5,-2);
 \node [above right] at (0.5,-1) {$m_2$};
\foreach \a in {1} {
 \begin{scope}[shift={(0.5*\a,-1)}]
 \node at (0,-0.01) {\textcolor{black}{\lamdot}};
 \end{scope}
 }
\foreach \a in {0,1,2} {
 \begin{scope}[shift={(0.5*\a,-1.5)}]
 \node at (0,-0.01) {\textcolor{black}{\lamdot}};
 \end{scope}
 }
\foreach \a in {-1,0,1,2,3} {
 \begin{scope}[shift={(0.5*\a,-2)}]
 \node at (0,-0.01) {\textcolor{black}{\lamdot}};
 \end{scope}
 }
\end{tikzpicture}
}

\newcommand{\quuz}{
\begin{tikzpicture}
\foreach \a in {1,2} {
 \begin{scope}[shift={(0.7*\a,0)}]
 \draw (0.3*\a,0) circle (0.3cm);
 \draw[black,thick] (0.3*\a+0.3,0)--(0.3*\a+0.7,0);
 \node [below] at (0.3*\a,-0.5) {$\a$};
 \end{scope}
 }
 \draw (3,0) circle (0.3cm);
 \node [below] at (3,-0.5) {$3$};
 \node at (1,0) {$1$};
 \node at (2,0) {$2$};
 \node at (3,0) {$1$};
 \draw[black,thick] (2,0.3) -- (2,0.7);
 \draw (1.7,0.7) rectangle (2.3,1.3);
 \node at (2,1) {$2$};
\end{tikzpicture}
}

\begin{figure}[t]
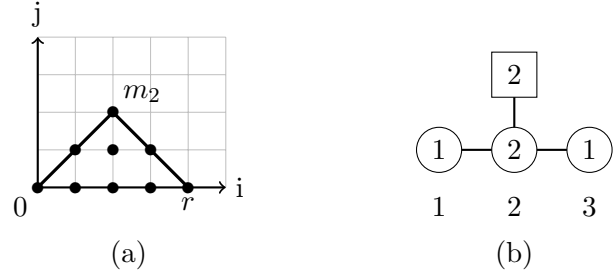
 \centering
 \begin{minipage}[b]{50mm}
 \centering \nuuz\\ (a)
 %\caption{} \label{fig:fig1b}
 \end{minipage}
\begin{minipage}[b]{50mm}
 \centering\quuz \\ (b)
 %\caption{} \label{fig:fig2b}
 \end{minipage}
 \caption{(a) shows the Newton polygon and (b) shows the quiver diagram for $\mu=(2,2)$, $\lambda=\varnothing$ and $r=4$. } \label{fig:muuuu}
\end{figure}

We consider a couple of explicit examples associated to the diagrams as shown on Figs.~\ref{fig:muuuu} and~\ref{fig:21}. The ranks in the gauged (circled) nodes on the right correspond to the heights of the profile on the left. The ranks of the framed (boxed) nodes on the right counting the number of fundamentals correspond to the (minus) change of the slope of the profile on the left in the corresponding corner (negative second difference),
 cf.~Section~\ref{sec:specdet} and \cite{Nekrasov:2012xe}.

\subsection[$\mu=(2,1)$ and $\lambda=\varnothing$]{$\boldsymbol{\mu=(2,1)}$ and $\boldsymbol{\lambda=\varnothing}$}

\begin{figure}[t]
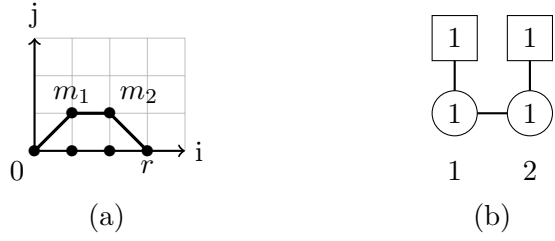
 \centering
 \begin{minipage}[b]{50mm}
 \centering
 \nuz \\ (a)
 %\caption{}
 %\label{fig:fig1}
 \end{minipage}
\begin{minipage}[b]{50mm}
 \centering\quz \\ (b)
% \caption{}
 %\label{fig:fig2}
 \end{minipage}
 \caption{(a) shows the Newton polygon and (b) shows the quiver diagram for $\mu=(2,1)$, $\lambda=\varnothing$ and $r=3$. The ranks in the gauged (circle) nodes on the right correspond to the heights of the profile on the left. The ranks of the framed (boxed) nodes on the right counting the number of fundamentals correspond to the change of the slope of the profile on the left.} \label{fig:21}
\end{figure}

The Lax matrix corresponding to the partitions $\mu=(2,1)$ and $\lambda=\varnothing$ can be obtained from~\eqref{eq:lax222111}. It contains two conjugate pairs of variables and reads
\begin{gather*}
 L_{(2,1)}(x;\underline{p},\underline{q})=\left(\begin{matrix}
 0&0&-{\rm e}^{-q_{3,3}}\\
 0&1&- p_{2,3}\\
 {\rm e}^{q_{3,3}}&q_{3,2}& x-p_{3,3}-q_{3,2}p_{2,3}
 \end{matrix}
\right) .
\end{gather*}
It resembles the structure of the DST and Toda chain. The Hamiltonians can be computed explicitly. The emerging Newton polygon and quiver diagram are depicted in Fig.~\ref{fig:21}, cf.~Section~\ref{sec:specdet}.

\subsection[$\mu=(2,2)$ and $\lambda=\varnothing$]{$\boldsymbol{\mu=(2,2)}$ and $\boldsymbol{\lambda=\varnothing}$}

The Lax matrix corresponding to the partitions $\mu=(2,2)$ and $\lambda=\varnothing$ can be obtained from~\eqref{eq:lax22alt}. They contain four conjugate pairs of variables and can be written as
\begin{gather*}
L_{(2,2)}(x;\underline{p},\underline{q})=\left(\begin{BMAT}[5pt]{c:c}{c:c}
0&\K_{(2,2)}\\
\Kb_{(2,2)}& x\ID -F_{(2,2)}
 \end{BMAT}\right) ,
\end{gather*}
where the $2\times 2$ block matrices are given by
\begin{gather*}
F_{(2,2)}=\left(\begin{matrix}
 p_{3,3}-q_{4,3}p_{3,4}&p_{3,4}\\
 {\rm e}^{q_{4,4}-q_{3,3}}p_{4,3}-q_{4,3}(p_{4,4}-p_{3,3}+q_{4,3}p_{3,4})& p_{4,4}+q_{4,3}p_{3,4}
 \end{matrix}
\right) ,
\\
\K_{(2,2)}=\left(\begin{matrix}
-{\rm e}^{-q_{3,3}}-{\rm e}^{-q_{4,4}}q_{3,4}q_{4,3}&{\rm e}^{-q_{4,4}}q_{3,4}\\
 {\rm e}^{-q_{4,4}}q_{4,3}&-{\rm e}^{-q_{4,4}}
 \end{matrix}
\right),
\\
\Kb_{(2,2)}=\left(\begin{matrix}
 {\rm e}^{q_{3,3}}&{\rm e}^{q_{3,3}}q_{3,4}\\
 {\rm e}^{q_{3,3}}q_{4,3}&{\rm e}^{q_{4,4}}+{\rm e}^{q_{3,3}}q_{3,4}q_{4,3}
 \end{matrix}
 \right).
\end{gather*}

\section{Cluster structures}\label{sec:clusterstructures}
In this section we elaborate on the cluster structure of the fusion procedure in Section~\ref{sec:fusel2}. As an example we study the ${\rm GL}_3$ case and introduce the Lax matrices
\begin{gather*}
 L_1(x)=\left(\begin{matrix}
 x+p_{12}''q_{21}''+p_{13}''q_{31}'' &-p_{12}''&-p_{13}''\\
 - q_{21}''&1&0\\
 - q_{31}''&0&1
 \end{matrix}\right) ,
\\
 L_2(x)=\left(\begin{matrix}
 1&-p_{12}'&0\\
 q_{21}'&x-q_{21}'p_{12}'+p_{23}'q_{32}'&-p_{23}'\\
 0&-q_{32}'&1
 \end{matrix}\right) ,
\\
 L_3(x)=\left(\begin{matrix}
 1&0&-p_{13}\\
 0&1&-p_{23}\\
 q_{31}&q_{32}&x-q_{31}p_{13}-q_{32}p_{23}
 \end{matrix}\right) .
\end{gather*}
Looking at the product
\begin{gather*}
 L_3(x-x_3)L_2(x-x_2)L_1(x-x_1)
\end{gather*}
there are two ways to proceed with the fusion. The standard way used in Section~\ref{sec:fusel2} first computes the Lax matrix resulting from the fusion $L_3(x-x_3)L_2(x-x_2)$ and then computes the final result by fusing the result with $L_1(x-x_1)$ from the right. It yields the matrix
\begin{gather*}
 L_{(32)1}(x)=Q(x-X-[PQ]_+)Q^{-1}
\end{gather*}
as introduced in Section~\ref{sec:fusel2}. Alternatively we can first fuse $L_2(x-x_2)L_1(x-x_1)$, bring it to the canonical form, and then multiply by $L_3(x-x_3)$ from the left. This procedure results in
\begin{gather*}
 L_{3(21)}=\tilde Q^{-1}\big(x-X-[\tilde Q\tilde P]_+\big)\tilde Q
\end{gather*}
with
\begin{gather*}
\tilde Q=\left(\begin{matrix}
 1&0&0\\
 -\tilde q_{21}&1&0\\
 -\tilde q_{31}&-\tilde q_{32}&1
 \end{matrix}
\right) ,\qquad \tilde P=\left(\begin{matrix}
 0&\tilde p_{12}&\tilde p_{13}\\
 0&0&\tilde p_{23}\\
 0&0&0
 \end{matrix}
\right).
\end{gather*}
The formulas are related by the canonical transformation
\begin{gather}\label{eq:notcanonic}
 \tilde p_{12}=p_{12}-p_{13}q_{32} ,\qquad \tilde p_{23}=p_{23}-q_{21}p_{13} ,\qquad \tilde q_{31}=q_{31}+q_{32}q_{21}.
\end{gather}

In other words, the fusion procedure to obtain the Darboux coordinates is not associative, and the failure of the associativity is described by the symplectomorphism (\ref{eq:notcanonic}).

\subsection*{Acknowledgements}
We would like to thank Chris Elliott and Alexei Sevastyanov for multiple helpful discussions. We further thank Oleksandr Tsymbaliuk and the very helpful anonymous referees for comments on the manuscript. R.F.~is supported by the {\small IH\'{E}S} visitor program. The research of V.P.\ on this project has received funding from the European Research Council (ERC) under the European Union's Horizon 2020 research and innovation program (QUASIFT grant agreement 677368), V.P.~also acknowledges grant RFBR 16-02-01021.

\pdfbookmark[1]{References}{ref}
\LastPageEnding

\end{document}